%% file: 0-main.tex
\documentclass[11pt]{article}
\usepackage[normalem]{ulem}
\usepackage{yaonan}

\input{macros}

\title{Query-Sample Tradeoffs for Revenue Maximization\thanks{This paper is an updated version of \href{https://arxiv.org/pdf/2602.10483}{arXiv:2602.10483}.}}

\author{
Yaonan Jin\thanks{Hong Kong University of Science and Technology. Email: {\tt jinyaonan@ust.hk}}
\and
Wei Tang\thanks{Chinese University of Hong Kong. Email: {\tt weitang@cuhk.edu.hk}}
\and
Yifan Wang\thanks{Google Research. Email: {\tt litcwyf@google.com}}
\and
Mengxiao Zhang\thanks{University of Iowa. Email: {\tt mengxiao-zhang@uiowa.edu}}
}
\date{}

\begin{document}

\maketitle

\thispagestyle{empty}

\begin{abstract}
\wtedit{We introduce and study the full query-sample tradeoff for revenue maximization with a single buyer whose value follows an unknown distribution. A learner may adaptively combine value samples, which reveal full buyer values, with pricing queries, which reveal only whether a fresh buyer accepts a selected price. The goal is to learn a price that achieves a \((1-\varepsilon)\)-multiplicative approximation to optimal revenue with high probability.  We ask when, and by how much, reducing the sample budget increases the number of pricing queries required.

We establish tight tradeoff bounds for monotone-hazard-rate (MHR), regular, and quasi-monotone-hazard-rate (quasi-MHR) distributions, and almost tight bounds for quasi-regular distributions. Our results reveal sharply different benefits from combining the two information sources. Quasi-MHR distributions admit a smooth intermediate tradeoff: additional samples can reduce the query requirement by polynomial factors. For MHR and regular distributions, however, a sufficiently small constant fraction of the sample-only requirement still leaves the learner facing the full asymptotic query cost attainable with one sample. Our upper bounds are achieved by algorithms that collect all value samples before making queries, while our lower bounds hold even under arbitrary adaptive interleaving of the two forms of access. Along the way, we also fully characterize the sample-only and single-sample query complexities for all four families, eliminating the extra logarithmic dependence on \(1/\eps\) in the literature.
}
\wtdelete{We study the PAC learnability of revenue maximization for a single buyer with an unknown value distribution, under both sample access and adaptive pricing-query access. The goal is to post a price that achieves a $(1 - \varepsilon)$-multiplicative approximation to the optimal revenue with probability $(1 - \delta)$.

Up to constant factors, we establish a tight characterization of the sample complexity, the (single-sample) query complexity, and the full query-sample tradeoff for the canonical families of \textit{$\lambda$-regular} and \textit{$\lambda$-quasi-regular} distributions ($0 \le \lambda \le 1$), including the four endpoint families: \textit{monotone-hazard-rate (MHR)}, \textit{regular}, \textit{quasi-monotone-hazard-rate (quasi-MHR)}, and \textit{quasi-regular} \cite{BMP63, M81, CR14, HR14, FJ25}. Our results tighten the known sample- and query-complexity bounds and, through the new perspective on the query-sample tradeoff, bridge the two previously parallel lines of research. \mz{not mentioning $\lambda$}}
\end{abstract}

\newpage
\setcounter{tocdepth}{2}
{\hypersetup{linkcolor=black}\tableofcontents}
\thispagestyle{empty}
\newpage
\setcounter{page}{1}

\input{Paper/intro}

\section{Preliminaries}
\label{sec:prelim}

\input{Paper/prelim-new}

\input{quasi-regular}

\input{regular}

\input{quasi-MHR}

\input{MHR}

\appendix
\newpage
\section{A Multiplicative-Accuracy DKW Inequality}
\label{sec:DKW}

\input{Paper/apx-DKW}

\section{Upper Bounds for Sample Complexity}
\label{sec:sample-complexity}
\input{Paper/apx-sample-UB}

\input{impossibility}

\paragraph{AI Methodology Statement.} The technical overview section of the paper (\Cref{subsec:overview}), which contains most of the main ideas of this paper, was developed and written by the authors. Based on this overview, the authors used Opus 5.5 to generate an initial draft of the proofs. The authors then revised and polished the proofs with assistance from GPT Astra, Gemini 4 Argon, and Opus 5.5. During this revision process, the AI tools were used primarily to improve the structure of the proofs and the clarity of the exposition.

\begin{flushleft}
\bibliographystyle{alphaurl}
\bibliography{Paper/main-new}
\end{flushleft}

\end{document}

%% file: macros.tex
\newcommand{\yjnote}[1]{\footnote{{\color{BurntOrange}Yaonan}: {#1}}}
\newcommand{\yj}[1]{{\color{BurntOrange} {{\bf Yaonan}: #1}}}

\newcommand{\mz}[1]{{\color{OliveGreen} {{\bf Mengxiao}: #1}}}

\usepackage[suppress]{color-edits}
\addauthor{wt}{red}

\renewcommand{\iff}{\Leftrightarrow}
\renewcommand{\implies}{\Rightarrow}

\newcommand{\DKL}{\mathsf{KL}}

\newcommand{\tTheta}{\Tilde{\Theta}}

\newcommand{\Rquantile}{R_{\mathrm{qt}}}
\newcommand{\Rprice}{R_{\mathrm{pr}}}

\newcommand{\Pcandidate}{\Hat{P_{\mathrm{cd}}}}

\newcommand{\SingleBuyerRevenueMaximization}{\textsc{Single-Buyer Revenue Maximization}}
\newcommand{\NoisyPriceLocation}{\textsc{Noisy Price Location}}
\newcommand{\NoisyBinarySearch}{\textsc{Noisy Binary Search}}
\newcommand{\NoisyConcaveOptimization}{\textsc{Noisy Concave Optimization}}
\newcommand{\BestArmIdentification}{\textsc{Best Arm Identification}}

\newcommand{\intlower}{a}
\newcommand{\intupper}{b}
\newcommand{\indexnum}{n}
\newcommand{\coordidx}{i}
\newcommand{\samplespace}{\Omega}
\newcommand{\outcome}{\omega}
\newcommand{\sigmaalgebra}{\calF}
\newcommand{\probmeasure}{\mathbb{P}}
\newcommand{\expectop}{\mathbb{E}}
\newcommand{\randvec}{X}
\newcommand{\randvar}{X}
\newcommand{\measurableset}{A}
\newcommand{\prior}{F}
\newcommand{\price}{p}
\newcommand{\val}{v}
\newcommand{\quantile}{q}
\newcommand{\cumhazard}{H}
\newcommand{\density}{f}
\newcommand{\virtualval}{\phi}
\newcommand{\hazardrate}{h}
\newcommand{\regularity}{\lambda}
\newcommand{\quasiregularbound}{B}
\newcommand{\convexfunc}{g}
\newcommand{\optrevenue}{r^{*}}
\newcommand{\learnalg}{\calA}
\newcommand{\roundidx}{t}
\newcommand{\sampleval}{s}
\newcommand{\queryfeedback}{z}
\newcommand{\failureprob}{\delta}
\newcommand{\samplecount}{S}
\newcommand{\querycount}{Q}
\newcommand{\bernparamone}{x}
\newcommand{\bernparamtwo}{y}
\newcommand{\distidx}{i}
\newcommand{\priceset}{P}
\newcommand{\transcriptevent}{\calE}
\newcommand{\samplerounds}{\calS}
\newcommand{\queryrounds}{\calQ}
\newcommand{\history}{\calH}

\newcommand{\tradeoffparam}{c}

\newcommand{\prob}[2][]{\mathbb{P}\ifthenelse{\not\equal{}{#1}}{_{#1}}{}\!\left[{\def\givenn{\middle|}#2}\right]}
\newcommand{\expect}[2][]{\mathbb{E}\ifthenelse{\not\equal{}{#1}}{_{#1}}{}\!\left[{\def\givenn{\middle|}#2}\right]}

\newcommand{\empCDF}{\hat{F}}
\newcommand{\relprocess}{X}
\newcommand{\expprocess}{Z}
\newcommand{\firstcrossing}{A}
\newcommand{\bernindicator}{B}
\newcommand{\centeredindicator}{Y}
\newcommand{\salequantile}{\Qsale^{-1}}

\newcommand{\Qsale}{Q}
\newcommand{\empQsale}{\hat{\Qsale}}
\newcommand{\empRprice}{\hat{R}_{\mathrm{pr}}}
\newcommand{\empprice}{\hat{p}}
\newcommand{\lowerprice}{\tilde{p}}

\newcommand{\monopolyprice}{p^{*}}
\newcommand{\monopolyquantile}{q^{*}}
\newcommand{\revenueloss}{\Delta}
\newcommand{\lossthreshold}{s}

%% file: Paper/intro.tex
\section{Introduction}
\label{sec:intro}
\wtdelete{First steps:
\begin{enumerate}
    \item Overview (only statement).
    \item Blocks in Sec 3 (only statement).
    \item ...
\end{enumerate}

Every section:

\begin{itemize}
    \item Section 2: Overview of the three parts technique: (1) the left-top corner; (2) the sample complexity of in the right bottom corner; (3) the lower bound.
    \item Section 3: DKW (a potential black box) there with the white box deferred to the appendix.
    \item Section 4: [Guiding subsection at the beginning of each section below.] Quasi-Regular Distribution [Query Upper Bound]; Most general one; introduce the building block that can also apply to the later sections (other distributions); Comment needs to emphasize that the common component is not the dominating term in the query/sample complexity (even for later distributions). S
    \item Section 5:
    \item Section 6: 
    \item Section 7: $\lambda$-Regular: $\eps^{-1.5}$; stronger than DKW; combine MHR with the DKW analysis.
    \item Appendix: Sample Complexity
    \begin{itemize}
        \item DKW
        \item Quasi-Regular Distributions and lambda-Quasi-Regular Distributions $\lambda\in[0,1)$.
        \item  lambda-Regular Distributions $\lambda\in[0,1)$
    \end{itemize}
\end{itemize}

\wtcomment{Questions that we want to address: 
\begin{itemize}
    \item The full query-sample tradeoff: 
    Suppose we have \(S\) full samples and then use \(Q\) adaptive queries. How does reducing \(S\) force us to increase \(Q\)?
\end{itemize}
Messages that we want to deliver
\begin{itemize}

    \item \textbf{MHR: essentially a threshold, not a smooth tradeoff}
    \begin{itemize}
        \item The sample optimum is
        \[
        S^*=\Theta\!\left(\epsilon^{-3/2}\log\frac{1}{\delta}\right),
        \]
        whereas the single-sample query optimum is
        \[
        Q^*=\Theta\!\left(\epsilon^{-2}\log\frac{1}{\delta}\right).
        \]

        \item Samples and queries are \textbf{not smoothly substitutable}: 
        with a sufficiently small constant fraction of $S^*$ samples, one still needs
        $\Omega(Q^*)$ queries.
        Thus, unless there are enough samples to solve the problem statistically,
        one essentially has to pay the full active-learning cost.
    \end{itemize}

    \item \textbf{Regular: the two information sources have the same standalone complexity, but still exhibit a threshold}
    \begin{itemize}
        \item Both the sample optimum and the single-sample query optimum are
        \[
        S^*=Q^*
        =\Theta\!\left(\epsilon^{-3}\log\frac{1}{\delta}\right).
        \]

        \item Nevertheless, combining the two sources does not lead to a smooth improvement:
        the sample and query lower bounds hold simultaneously.
        A subcritical amount of one information source still requires essentially
        the full amount of the other.
    \end{itemize}

    \item \textbf{Quasi-MHR: a genuine and smooth sample--query tradeoff}
    \begin{itemize}
        \item The two endpoints are
        \[
        S^*=\Theta\!\left(\epsilon^{-2}\log\frac{1}{\delta}\right),
        \qquad
        Q^*=\Theta\!\left(\epsilon^{-3}\log\frac{1}{\delta}\right).
        \]

        \item More generally, for every $c\in[0,1]$, a sample budget
        \[
        S=\Theta\!\left(
        \epsilon^{-(2-c)}\log\frac{1}{\delta}
        \right)
        \]
        corresponds to the tight query requirement
        \[
        Q=\Theta\!\left(
        \epsilon^{-(2+c)}\log\frac{1}{\delta}
        \right).
        \]

        \item Thus, unlike MHR and regular distributions,
        samples and queries are \textbf{smoothly substitutable}.
        Ignoring the $\log(1/\delta)$ factor, the two polynomial exponents add up to $4$:
        every polynomial reduction in the query burden requires a corresponding
        polynomial increase in the number of samples.
    \end{itemize}

    \item \textbf{Quasi-regular: almost no useful polynomial interpolation}
    \begin{itemize}
        \item The sample optimum is
        \[
        S^*=\Theta\!\left(\epsilon^{-3}\log\frac{1}{\delta}\right),
        \]
        while the single-sample query optimum is
        \[
        Q^*=\Theta\!\left(
        \epsilon^{-3}\log\frac{1}{\epsilon}\log\frac{1}{\delta}
        \right).
        \]

        \item If the number of samples is reduced from $S^*$ by any polynomial factor,
        for example to
        \[
        O\!\left(
        \epsilon^{-(3-c)}\log\frac{1}{\delta}
        \right)
        \qquad \text{for any constant } c>0,
        \]
        the learner still needs
        \[
        \Omega\!\left(
        \epsilon^{-3}\log\frac{1}{\epsilon}\log\frac{1}{\delta}
        \right)
        \]
        queries.

        \item Hence, a polynomially large number of additional samples does essentially
        nothing to reduce the query complexity until the sample budget approaches
        the standalone sample complexity.
    \end{itemize}
\end{itemize}
}}A central question in learning in mechanism design is how much information about buyers' values a seller needs to obtain approximately optimal revenue. 
We study this question in {\SingleBuyerRevenueMaximization} \cite{M81}, where a seller offers an indivisible item to a single buyer whose value follows an unknown prior value distribution on $\left[0,\infty\right)$, whose cumulative density function is denoted by $\prior$. The optimal expected revenue under full knowledge of the prior is $\optrevenue\triangleq\sup\nolimits_{\price\geq0}\price\left(1-\prior\left(\price\right)\right)$, which we assume is finite. We adopt the probably approximately correct (PAC) learning framework \cite{V84}. For an accuracy parameter $\eps\in\left(0,1\right)$ and a failure probability $\failureprob\in\left(0,1\right)$, a learning algorithm $\learnalg$ must output a price $\price_{\learnalg}$ satisfying
\begin{align*}
    \prob[\prior,\learnalg]{\price_{\learnalg}\left(1-\prior\left(\price_{\learnalg}\right)\right)\geq\left(1-\eps\right)\optrevenue}
    &\geq1-\failureprob~,
\end{align*}
where the probability is over the observations available to the algorithm and its internal randomness. 
We consider two forms of information access:
\begin{itemize}
    \item \textbf{Sample access.}
    The algorithm observes an independent value sample $\val_{\roundidx}\sim\prior$ at each sample round. Each observation reveals the buyer's full value, providing passive information about the unknown prior.

    \item \textbf{Pricing-query access.}
    The algorithm selects a price $\price_{\roundidx}$ and observes only the purchase outcome $\indicator\left[\val_{\roundidx}\geq\price_{\roundidx}\right]$, where the buyer's value $\val_{\roundidx}\sim\prior$ is drawn independently for each query. The query price $\price_{\roundidx}$ may depend on all previously observed samples, query prices, and purchase outcomes. This access model therefore permits adaptive exploration.
\end{itemize}

A full value sample determines whether a sale would occur at every price, whereas a pricing query reveals the outcome at one selected price. 
Although samples provide richer information, full value observations may be costly to obtain, and a learner may wish to supplement an initial collection of samples with pricing experiments. 
There has been an extensive literature studying the sample complexity of revenue maximization (see, e.g., \cite{BBHM08,DRY15,HMR18,FJ25}), 
and an active parallel line studying its pricing-query complexity (see, e.g., \cite{LSTW23ec, LSTW23soda, GKSW24, SW24, TW25, CJLZ26www}).\footnote{We also note that previous sample-complexity upper bounds leave a gap on an extra logarithmic dependence on $1/\eps$ \cite{DRY15, HMR18, FJ25}. Such logarithmic gaps also remain between the best-known pricing-query upper and lower bounds \cite{LSTW23ec,LSTW23soda,SW24,CJLZ26www}. 
Moreover, the cited pricing-query results concern $\eps$-additive approximation for distributions supported on \([0,1]\), so they do not directly characterize the query endpoint under our multiplicative PAC objective.}
Despite substantial progress in both directions, the inherent tradeoff between these two forms of information access remains much less understood.
\wtedit{The separate complexity bounds do not determine how value samples and pricing queries can substitute for one another when a learner may adaptively combine the two forms of access. 
We study this interaction through the following question:}

\begin{quote}
\textbf{The full query-sample tradeoff.}
{\em \wtedit{A learner may adaptively interleave value samples and pricing queries, and can choose the access type and each query price based on previous observations. 
Which combinations of the sample budget \(\samplecount\) and query budget \(\querycount\) suffice to learn a \((1-\eps)\)-approximately optimal price with probability \(1-\failureprob\)? In particular, when, and by how much, must the query budget \(\querycount\) increase as the sample budget \(\samplecount\) decreases?}
}
\end{quote}
A characterization of this tradeoff must explain how the two information sources interact. 
Additional samples might steadily reduce the number of queries needed, or they might provide little asymptotic benefit until the sample budget approaches the amount needed to learn without queries. 
We consider the multiplicative objective throughout this work.
We first show that pricing-query access alone cannot provide a multiplicative PAC guarantee with any finite, distribution-independent query budget, \wtedit{even for a very ``well-behaved'' distribution like a uniform distribution supported within $\left[0,1\right]$ (\Cref{sec:impossibility})}.
\wtedit{This impossibility shows the learner's initial uncertainty about the scale of buyer values. For the distribution families studied here, our algorithms overcome this obstacle using a single value sample to initialize scale estimation, after which pricing queries would suffice for PAC learning. 
We call the corresponding query requirement the {\em single-sample query complexity}. This model provides a natural endpoint with the fewest value samples, while the sample-only model provides the other endpoint with no pricing queries.}

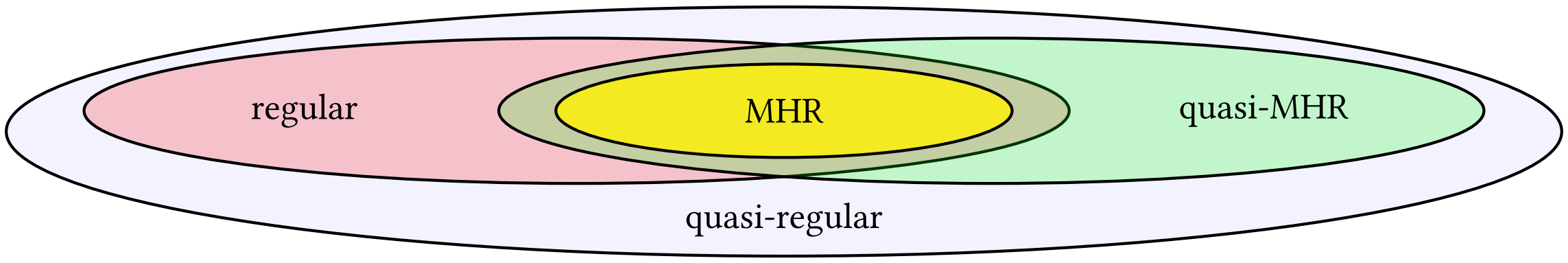
\begin{figure}[t]
    \centering\small
    \begin{tikzpicture}
        \draw[thick, fill = blue, fill opacity = 0.05]
            (0, -0.2) ellipse (7.5cm and 1.2cm);
        \draw[thick, fill = red, fill opacity = 0.2]
            (-2, 0) ellipse (4.75cm and 0.7cm);
        \draw[thick, fill = green, fill opacity = 0.2]
            (2, 0) ellipse (4.75cm and 0.7cm);
        \draw[thick, fill = yellow, fill opacity = 0.8]
            (0, 0) ellipse (2.2cm and 0.45cm);
        \node at (0, -1.05) {quasi-regular};
        \node at (-4.625, 0) {regular};
        \node at (4.625, 0) {quasi-MHR};
        \node at (0, 0) {MHR};
    \end{tikzpicture}
    \caption{\label{fig:distribution-hierarchy}
    A Venn diagram of the four distribution families:
    regular,
    MHR,
    quasi-regular, and
    quasi-MHR.}
\end{figure}

\wtedit{Without distributional assumptions on the prior $\prior$, PAC learning for {\SingleBuyerRevenueMaximization} is impossible, even when both value samples and pricing queries are available \cite[Thm.~4.6]{HMR18}. We therefore study four distribution families: monotone-hazard-rate (MHR) distributions \cite{BMP63}, regular distributions \cite{M81}, quasi-monotone-hazard-rate (quasi-MHR) distributions \cite{BEM66,FJ25}, and quasi-regular distributions \cite{HR14,FJ25}. The latter two families relax MHR and regularity, respectively. 
Every MHR distribution is both regular and quasi-MHR, and both regular and quasi-MHR distributions are quasi-regular. \Cref{fig:distribution-hierarchy} illustrates these relationships, and \Cref{subsec:prob} provides the formal definitions.}
Our results show that the value of combining samples and queries depends sharply on the distribution family. Quasi-MHR distributions admit a smooth intermediate tradeoff: increasing the sample budget can substantially reduce the query requirement. For MHR and regular distributions, however, a sufficiently small constant fraction of the sample-only requirement still leaves the learner facing the full asymptotic single-sample query cost. We establish tight tradeoff bounds for these three families and almost tight bounds for quasi-regular distributions. 
\wtedit{Our upper bounds are achieved by algorithms that collect their value samples before making pricing queries, while our lower bounds hold even when the learner can arbitrarily and adaptively interleave the two forms of access.}
Along the way, we also characterize the sample-only and single-sample query complexities up to constant factors for all four families, which closes the remaining logarithmic gaps in the literature.
\wtdelete{Over the past two decades, \textit{learning in mechanism design} has become a core subfield at the intersection of computer science, economics, and statistics, for which the dominant framework is \textit{probably approximately correct (PAC) learning} \cite{V84}. In the most basic {\SingleBuyerRevenueMaximization} problem \cite{M81}, a seller aims to sell an indivisible item to a single buyer whose value is drawn from a prior $F$. If the seller had full knowledge of $F$, he/she would post the revenue-optimal price $p^{*} \eqdef \argmax_{p > 0} p \cdot (1 - F(p))$ to the buyer. When $F$ is initially unknown, the PAC learning framework asks the following question: given $\eps, \delta \in (0, 1)$, how much information about $F$ is sufficient and necessary for an algorithm $\calA$ to learn an $(1 - \eps)$-approximately optimal price $p^{\calA}$ with probability $(1 - \delta)$?
\begin{align*}
    \Pr_{F, \calA}\bigl[p^{\calA} \cdot (1 - F(p^{\calA})) \ge (1 - \eps) \cdot r^{*}\bigr] ~\ge~ 1 - \delta,
\end{align*}
where $r^{*} \eqdef p^{*} \cdot (1 - F(p^{*})) = \max_{p > 0}  p \cdot (1 - F(p))$ is the optimal revenue\ignore{ and the probability is taken over the observations generated from $F$ and any internal randomness of $\calA$}.
There are two canonical types of information access, and the literature has correspondingly developed along two lines:
\begin{enumerate}
    \item \textbf{Sample Access:}
    At each time $t = 1, 2, \dots$, the algorithm $\calA$ receives an independent sample $v_{t} \sim F$.

    Such algorithms have no control over the information-generation process and fall into the \textit{passive learning} category. Without loss of generality, we can concentrate on \textit{nonadaptive} algorithms that first collect a prescribed number of samples in a batch and then compute a desired price $p^{\calA}$.

    \item \textbf{(Pricing) Query Access:}
    At each time $t = 1, 2, \dots$, the algorithm $\calA$ determines a query price $p_{t}$, based on all knowledge thus far, and only observes the sold/unsold feedback $\indicator[v_{t} \ge p_{t}] \in \{0, 1\}$.\footnote{More rigorously, this is the \textit{adaptive query} model. As all previous works in this line \cite{LSTW23ec, LSTW23soda, GKSW24, SW24, TW25, CJLZ26www} allow (arbitrary) adaptivity, we adopt the same convention. Arguably, this is the more natural formulation, given the \textit{online learning} nature of the problem.}

    Such algorithms do have control over the information-generation process and fall into the category of \textit{active learning}. Crucially, they can operate \textit{adaptively}---each query price $p_{t}$ may depend on the entire history of previous queries and their resulting feedback.
\end{enumerate}
(Information-theoretically, ``expensive'' sample access $v_{t}$ directly implies ``cheap'' query access $\indicator[v_{t} \ge p_{t}]$.)
These distinctions have caused the literature to split into two parallel lines. Despite the vast body of work on each front, however, a fundamental unifying question---as well as its two classical endpoints---has not been systematically resolved:
\begin{enumerate}
    \item \textbf{Query-Sample Tradeoffs:}
    While previous research has considered a single type of information access---either samples or queries---their combination naturally arises in both theory and practice.
    For instance, an algorithm $\calA$ may first use expensive samples to obtain a ``coarse-grained'' sketch of the prior $F$, and then use cheap queries to locate a ``fine-grained'' price $p^{\calA}$.
    Accordingly, a complete characterization of PAC learnability requires a unified understanding of how these two qualitatively different information sources can substitute for one another, i.e., the \textit{query-sample tradeoffs}.
    
    \item \textbf{Sample Complexity Alone:}
    Even if we restrict attention to sample complexity, there are long-standing polylogarithmic gaps between the best-known upper and lower bounds even for the above {\SingleBuyerRevenueMaximization} problem \cite{BBHM08, DRY15, HMR18, FJ25}. Hence, the PAC learnability of this most basic setting is not yet fully understood from the sample-complexity perspective alone.

    \item \textbf{Query Complexity Alone:}
    The situation is similar for query complexity---persistent gaps remain between the best-known upper and lower bounds \cite{LSTW23soda, LSTW23ec, SW24, CJLZ26www}. More importantly, due to the nature of (pricing) query access, previous works assume that the prior $F$ is $[0, 1]$-supported and relax the PAC learning requirement from a $(1 - \eps)$-multiplicative approximation to an $\eps$-additive approximation. Although standard, these simplifications fundamentally alter the problem's intrinsic scaling and preclude a direct comparison with the sample-based setting.
\end{enumerate}
In this work, we address these issues and prove tight bounds, up to \textbf{constant factors}, on the sample complexity, the query complexity, and the full query-sample tradeoffs, thus providing a unified characterization of the PAC learnability of {\SingleBuyerRevenueMaximization}.}

\subsection{PAC Learnability of Single-Buyer Revenue Maximization}
\label{subsec:result}

\input{Paper/contributions-new}

\subsection{Technical Overview}
\label{subsec:overview}

To fully characterize the query-sample tradeoffs, we have developed an arsenal of technical ingredients that are of independent interest and should find applications in \textit{learning in mechanism design} and other related fields. Below, we highlight the most significant ones, mainly for constructing the complexity upper bounds, and their underlying intuitions. For readability, we have hidden many minor technical details; for example, some parameters given below differ slightly from their actual values in the formal version of the proof.

\input{Paper/tech-overview-quasiregular}

\paragraph{Query-Sample Tradeoff of the Quasi-MHR Family.}
Recall \Cref{fig:result}---provided all results thus far, we are left with only one upper bound: under \uline{quasi-MHR} and a specific number of samples $\Theta\bigl(\frac{1}{\eps^{2 - c}}\log(\frac{1}{\delta})\bigr)$ for $c \in [0, 1]$, how many queries are sufficient. A query-optimal algorithm for this must be a substantial extension---aligned with the current sample access $\Theta\bigl(\frac{1}{\eps^{2 - c}}\log(\frac{1}{\delta})\bigr)$---of the one described in \Cref{rmk:overview:quasi-MHR}. It turns out that $O\bigl(\frac{1}{\eps^{2 + c}}\log(\frac{1}{\delta})\bigr)$ queries are sufficient, by modifying Module~\ref{overview:quasi-regular:2} into Module~\ref{overview:quasi-MHR:2}:
\begin{enumerate}[label = C\arabic*., ref = C\arabic*]
\setcounter{enumi}{1}
    \item[C\textsuperscript{$\dagger$}.] \textit{Running Module~\ref{overview:quasi-regular:1} with $\Theta\bigl(\log(\frac{1}{\delta})\bigr)$ queries, and obtain $\Hat{r^{*}} \in [r^{*}, \frac{11}{10}r^{*}]$.}
    
    \item\label{overview:quasi-MHR:2}
    \textit{Further sparsifying the candidate price set $\Pcandidate$ using samples:}
    \\
    Here, we aim to use the $\Theta\bigl(\frac{1}{\eps^{2 - c}}\log(\frac{1}{\delta})\bigr)$ samples to construct a size‑$\Theta\bigl(\frac{1}{\eps^{c}}\bigr)$ candidate set $\Pcandidate$ that, with probability $(1 - \delta)$, contains a $(1 - \eps)$-approximately optimal price.\footnote{As a sanity check, when $c = 0$, we already have sufficient samples $\Theta\bigl(\frac{1}{\eps^{2}}\log(\frac{1}{\delta})\bigr)$ to learn a $(1 - \eps)$-approximately optimal price $\Hat{p^{*}}$; we may then simply take the singleton $\Pcandidate = \{\Hat{p^{*}}\}$.}
    (This further sparsifies the size-$\Theta\bigl(\frac{1}{\eps}\bigr)$ grid used in Module~\ref{overview:quasi-regular:2}.)
    The quasi-MHR condition implies that the monopoly price $p^{*}$ lies in $[r^{*}, e r^{*}] \subseteq [\underline{p}, \Bar{p}] \eqdef [\frac{10}{11}\Hat{r^{*}}, e\Hat{r^{*}}]$.
    We divide $[\underline{p}, \Bar{p}]$ evenly into $\Theta\bigl(\frac{1}{\eps^{c}}\bigr)$ subintervals. Inside each subinterval $1 \le j \le \Theta\bigl(\frac{1}{\eps^{c}}\bigr)$, we place a finer-grained geometric $(1 + \Theta(\eps))$-grid and choose the grid price $\Hat{p_{j}}$ that maximizes the \textit{empirical revenue} $\Hat{\Rprice}(p)$ (evaluated on the samples). The set of all $\Hat{p_{j}}$'s constitutes $\Pcandidate$, which directly satisfies $|\Pcandidate| = \Theta\bigl(\frac{1}{\eps^{c}}\bigr)$.
    Further, a technically involved probabilistic argument yields a \textit{uniform convergence} result: on any specific subinterval $j$, the empirical revenue $\Hat{\Rprice}(p)$ is uniformly close to the true revenue $\Rprice(p)$ over that subinterval, with high probability.
    So, for the particular subinterval $j^{*}$ containing $p^{*}$, the chosen grid price $\Hat{p_{j^{*}}}$ is sufficiently near‑optimal. That is, our new set $\Pcandidate$ again fulfills \eqref{eq:overview:quasi-MHR:2} with probability $(1 - \delta)$:
    \begin{align}
        \textstyle
        \max_{p \in \Pcandidate} \Rprice(p)
        ~\ge~ (1 - \eps) \cdot \bigl(\max_{p \in [\underline{p}, \Bar{p}]} \Rprice(p)\bigr)
        ~=~ (1 - \eps) \cdot r^{*}.
        \label{eq:overview:quasi-MHR:2}\tag{$\heartsuit$}
    \end{align}
    {Again, this further sparsification is in a similar spirit to previous work \cite{LSTW23ec} on ``the description complexity of distributions'' and should be of independent interest.}

    \item[C\textsuperscript{$\ddagger$}.] \textit{Adapting Modules~\ref{overview:quasi-regular:3} and \ref{overview:quasi-regular:4} in alignment with Module~\ref{overview:quasi-MHR:2}:}\\
    The adaptations are almost the same as those described in \Cref{rmk:overview:quasi-MHR}, except for a sparser candidate price set $|\Pcandidate| = \Theta\bigl(\frac{1}{\eps^{c}}\bigr)$, so the query complexity improves correspondingly to $O\bigl(\frac{1}{\eps^{2 + c}}\log(\frac{1}{\delta})\bigr)$.
\end{enumerate}

\paragraph{Sample complexity upper bounds: a multiplicative-accuracy DKW inequality, and more.}
Regarding the\ignore{ (no-query)} sample complexity, as the dependency on the $\poly(\eps^{-1})$ is well understood, our main contribution is to provide a tighter dependency on logarithmic terms. The previously best-known sample compelxity upper bounds are $O\bigr(\frac{1}{\eps^{3}}\log(\frac{1}{\eps \delta})\bigr)$ for the regular and quasi-regular families \cite[Thm~4.1]{DRY15} \cite[Thm~20]{FJ25}, $O\bigr(\frac{1}{\eps^{2}}\log(\frac{1}{\eps \delta})\bigr)$ for the quasi-MHR family \cite[Thm~20]{FJ25}, and $O\bigr(\frac{1}{\eps^{3 / 2}}\log(\frac{1}{\eps \delta})\bigr)$ for the MHR family \cite[Thm~3.2]{HMR18}.

First, we observe that it is simple to improve the \uline{quasi-MHR} upper bound to $O\bigr(\frac{1}{\eps^{2}}\log(\frac{1}{\delta})\bigr)$ [Thm~\ref{thm:MHR:UB-sample}]. Concretely, it is known that the monopoly quantile $q^{*}$ (i.e., the revenue-optimal sale probability) of a quasi-MHR distribution satisfies $q^{*} \ge \frac{1}{e}$ \cite[Lemma~3.6]{FJ25}. So, distinguishing between multiplicative vs.\ additive accuracies only affects the constant factors in the sample complexity, and the upper bound of $O\bigr(\frac{1}{\eps^{2}}\log(\frac{1}{\delta})\bigr)$ follows as a direct consequence of the standard \textit{additive-accuracy} DKW inequality \cite{M90}.

The circumstances of \uline{regular} and \uline{quasi-regular} distributions are entirely different. Now the monopoly quantile $q^{*}$ can be zero (and unique); also, it is possible that all $(1 - \eps)$-approximately optimal monopoly quantiles are at most $O(\eps)$. Consequently, the standard DKW inequality is inapplicable. Instead, previous works \cite{DRY15, FJ25} adopt the Chernoff bound or Bernstein's inequality, in combination with the union bound. Such methods, unfortunately, introduce a loss of $\log(\frac{1}{\eps})$ in the upper bounds.

To the rescue, we develop a \textit{multiplicative-accuracy} version of the DKW inequality [Thm~\ref{thm:multiplicative-DKW}]. Denote by $Q(x) \eqdef 1 - F(x)$ the quantile function: given any $q, \eps', \delta' \in (0, 1)$ and \uline{any} distribution, the empirical quantile function $\Hat{Q}(x)$ built from $O\bigl(\frac{1}{q \cdot \eps'^{2}} \log(\frac{1}{\delta'})\bigr)$ samples satisfies the \textit{uniform convergence} guarantee
\begin{align*}
    \textstyle
    \Pr\Bigl[ \sup_{x \le Q^{-1}(q)} \Bigl| \frac{\Hat{Q}(x)}{Q(x)} - 1 \Bigr| > \eps' \Bigr] ~\le~ \delta'.
\end{align*}
The proof first observes that $\Hat{Q}(x) / Q(x)$ is a reverse martingale. Doob's martingale inequality \cite{D19} then reduces the continuum supremum to a single tail event at the threshold quantile $q$, and a moment-generating-function-based argument completes the analysis.
Setting $q = \frac{\eps}{2}$, $\eps' = \frac{\eps}{2}$, and $\delta' = \delta$ results in a sample complexity of $O\bigr(\frac{1}{\eps^{3}}\log(\frac{1}{\delta})\bigr)$, which directly implies our upper bounds for \uline{regular} and \uline{quasi-regular} distributions [Thms~\ref{thm:quasi-regular:UB-sample} \& \ref{thm:regular:UB-sample}].

The circumstances of \uline{MHR} distributions are more severe. The new inequality alone is insufficient; the previous $O\bigr(\frac{1}{\eps^{3 / 2}}\log(\frac{1}{\eps \delta})\bigr)$ upper bound \cite[Thm~3.2]{HMR18} already polynomially exceeds the DKW scope, as a consequence of two salient MHR properties: (i)~The monopoly quantile is at least $q^{*} \ge \frac{1}{e}$, making a single-price estimation more sample-efficient. (ii)~The revenue-quantile curves $\Rquantile(q)$ are strongly concave, and we can filter prices more sample-efficiently. (Instead, the other three distribution families each violate at least one property.) While the previous work already leveraged both properties to achieve a polynomial improvement, their ``Chernoff bound plus union bound'' approach again incurs a loss of $\log(\frac{1}{\eps})$.

Our upper-bound proof incorporates both properties into the multiplicative-accuracy DKW analysis above. Concretely, except with negligible probability, property~(i) allows us to restrict attention to prices whose empirical quantiles are at least $\frac{2}{3e}$ (say); for any such price $p$, we also have $Q(p) \ge \frac{1}{3e}$. Moreover, property~(ii) implies that the revenue-quantile curve decreases quadratically $\Rquantile(q) \le (1 - \frac{1}{4} \cdot (q - q^{*})^{2}) \cdot r^{*}$ \cite[Lem~3.3]{HMR18}; thus, all $(1 - \eps)$-approximately optimal quantiles lie in $q \in [q^{*} \pm 2\sqrt{\eps}]$. By adapting the DKW analysis, we obtain the following \textit{local stability} result: given any $p \le Q^{-1}(\frac{1}{3e})$ and $\eta', \eps', \delta' \in (0, 1)$, the empirical quantile function $\Hat{Q}(p)$ built from $O\bigl(\bigl(\frac{\eta'}{\eps'^{2}} + \frac{1}{\eps'}\bigr)\log(\frac{1}{\delta'})\bigr)$ samples satisfies
\begin{align*}
    \textstyle
    \Pr\Bigl[ \sup_{x:\ |Q(x) - Q(p)| \le \eta'} \Bigl| \frac{\Hat{Q}(x)}{Q(x)} - \frac{\Hat{Q}(p)}{Q(p)} \Bigr| > \eps' \Bigr] ~\le~ \delta'.
\end{align*}
Setting $p = p^{*}$ (note that $p^{*} \le Q^{-1}(\frac{1}{e}) \le Q^{-1}(\frac{1}{3e})$), $\eta' = \Theta(\sqrt{\eps})$, $\eps' = \Theta(\eps)$, and $\delta' = \Theta(\delta)$ gives a sample complexity of $O\bigr(\frac{1}{\eps^{3 / 2}}\log(\frac{1}{\delta})\bigr)$. After a moderate amount of additional work, this translates into our upper bound for the MHR family [Thm~\ref{thm:MHR:UB-sample}].

Note that these upper bounds all admit matching lower bounds \cite[Thms~4.8 \& 4.12]{HMR18} [Thm~\ref{thm:quasi-MHR:LB-sample}]. Altogether, we now acquire a tight characterization of the sample complexity of revenue maximization.

\paragraph{Lower Bounds.}
Sample complexity lower bounds and pricing query complexity lower bounds for revenue maximization have been studied in~\cite{HMR18, GHZ19} and~\cite{LSTW23soda, TW25, CJLZ26www}, respectively.
These lower bounds follow a common template: one constructs hard distributions whose sets of $(1-\eps)$-optimal prices are disjoint, so that a learning algorithm must identify the distribution it faces, and then bounds the information that each value sample or pricing query reveals about this distribution.
Since our query-sample tradeoffs combine both kinds of feedback, we generalize the information argument of~\cite{CJLZ26www} from transcripts of pricing queries to transcripts that interleave value samples and pricing queries (\Cref{lem:kl-bound}).
We then construct tailor-made hard distributions for different distribution families, so that each lower bound matches the corresponding upper bound. {One subtlety is that~\cite{CJLZ26www} chooses the hard distribution \textit{uniformly} at random from their constructed class and proves that this distribution is hard to identify.}
In our quasi-regular construction, a pricing query is informative only about the hard distribution whose price interval contains its price, and the information it reveals grows linearly with the index of that interval.
Under the uniform random choice, an algorithm can therefore place its queries in the most informative intervals, and the resulting lower bound loses the $\log\frac1\eps$ factor.
We instead make the random choice \textit{nonuniform}: the $i$-th hard distribution is chosen with probability $\frac{1/i}{\sum_{i'\in[n]}1/i'}$ for $i\in[n]$ {and $n = \eps^{-\Omega(1)}$}.
Under these weights, every pricing query contributes the same weighted information regardless of the interval containing its price, and the normalization $\sum_{i'\in[n]}1/i'=\Theta(\log n)$ yields the $\log\frac1\eps$ factor.






\subsection{Further Related Works}
\label{subsec:related-work}

\input{Paper/related-work}

%% file: Paper/contributions-new.tex

We denote the numbers of value samples and pricing queries by $\samplecount$ and $\querycount$, respectively. The sample-only endpoint concerns learning without pricing queries. At the other endpoint, we study the \textit{single-sample query complexity}: the number of pricing queries needed after observing one value sample. This sample provides the scale information needed by our query algorithms; without any value samples, no finite query budget suffices for multiplicative PAC learning, even for MHR distributions [\Cref{thm:impossibility}].

\paragraph{Query-sample tradeoffs.}
Our main contribution is a characterization of how value samples and pricing queries can substitute for one another. We establish tight tradeoff bounds for MHR, regular, and quasi-MHR distributions, and almost tight bounds for quasi-regular distributions. These results reveal substantially different behavior across the four families, as illustrated in \Cref{fig:result}.

\afterpage{
\begin{figure}[t]
    \centering
    \subfloat[\label{fig:result-MHR}
    MHR]{
        \includegraphics[width = .48\textwidth]
        {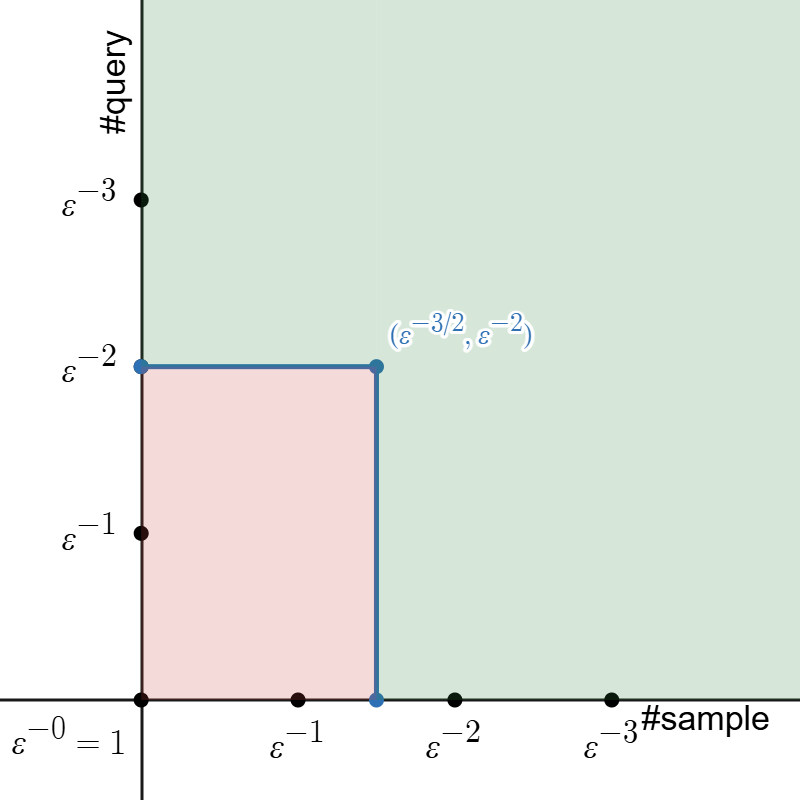}}
    \hfill
    \subfloat[\label{fig:result-regular}
    Regular]{
        \includegraphics[width = .48\textwidth]
        {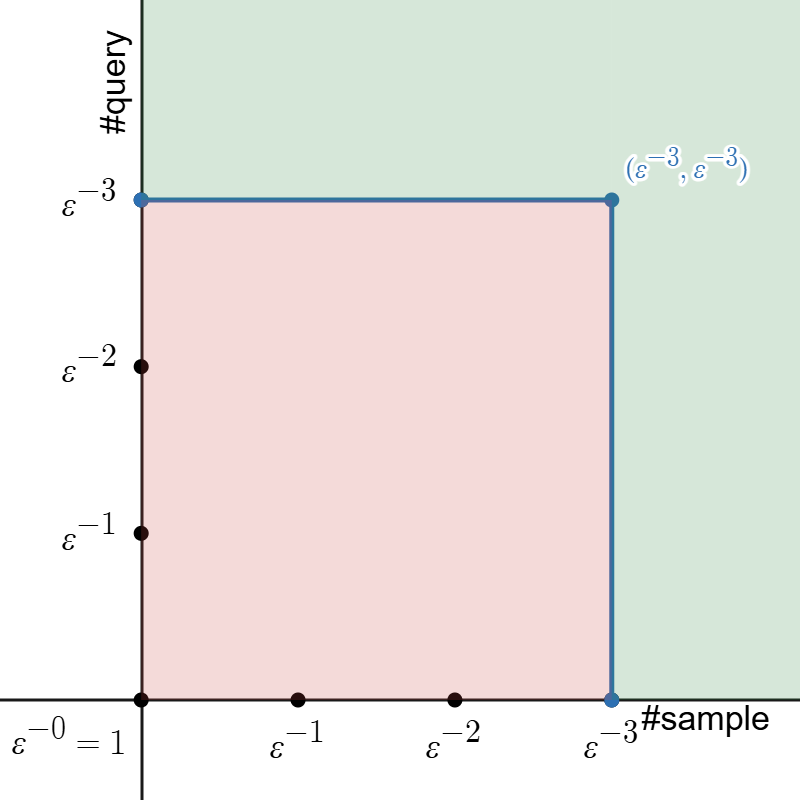}}
    \\
    \subfloat[\label{fig:result-quasi-MHR}
    Quasi-MHR]{
        \includegraphics[width = .48\textwidth]
        {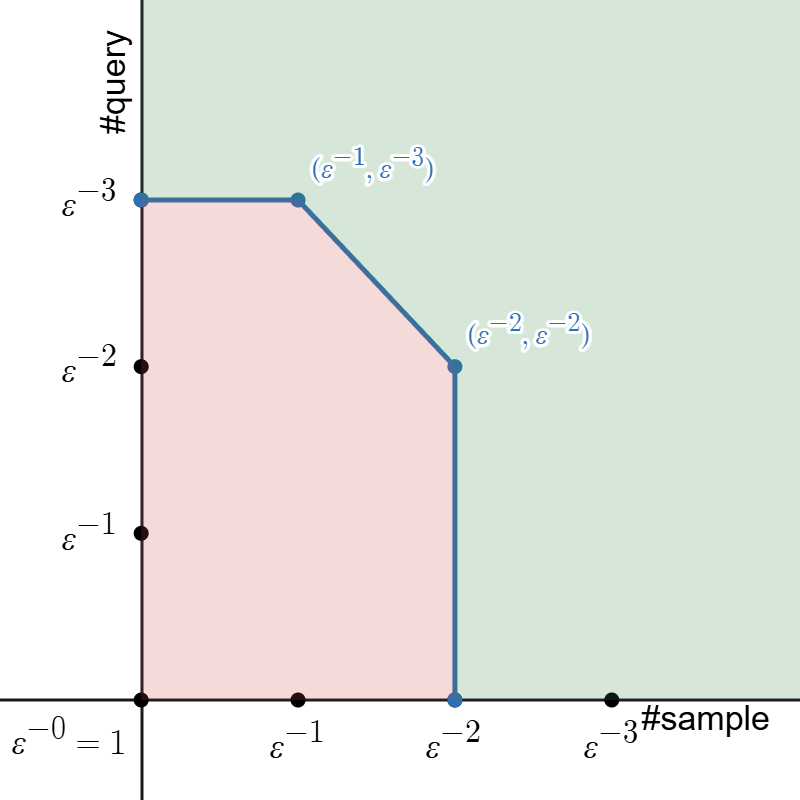}}
    \hfill
    \subfloat[\label{fig:result-quasi-regular}
    Quasi-regular]{
        \includegraphics[width = .48\textwidth]
        {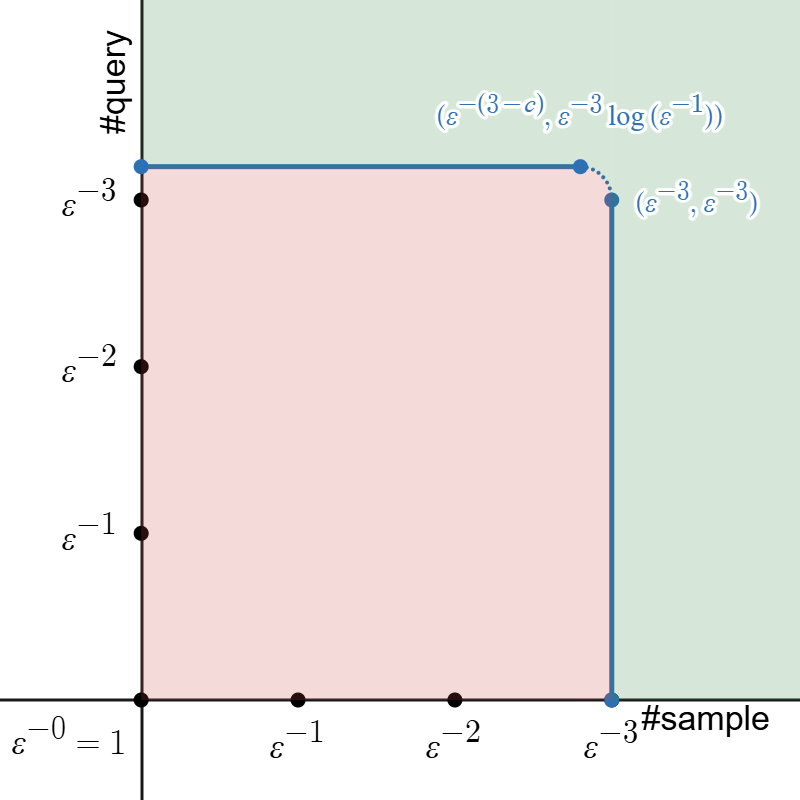}}
    \caption{Diagrams of \textbf{log-log plots} of the query-sample tradeoffs---the \textbf{red} regions refer to the \textbf{unlearnable} regimes and the \textbf{green} regions refer to the \textbf{learnable} regimes---for all considered families: 
    \Cref{fig:result-MHR} is for MHR; 
    \Cref{fig:result-regular} is for regular;
    \Cref{fig:result-quasi-MHR} is for quasi-MHR;
    and \Cref{fig:result-quasi-regular} is for quasi-regular. Constant factors and $\log (1/\delta)$ term are suppressed.
    \label{fig:result}}
\end{figure}
\clearpage}

The quasi-MHR family admits a smooth tradeoff between the two information sources (\Cref{fig:result-quasi-MHR}). For every fixed parameter $\tradeoffparam\in\left[0,1\right]$, we construct a PAC learning algorithm with sample and query budgets
\begin{align*}
    \samplecount=O\!\left(\eps^{-\left(2-\tradeoffparam\right)}\log\frac{1}{\failureprob}\right)~,
    \quad
    \querycount=O\!\left(\eps^{-\left(2+\tradeoffparam\right)}\log\frac{1}{\failureprob}\right)~.
\end{align*}
Conversely, any PAC learning algorithm using at most a sufficiently small constant multiple of $\eps^{-\left(2-\tradeoffparam\right)}\log\left(1/\failureprob\right)$ samples requires $\Omega\left(\eps^{-\left(2+\tradeoffparam\right)}\log\left(1/\failureprob\right)\right)$ queries on some quasi-MHR distribution [\Cref{thm:quasi-MHR:UB-tradeoff,thm:quasi-MHR:LB-tradeoff}]. Thus, throughout the intermediate regime, a polynomial increase in the sample budget permits a corresponding polynomial decrease in the query budget. The case $\tradeoffparam=1$ matches the single-sample query complexity, so even a sample budget of order $\eps^{-1}\log\left(1/\failureprob\right)$, with a sufficiently small constant, cannot improve that query rate. The case $\tradeoffparam=0$ reaches the sample-only scale: a sufficiently large constant multiple of $\eps^{-2}\log\left(1/\failureprob\right)$ samples eliminates the need for queries altogether.

The MHR and regular families exhibit a different behavior (\Cref{fig:result-MHR,fig:result-regular}). For MHR distributions, any PAC learning algorithm using at most a sufficiently small constant multiple of $\eps^{-3/2}\log\left(1/\failureprob\right)$ samples requires $\Omega\left(\eps^{-2}\log\left(1/\failureprob\right)\right)$ queries [\Cref{thm:MHR:LB-tradeoff}]. For regular distributions, a sample budget of at most a sufficiently small constant multiple of $\eps^{-3}\log\left(1/\failureprob\right)$ still requires $\Omega\left(\eps^{-3}\log\left(1/\failureprob\right)\right)$ queries [\Cref{thm:regular:LB-tradeoff}]. These query lower bounds match the corresponding single-sample query complexities. Hence, a sufficiently small constant fraction of the sample-only requirement leaves the learner facing the full asymptotic query cost.

The quasi-regular family exhibits a similar limitation over a broad range of sample budgets (\Cref{fig:result-quasi-regular}). For every fixed constant $\tradeoffparam>0$, any PAC learning algorithm using at most a sufficiently small constant multiple of $\eps^{-\left(3-\tradeoffparam\right)}\log\left(1/\failureprob\right)$ samples requires $\Omega\left(\eps^{-3}\log\left(1/\eps\right)\log\left(1/\failureprob\right)\right)$ queries, where the hidden constant may depend on $\tradeoffparam$ [\Cref{thm:quasi-regular:LB-tradeoff}]. This query bound matches the single-sample query complexity. Thus, whenever the sample budget falls below the sample-only complexity by a fixed polynomial factor, additional samples do not reduce the asymptotic query requirement.

\wtedit{En route, our characterization also determines the optimal complexity at the two endpoints: learning from value samples alone and learning from pricing queries after one initial sample. We next present these endpoint bounds and their connections to prior work.}

\paragraph{Sample complexity alone.}
\wtedit{We establish tight sample complexity bounds for all four distribution families, closing the gaps between the previously known upper and lower bounds. 
The following table summarizes these improvements.}

\begin{center}
\small
\setlength{\tabcolsep}{5pt}
\renewcommand{\arraystretch}{1.35}
\begin{tabular}{@{}lccc@{}}
\hline
Distribution family & Previous upper bound & Previous lower bound & Our tight bound \\
\hline
MHR
& \shortstack{$O\left(\eps^{-3/2}\log\left(1/\left(\eps\failureprob\right)\right)\right)$\\\cite{HMR18}}
& \shortstack{$\Omega\left(\eps^{-3/2}\log\left(1/\failureprob\right)\right)$\\\cite{HMR18}}
& \shortstack{$\Theta\left(\eps^{-3/2}\log\left(1/\failureprob\right)\right)$\\{[\Cref{thm:MHR:UB-sample}]}} \\
Regular
& \shortstack{$O\left(\eps^{-3}\log\left(1/\left(\eps\failureprob\right)\right)\right)$\\\cite{DRY15}}
& \shortstack{$\Omega\left(\eps^{-3}\log\left(1/\failureprob\right)\right)$\\\cite{HMR18}}
& \shortstack{$\Theta\left(\eps^{-3}\log\left(1/\failureprob\right)\right)$\\{[\Cref{thm:regular:UB-sample}]}} \\
Quasi-MHR
& \shortstack{$O\left(\eps^{-2}\log\left(1/\left(\eps\failureprob\right)\right)\right)$\\\cite{FJ25}}
& \shortstack{$\Omega\left(\eps^{-3/2}\log\left(1/\failureprob\right)\right)$\\{[by inclusion]}}
& \shortstack{$\Theta\left(\eps^{-2}\log\left(1/\failureprob\right)\right)$\\{[\Cref{thm:quasi-MHR:LB-sample,thm:quasi-MHR:UB-sample}]}} \\
Quasi-regular
& \shortstack{$O\left(\eps^{-3}\log\left(1/\left(\eps\failureprob\right)\right)\right)$\\\cite{FJ25}}
& \shortstack{$\Omega\left(\eps^{-3}\log\left(1/\failureprob\right)\right)$\\{[by inclusion]}}
& \shortstack{$\Theta\left(\eps^{-3}\log\left(1/\failureprob\right)\right)$\\{[\Cref{thm:quasi-regular:UB-sample}]}} \\
\hline
\end{tabular}
\end{center}

Our upper bounds replace the logarithmic factor $\log\left(1/\left(\eps\failureprob\right)\right)$ with $\log\left(1/\failureprob\right)$ in all four families \cite[Thm.~4.1]{DRY15} \cite[Thm.~3.2]{HMR18} \cite[Thm.~20]{FJ25}. These bounds match the existing lower bounds for MHR, regular, and quasi-regular distributions \cite[Thms.~4.8 and~4.12]{HMR18}. For quasi-MHR distributions, we establish a new lower bound of $\Omega\left(\eps^{-2}\log\left(1/\failureprob\right)\right)$, which also matches our upper bound [\Cref{thm:quasi-MHR:LB-sample}]. 
Our analysis develops a multiplicative-accuracy version of the Dvoretzky--Kiefer--Wolfowitz (DKW) inequality and a local stability argument for MHR distributions; \Cref{subsec:overview} explains these techniques.
\wtedit{To our knowledge, we are the first to provide tight characterization of sample complexity for all four distribution families, which eliminates the extra logarithmic dependence on $1/\eps$ in previous upper bounds and matches the lower bounds up to constant factors.}

\paragraph{Single-sample query complexity.}
\wtedit{We now turn to the other endpoint of our tradeoff characterization, where the learner receives one initial value sample and obtains all subsequent information through pricing queries. 
The following table summarizes our tight query complexity bounds for all four distribution families.}

\begin{center}
\small
\renewcommand{\arraystretch}{1.35}
\begin{tabular}{@{}lc@{}}
\hline
Distribution family & Single-sample query complexity \\
\hline
MHR
& $\Theta\left(\eps^{-2}\log\left(1/\failureprob\right)\right)$ [\Cref{thm:MHR:UB-query,thm:MHR:LB-tradeoff}] \\
Regular
& $\Theta\left(\eps^{-3}\log\left(1/\failureprob\right)\right)$ [\Cref{thm:regular:UB-query,thm:regular:LB-tradeoff}] \\
Quasi-MHR
& $\Theta\left(\eps^{-3}\log\left(1/\failureprob\right)\right)$ [\Cref{thm:quasi-MHR:UB-query,thm:quasi-MHR:LB-tradeoff}] \\
Quasi-regular
& $\Theta\left(\eps^{-3}\log\left(1/\eps\right)\log\left(1/\failureprob\right)\right)$ [\Cref{thm:quasi-regular:UB-query,thm:quasi-regular:LB-tradeoff}] \\
\hline
\end{tabular}
\end{center}

The comparison with sample complexity yields three conclusions. For MHR and quasi-MHR distributions, the single-sample query complexity exceeds the sample complexity by polynomial factors in $1/\eps$. For regular distributions, the two complexities coincide up to constant factors. For quasi-regular distributions, they differ by a factor of $\log\left(1/\eps\right)$. Together with our tradeoff bounds, these results identify both the relative efficiency of the two information sources and the extent to which combining them can reduce the learning cost.

Previous work on pricing-query access assumes that the prior $\prior$ is supported on $\left[0,1\right]$ and seeks an $\eps$-additive approximation \cite{LSTW23ec,LSTW23soda,GKSW24,SW24,TW25,CJLZ26www}. These assumptions avoid the scale difficulty that prevents query-only multiplicative PAC learning. The following table summarizes the previous bounds in that setting.

\begin{center}
\small
\setlength{\tabcolsep}{5pt}
\renewcommand{\arraystretch}{1.35}
\begin{tabular}{@{}lcc@{}}
\hline
Distribution family & Previous upper bound & Previous lower bound \\
\hline
MHR / Regular
& \shortstack{$O\left(\eps^{-2}\log^{3}\left(1/\eps\right)\log\left(\left(1/\failureprob\right)\log\left(1/\eps\right)\right)\right)$\\\cite{LSTW23soda}}
& \shortstack{$\Omega\left(\eps^{-2}\log\left(1/\failureprob\right)\right)$\\\cite{LSTW23soda}} \\
Quasi-MHR / Quasi-regular
& \shortstack{$O\left(\eps^{-3}\log\left(1/\left(\eps\failureprob\right)\right)\right)$\\\cite{LSTW23soda}}
& \shortstack{$\Omega\left(\eps^{-3}\log\left(1/\failureprob\right)\right)$\\\cite{CJLZ26www}} \\
\hline
\end{tabular}
\end{center}

Our techniques also apply to this additive-ap\-prox\-imation setting, yielding tight bounds of $\Theta\left(\eps^{-2}\log\left(1/\failureprob\right)\right)$ for MHR and regular distributions and $\Theta\left(\eps^{-3}\log\left(1/\failureprob\right)\right)$ for quasi-MHR and quasi-regular distributions. \Cref{subsec:overview} describes the structural and algorithmic ingredients underlying our query upper bounds.

\begin{remark}[The Single Sample and Alternative Predictions]
Our single-sample query algorithms use the initial sample to obtain coarse information about the scale of the prior $\prior$. A value sample is a natural source of this information in our query-sample framework, but 
our techniques can be extended straightforwardly to accommodate alternative predictions; two reasonable alternatives are:\\
(i)~A not-too-bad prediction to the optimal revenue $\Hat{r^{*}} \in [2^{-\poly(1 / \eps)} \cdot r^{*},\ 2^{\poly(1 / \eps)} \cdot r^{*}]$.\\
(ii)~A not-too-bad prediction to the prior $F$'s median $\Hat{\mathrm{med}} \in [F^{-1}(2^{-\poly(1 / \eps)}),\ F^{-1}(1 - 2^{-\poly(1 / \eps)})]$.
\end{remark}

%% file: Paper/tech-overview-quasiregular.tex

\paragraph{Query complexity upper bounds: structural results and reductions to online learning.}
Regarding query complexity, we examine \textit{single-sample multiplicative-accuracy} learnability, rather than no-sample additive-accuracy learnability as in previous works \cite{LSTW23ec, LSTW23soda, GKSW24, SW24, TW25, CJLZ26www}. (Nonetheless, we can seamlessly extend our techniques to avoid the $\polylog(\frac{1}{\eps})$ losses in previous works.) Specifically, we show an upper bound of $O\bigl(\frac{1}{\eps^{3}}\log(\frac{1}{\eps})\log(\frac{1}{\delta})\bigr)$ for the quasi-regular family, $O\bigl(\frac{1}{\eps^{3}}\log(\frac{1}{\delta})\bigr)$ for the regular and quasi-MHR families, and $O\bigl(\frac{1}{\eps^{2}}\log(\frac{1}{\delta})\bigr)$ for the MHR family.

For ease of presentation, we elaborate more carefully on our quasi-regular and regular algorithms---the quasi-MHR and MHR algorithms are largely their baby versions, so we merely sketch the adaptations. For the formal proofs, however, we recommend that the reader first consult the more accessible baby versions.

Firstly, our \uline{quasi-regular} algorithm consists of the following {four} modules:
\begin{enumerate}[label = A\arabic*., ref = A\arabic*]
    \item\label{overview:quasi-regular:1}
    \textit{Estimating the monopoly revenue $\Hat{r^{*}} \approx r^{*}$ up to constant factors:}\\
    Here, we aim to derive an estimate $\Hat{r^{*}} \in [r^{*}, \frac{11}{10}r^{*}]$ with probability $(1 - \delta)$ using $O\bigl(\log(\frac{1}{\delta})\bigr)$ queries.
    {To do so, we first solve the {\NoisyPriceLocation} subproblem: given $\kappa, \eta, \delta' \in (0, 1)$, find a price interval $[p', p'']$ such that $p'' / p' \le 1 + \kappa$ and $[Q(p''), Q(p')] \cap [\eta \pm \frac{1}{3}\eta] \ne \emptyset$, with probability $(1 - \delta')$.

    Using quasi-regularity, we first show that the target quantile $\eta$ lies inside a price interval $[\underline{p'}, \Bar{p''}]$ whose ratio is $\Bar{p''} / \underline{p'} = O\bigl(\frac{1}{\delta'^{2}}\bigr)$.\footnote{How to choose this price interval $[\underline{p'}, \Bar{p''}]$ is the only place throughout our algorithm that (explicitly) uses the single sample.}
    Discretizing it into a geometric $(1 + \kappa)$-grid $G$ reduces our subproblem to the {\NoisyBinarySearch} problem: a failed-sale indicator at price $p \in G$ is a coin flip with head probability $F(p)$, and these probabilities are nondecreasing in $p$.
    Applying \cite[Thm~1]{GP24} with target $1 - \eta$ and tolerance $\eta / 3$ finds the desired adjacent prices using $O\bigl(\frac{1}{\eta}\log(\frac{1}{\kappa\delta'})\bigr)$ queries.}

    {By taking} $\kappa = 0.01$, $\eta = 0.015$, and $\delta' = \delta / 2$, we find an anchor price $p'$ such that $[Q(1.01p'), Q(p')] \cap [0.01, 0.02] \ne \emptyset${. Consequently,} $\Rprice(p) \le \frac{1.01}{1 - 0.02}\Rprice(p')$ for $p \ge p'$ (by quasi‑regularity) and $\Rprice(p) \le \Rprice(p')$ for $p \le 0.01p'$ (trivial). Thus, we can focus on the price interval $[0.01p', p']$, discretize it into a geometric $1.01$-grid $G$, and obtain an estimate $\Hat{r_{i}} \in [\Rprice(p_{i}) \pm 0.01\Rprice(p_{i})]$ for each grid price $p_{i}$ with probability $(1 - \delta / (2|G|))$, using $O(\log(2|G| / \delta))$ queries.
    Then, the maximum scaled estimate $\Hat{r^{*}} \eqdef \max_{{p_i \in G}} \frac{\Hat{r_{i}}}{0.95}$ lies in $\Hat{r^{*}} \in [r^{*}, \frac{11}{10}r^{*}]$ with probability $(1 - \delta)$, as required, and Module~\ref{overview:quasi-regular:1} has a query complexity of $O\bigl(\frac{1}{\eta}\log(\frac{1}{\kappa\delta'})\bigr) + |G| \cdot O(\log(2|G| / \delta)) = O\bigl(\log(\frac{1}{\delta})\bigr)$.
    
    \item\label{overview:quasi-regular:2}
    \textit{Sparsifying the candidate price set $\Pcandidate$ via quasi-regularity:}\\
    Here, we aim to construct a size-$\Theta\bigl(\frac{1}{\eps}\bigr)$ discrete set $\Pcandidate$ that contains a $(1 - \eps)$-approximately optimal price $\max_{p \in \Pcandidate} \Rprice(p) \ge (1 - \eps) \cdot r^{*}$. Given the above estimate $\Hat{r^{*}} \in [r^{*}, \frac{11}{10}r^{*}]$, we can concentrate on the price interval $[\underline{p}, \Bar{p}] \eqdef [\frac{10}{11}\Hat{r^{*}}, \frac{2}{\eps}\Hat{r^{*}}]$; quasi-regularity guarantees $\max_{p \in [\underline{p}, \Bar{p}]} \Rprice(p) \ge (1 - \frac{\eps}{2}) \cdot r^{*}$ \cite[Sec~8.2]{FJ25}. It remains to find a size-$\Theta\bigl(\frac{1}{\eps}\bigr)$ discrete set $\Pcandidate$ such that
    \begin{align}
        \textstyle
        \max_{p \in \Pcandidate} \Rprice(p)
        ~\ge~ (1 - \frac{\eps}{2}) \cdot \bigl(\max_{p \in [\underline{p}, \Bar{p}]} \Rprice(p)\bigr)
        ~\ge~ (1 - \eps) \cdot r^{*}.
        \label{eq:overview:quasi-regular:2}\tag{$\heartsuit$}
    \end{align}
    The straightforward approach is to consider the \textit{geometric} $(\frac{1}{1 - \eps / 2})$-grid over $[\underline{p}, \Bar{p}]$. Unfortunately, this grid has size $\Theta\bigl(\frac{1}{\eps}\log(\frac{1}{\eps})\bigr)$ and results in a query complexity of $O\bigl(\frac{1}{\eps^{4}}\log(\frac{1}{\delta})\bigr)$ in the bottleneck module (Module~\ref{overview:quasi-regular:4}). To circumvent this, we prove a new structural result for quasi-regularity: the following size-$\Theta\bigl(\frac{1}{\eps}\bigr)$ \textit{reciprocal} grid also fulfills \eqref{eq:overview:quasi-regular:2}.
    \begin{align*}
        & \Pcandidate ~\eqdef~ \bigl\{p_{i} \eqdef \Bar{p} / i\bigr\}_{i \in [\Bar{i}]},
        &&\textstyle \text{where $\Bar{i} \eqdef \lceil \Bar{p} / \underline{p} \rceil = \Theta\bigl(\frac{1}{\eps}\bigr)$}.
    \end{align*}
    This sparsification is in a similar spirit to previous work \cite{LSTW23ec} on ``the description complexity of distributions'' and should be of independent interest.\footnote{In more detail, \cite{LSTW23ec} showed that $\tTheta\bigl(\frac{1}{\eps^{1 / 2}}\bigr)$ bits are sufficient and necessary---tight up to polylogarithmic factors---to sparsely describe a $[0, 1]$-supported regular distribution within L\'{e}vy distance $\eps \in (0, 1)$.}
    Clearly, Module~\ref{overview:quasi-regular:2} makes no queries.
    
    \item\label{overview:quasi-regular:3}
    \textit{Finding a threshold index $\underline{i} \in [\Bar{i}]$:}\\
    For simplicity, denote $r_{i} \eqdef \Rprice(p_{i})$ and $q_{i} \eqdef Q(p_{i})$ for each $i \in [\Bar{i}]$. Here, we aim to find a threshold index $\underline{i} \in [\Bar{i}]$ such that $r_{\underline{i}} \ge {0.55}\Hat{r^{*}}$ and $r_{i} \le {0.65}\Hat{r^{*}}$ for each $i \in [\underline{i} - 1]$ (if existential), with probability $(1 - \delta)$; its existence is guaranteed by \eqref{eq:overview:quasi-regular:2} and the fact $r^{*} \in [\frac{10}{11}\Hat{r^{*}}, \Hat{r^{*}}]$.
    With tolerance ${0.65}\Hat{r^{*}} - {0.55}\Hat{r^{*}} = {0.1}\Hat{r^{*}}$, Module~\ref{overview:quasi-regular:3} can be implemented in a standard way (i.e., {Bernstein's inequality} plus the union bound), using $O\bigl(\frac{1}{\eps}\polylog\bigl(\frac{1}{\eps}\bigr)\log(\frac{1}{\delta})\bigr)$ queries.
    The consequences are threefold:\\
    (i)~Every before-threshold index $i \in [\underline{i} - 1]$ is suboptimal $r_{i} \le {0.65}\Hat{r^{*}} \le 0.8r^{*}$ and can be eliminated.\\
    (ii)~The threshold revenue and quantile are bounded: ${0.55}\Hat{r^{*}} \le r_{\underline{i}} \le \Hat{r^{*}}$ and $q_{\underline{i}} = r_{\underline{i}} / p_{\underline{i}} \ge \frac{\eps}{4} \cdot \underline{i}$.\\
    (iii)~Quasi-regularity extends (ii) approximately: $0.3\Hat{r^{*}} \le r_{i} \le \Hat{r^{*}}$ and $q_{i} \ge \frac{\eps}{6} \cdot i$ for each $i \in [\underline{i}: \Bar{i}]$.

    \item\label{overview:quasi-regular:4}
    \textit{Formulating and solving a modified {\BestArmIdentification} problem:}\\
    Now, we are facing a modified {\BestArmIdentification} problem: there are $(\Bar{i} - \underline{i} + 1) = O\bigl(\frac{1}{\eps}\big)$ arms; each arm $i \in [\underline{i}: \Bar{i}]$ has a bounded expected revenue $0.3\Hat{r^{*}} \le r_{i} \le \Hat{r^{*}}$ and, given $\eps', \delta' \in (0, 1)$, can be estimated within $\Hat{r_{i}} \in [r_{i} \pm \eps' r_{i}]$ with probability $(1 - \delta')$ using $O\bigl(\frac{1}{\eps'^{2} \cdot q_{i}}{\log(\frac{1}{\delta'})}\bigr)$ queries. We adapt the Median Elimination algorithm \cite{EMM06} (query-optimal for the \textit{original} {\BestArmIdentification} problem) in a non-trivial way to solve our modified problem.

    {To be more specific, we divide the remaining indices into $O\bigl(\log(\frac{1}{\eps})\bigr)$ groups between consecutive powers of two and search each group using $O\bigl(\frac{1}{\eps^{3}}\log(\frac{1}{\delta})\bigr)$ queries; only the group containing an optimal candidate needs to succeed.
    Then, we compare the returned prices using fresh estimates with geometrically decreasing failure probabilities, giving a total query complexity of $O\bigl(\frac{1}{\eps^{3}}\log(\frac{1}{\eps})\log(\frac{1}{\delta})\bigr)$.}
\end{enumerate}
To conclude, we can find a $(1 - O(\eps))$-approximately optimal price $\Hat{p^{*}}$ with probability $(1 - O(\delta))$, using $O\bigl(\frac{1}{\eps^{3}}\log(\frac{1}{\eps})\log(\frac{1}{\delta})\bigr)$ queries. Thus, scaling $\eps, \delta \in (0, 1)$ appropriately completes the analysis.

\begin{remark}[The \uline{Quasi-MHR} Algorithm]
\label{rmk:overview:quasi-MHR}
Under quasi-MHR, the bounded monopoly quantile $q^{*} \ge \frac{1}{e}$ enables several simplifications.
(Module~\ref{overview:quasi-regular:2}) The monopoly price $p^{*}$ lies in $[r^{*}, e r^{*}] \subseteq [\underline{p}, \Bar{p}] \eqdef [\frac{10}{11}\Hat{r^{*}}, e\Hat{r^{*}}]$, and a straightforward geometric grid of $[\underline{p}, \Bar{p}]$ already works.
(Module~\ref{overview:quasi-regular:3}) The threshold index $\underline{i} \in [\Bar{i}]$ satisfies a better quantile bound $q_{\underline{i}} \ge \frac{1}{3e}$.
(Module~\ref{overview:quasi-regular:4}) The \textit{original} Median Elimination algorithm \cite{EMM06} already works. Altogether, the query complexity improves to $O\bigl(\frac{1}{\eps^{3}}\log(\frac{1}{\delta})\bigr)$.
\end{remark}

Secondly, our \uline{regular} algorithm only modifies Module~\ref{overview:quasi-regular:4} into Module~\ref{overview:regular:4}:
\begin{enumerate}[label = B\arabic*., ref = B\arabic*]
\setcounter{enumi}{3}
    \item[B\textsuperscript{$\dagger$}.] \textit{Running Modules~\ref{overview:quasi-regular:1} to \ref{overview:quasi-regular:3} with $O\bigl(\frac{1}{\eps}\polylog\bigl(\frac{1}{\eps}\bigr)\log(\frac{1}{\delta})\bigr)$ queries.}

    \item\label{overview:regular:4}
    \textit{Formulating and solving a {\NoisyConcaveOptimization} problem:}\\
    By regularity, the revenue-quantile curve $\Rquantile(q)$ is concave, so the polyline specified by the points $\{(q_{i}, r_{i})\}_{i \in [\underline{i}, \Bar{i}]}$ is also concave. For each $i \in [\underline{i}, \Bar{i}]$, the slope $r_{i} / q_{i} = p_{i}$ is known, but the point itself, for $\eps', \delta' \in (0, 1)$, can only be estimated as $(\Hat{q_{i}}, \Hat{r_{i}}) \in [q_{i} \pm \eps' q_{i}] \times [r_{i} \pm \eps' r_{i}]$ with probability $(1 - \delta')$, using $O\bigl(\frac{1}{\eps'^{2} \cdot q_{i}}{\log(\frac{1}{\delta'})}\bigr)$ queries. This gives a specific {\NoisyConcaveOptimization} problem.\\
    We adopt a noise‑tolerant ternary search over $[\underline{i}: \Bar{i}]$: in each iteration, we estimate both trisection indices $i \in \bigl\{\frac{1}{3}(2\underline{i} + \Bar{i}),\ \frac{1}{3}(\underline{i} + 2\Bar{i})\bigr\}$ with $\eps' = \Theta(\eps)$ and $\delta' = \Theta(\delta / i^{2})$, compare the two estimated revenues, and eliminate one third of the indices. Throughout, concavity guarantees that a $(1 - O(\eps))$-approximately optimal price is preserved (even if one or more incorrect eliminations occur), with probability $1 - \sum_{i \in [\underline{i}: \Bar{i}]} \Theta(\delta / i^{2}) = 1 - O(\delta)$. Further, the number of queries across iterations decays geometrically (precisely because the remaining interval decreases geometrically), resulting in a total query complexity of $O\bigl(\frac{1}{\eps^{3}}\log(\frac{1}{\delta})\bigr)$. Scaling $\eps, \delta \in (0, 1)$ completes the analysis.
\end{enumerate}

\begin{remark}[The \uline{MHR} Algorithm]
\label{rmk:overview:MHR}
Under MHR, (i)~the bounded monopoly quantile $q^{*} \ge \frac{1}{e}$ again enables the simplifications described in \Cref{rmk:overview:quasi-MHR}, and (ii)~the cumulative hazard function $H(x) \eqdef -\ln(1 - F(x))$ is \textit{convex}, allowing a ternary search akin to the one described in Module~\ref{overview:regular:4}. Combining both (i) and (ii), the query complexity improves to $O\bigl(\frac{1}{\eps^{2}}\log(\frac{1}{\delta})\bigr)$.
\end{remark}

%% file: Paper/related-work.tex
\paragraph{Sample complexity of revenue maximization}
Learning revenue-maximizing mechanisms from value samples has been studied in single-parameter auctions \cite{CR14,GHZ19}, multi-item mechanisms \cite{MR16,GW21,BCD20}, and single-item posted pricing and reserve-price learning \cite{JLX23,JKMS24}. For a single buyer, \cite{HMR18} establish sample complexity bounds that are tight up to logarithmic factors: $\tilde{\Theta}(\varepsilon^{-3/2})$ for MHR distributions and $\tilde{\Theta}(\varepsilon^{-3})$ for regular distributions, at constant confidence. Our sample-only results sharpen these bounds by removing the remaining logarithmic dependence on $1/\varepsilon$ and form one endpoint of our query-sample tradeoffs. \cite{HKMV26} take a complementary perspective by studying how revenue-learning error decays for a fixed underlying distribution. We study PAC guarantees that hold uniformly over each specified distribution family and quantify the resources required under two forms of information access.
\cite{FJ25} study quasi-regular and quasi-MHR distributions as relaxations of regularity and MHR, and extend numerous revenue-approximation and learning guarantees to these broader families. In particular, their results imply sample upper bounds of $\tilde{O}(\varepsilon^{-3})$ and $\tilde{O}(\varepsilon^{-2})$ for single-buyer revenue maximization under quasi-regularity and quasi-MHR, respectively. 
Further developments in this line of research study learning revenue-maximizing mechanisms from samples across a broad range of auction environments and distributional assumptions \cite{BBHM08, DRY15, FJ25, HMR18, MR15, CGM15, MM16, DHP16, RS16, GN17, S17, HT19, CHMY23}.

Our results sharpen these sample bounds and examine how the weaker structural assumptions affect learning from pricing queries. Regular and quasi-regular distributions have the same optimal sample complexity but different single-sample query complexities. Moreover, quasi-MHR distributions exhibit a gradual tradeoff between samples and queries, whereas MHR and regular distributions exhibit threshold behavior. These distinctions show that guarantees under either access model alone do not determine the benefits of combining them.

\paragraph{Pricing queries and censored feedback}
\cite{LSTW23soda} study revenue maximization from binary purchase feedback. For distributions supported on $[0,1]$, they establish query complexity bounds of $\tilde{\Theta}(\varepsilon^{-3})$ for general distributions and $\tilde{\Theta}(\varepsilon^{-2})$ for regular and MHR distributions, under an additive revenue guarantee. Related work considers threshold learning with latent values and censored rewards \cite{ZLZFTD24}, unit-demand pricing with independent item values \cite{TW25}, and uniform pricing with multiple buyers \cite{CJLZ26www}. We study a multiplicative revenue guarantee, which is invariant under rescaling the values and requires accuracy relative to the unknown optimal revenue. This requirement makes information about the distribution's scale essential: as shown in \Cref{sec:impossibility}, no finite, distribution-independent pricing-query budget suffices without value samples, even for MHR distributions supported on $[0,1]$. Our framework quantifies how value samples enable and improve learning from pricing queries.

\paragraph{Revenue maximization from few samples}
\cite{DRY15} show that posting a single sampled value yields, in expectation, half the optimal revenue for regular distributions.
{For MHR distributions, this guarantee can be improved through a scaled pricing rule \cite{HMR18}. Subsequent work extends this line of research from various perspectives \cite{FILS15, BGMM18, DZ20, RTY20, ABB22, FHL21, FJ25}.}
\ignore{For MHR distributions, \cite{HMR18} improve this guarantee to $0.589$ through a scaled pricing rule. Subsequent work studies the revenue gains obtainable from two samples \cite{BGMM18,DZ20}.}
These results quantify the approximation achievable from a fixed, small number of value observations. In contrast, our single-sample endpoint asks how many additional pricing queries are needed to obtain a $(1-\varepsilon)$-approximation with probability at least $1-\delta$. More generally, our tradeoff results determine when additional value samples can reduce the required query budget.

\paragraph{Connections to bandit learning}
A pricing query reveals the realized revenue at the selected price, while a value sample determines the counterfactual revenue at every price for the same realization. This observation connects our two access models to bandit and full-information feedback. \wtdelete{including their study over continuous action spaces \cite{KSU19}.\yjnote{Remove ``including their study over continuous action spaces \cite{KSU19}''?} }
In online pricing, \cite{SW24} exploit half-concavity to obtain regret bounds of $\tilde{O}(\sqrt{T})$ for regular distributions and $\tilde{O}(T^{2/3})$ for general distributions, for a fixed number of buyers. Our objective is to identify a near-optimal price with prescribed confidence. Our algorithms also connect to stochastic optimization: for MHR distributions, negative log-revenue is convex in log-price coordinates, allowing us to apply the sharp one-dimensional stochastic convex-optimization guarantee of \cite{CRTA26} after constructing an appropriate observation from pricing queries.
\wtedit{More broadly, our techniques for combining value samples with adaptive pricing queries may also shed light on the information requirements for learning fixed-price mechanisms in bilateral trade \cite{CCCFL24mor, CCCFL24jmlr, AFF24, BCCF24, CJLZ26icalp, LCM26soda, CLM26}, particularly the tradeoffs between full value observations and pricing feedback.}

%% file: Paper/prelim-new.tex
We denote the integer interval by $\left[\intlower:\intupper\right]\triangleq\left\{\intlower,\intlower+1,\ldots,\intupper\right\}$, with $\left[\intlower:\intupper\right]=\varnothing$ when $\intlower>\intupper$. In particular, we write $\left[\indexnum\right]\triangleq\left[1:\indexnum\right]$ for every nonnegative integer $\indexnum$. All logarithms are natural unless a base is specified.

\subsection{Probability and Distribution}
\label{subsec:prob}
\wtdelete{We consider a probability space $\left(\samplespace,\sigmaalgebra,\probmeasure\right)$. For a finite vector of real-valued random variables $\randvec=\left(\randvar_{\coordidx}\right)_{\coordidx\in\left[\indexnum\right]}$, we denote its pushforward probability measure by $\probmeasure_{\randvec}$. 
For every measurable set $\measurableset$, this measure satisfies
\begin{align*}
    \probmeasure_{\randvec}\left(\measurableset\right)
    &\triangleq \prob{\left\{\outcome\in\samplespace:\randvec\left(\outcome\right)\in\measurableset\right\}}~.
\end{align*}
For a sub-$\sigma$-algebra $\sigmaalgebra'\subseteq\sigmaalgebra$, we similarly denote the conditional law by
$\probmeasure_{\randvec\mid\sigmaalgebra'}\left(\measurableset\right)
\triangleq \prob{\randvec\in\measurableset\mid\sigmaalgebra'}$.
This notation applies to discrete, continuous, and mixed distributions.}
We consider value distributions supported on $\left[0,\infty\right)$. We use the same notation $\prior$ for a distribution and its left-continuous cumulative distribution function (CDF). The CDF and the associated sale-probability function $\Qsale$ are
\begin{align*}
    \prior\left(\price\right)\triangleq\prob[\val\sim\prior]{\val<\price}~,
    \quad
    \Qsale\left(\price\right)\triangleq1-\prior\left(\price\right)=\prob[\val\sim\prior]{\val\ge\price}~,
    \quad
    \price\ge0~.
\end{align*}
This convention includes a buyer whose value equals the posted price among those willing to purchase. We denote the right limit of the CDF by $\prior^{+}\left(\price\right)\triangleq\prob[\val\sim\prior]{\val\le\price}$; thus, the probability mass at the value $\price$ is $\prior^{+}\left(\price\right)-\prior\left(\price\right)$.\footnote{The auxiliary empirical CDF of uniform random variables in \Cref{sec:DKW} uses the usual right-continuous convention, as specified there.}

We define the generalized inverse of the sale-probability function and the associated revenue-quantile curve by
\begin{align*}
    \salequantile\left(\quantile\right)\triangleq\sup\nolimits\left\{\price\ge0:\Qsale\left(\price\right)\ge\quantile\right\}~,
    \quad
    \Rquantile\left(\quantile\right)\triangleq\quantile\salequantile\left(\quantile\right)~,
    \quad
    \quantile \in\left(0,1\right]~.
\end{align*}
We extend the revenue-quantile curve to $\quantile=0$ by its right-hand limit whenever this endpoint is needed. This limit exists for the distribution families considered below. In particular, its value need not be zero, as illustrated by the equal-revenue distribution.
\wtdelete{We also define the cumulative hazard function by
\begin{equation}
\label{eq:cum-hazard}
    \cumhazard\left(\price\right)\triangleq
    -\log\Qsale\left(\price\right)~,
\end{equation}
with $\cumhazard\left(\price\right)=+\infty$ when $\Qsale\left(\price\right)=0$.}
Where a distribution is absolutely continuous, we denote its probability density function by $\density$. The following structural definitions also specify our conventions for distributions with atoms.

\begin{definition}[{Regular Distributions \cite{M81}}]
\label{def:regular}
A distribution $\prior$ is regular if its revenue-quantile curve $\Rquantile$ is concave. For a distribution with positive density on the interior of its support, the equivalent density formulation requires the virtual value function
$\virtualval\left(\price\right)\triangleq\price-\frac{\Qsale\left(\price\right)}{\density\left(\price\right)}$
to be nondecreasing on the interior of the support.
\end{definition}

\begin{definition}[{MHR Distributions \cite{BMP63}}]
\label{def:MHR}
A distribution $\prior$ satisfies the monotone-hazard-rate (MHR) condition if its cumulative hazard function, \wtedit{denoted by $\cumhazard(\price)\triangleq -\log\Qsale\left(\price\right)$,} is convex. For a distribution with positive density on the interior of its support, the equivalent density formulation requires the hazard rate function
$\hazardrate\left(\price\right)\triangleq\frac{\density\left(\price\right)}{\Qsale\left(\price\right)}$
to be nondecreasing wherever $\Qsale\left(\price\right)>0$.
\end{definition}

\begin{definition}[{$\regularity$-Regular Distributions \cite{CR14}}]
\label{def:lambda-regular}
For a parameter $\regularity\in\left[0,1\right]$, a distribution $\prior$ is $\regularity$-regular if the function
\begin{align*}
    \price&\longmapsto
    \begin{cases}
        \Qsale\left(\price\right)^{-\regularity}~, & \regularity\in\left(0,1\right]~,\\
        -\log\Qsale\left(\price\right)~, & \regularity=0
    \end{cases}
\end{align*}
is convex, with value $+\infty$ when $\Qsale\left(\price\right)=0$. For a distribution with positive density on the interior of its support, the equivalent density formulation requires the $\regularity$-virtual value function
$\virtualval_{\regularity}\left(\price\right)\triangleq\regularity\price-\frac{\Qsale\left(\price\right)}{\density\left(\price\right)}$
to be nondecreasing on the interior of the support.
\end{definition}

\begin{definition}[{$\regularity$-Quasi-Regular Distributions \cite{FJ25}}]
\label{def:lambda-quasi-regular}
For a parameter $\regularity\in\left[0,1\right]$, a distribution $\prior$ is $\regularity$-quasi-regular if the function
\begin{align*}
    \price&\longmapsto
    \begin{cases}
        \dfrac{\Qsale\left(\price\right)^{-\regularity}-1}{\price}~, & \regularity\in\left(0,1\right]~,\\[6pt]
        \dfrac{-\log\Qsale\left(\price\right)}{\price}~, & \regularity=0
    \end{cases}
\end{align*}
is nondecreasing for $\price>0$, with value $+\infty$ when $\Qsale\left(\price\right)=0$.
\end{definition}

The parameter values $\regularity=0$ and $\regularity=1$ recover MHR and regular distributions, respectively, in \Cref{def:lambda-regular}, and quasi-MHR and quasi-regular distributions, respectively, in \Cref{def:lambda-quasi-regular}. In particular, quasi-MHR requires the ratio $-\log\Qsale\left(\price\right)/\price$ to be nondecreasing. Quasi-regularity equivalently requires the function
$\quasiregularbound\left(\price\right)\triangleq\frac{\price\Qsale\left(\price\right)}{\prior\left(\price\right)}$
for $\price > 0$ to be nonincreasing, with $\quasiregularbound\left(\price\right)=+\infty$ when $\prior\left(\price\right)=0$.

Every MHR distribution is regular and quasi-MHR, and both regular and quasi-MHR distributions are quasi-regular. The converse implications need not hold. For example, the equal-revenue distribution, whose CDF satisfies $\prior\left(\price\right)=0$ for $0\le\price\le1$ and $\prior\left(\price\right)=1-1/\price$ for $\price>1$, is regular but not MHR. Further characterizations appear in \cite{H13,FJ25}.

We will also use the following standard inequality.

\begin{claim}[{Jensen's Inequality \cite[Theorem~2.6.2]{CT06}}]
\label{cla:Jensen}
An integrable random variable $\randvar$ and a convex function $\convexfunc$ satisfy
$\convexfunc\left(\expect{\randvar}\right)\le\expect{\convexfunc\left(\randvar\right)}$
whenever the expectation on the right is well-defined.
\end{claim}

\subsection{Single-Buyer Revenue Maximization}
We consider a seller who offers an indivisible item to a single buyer whose value follows an unknown distribution $\prior$ on $\left[0,\infty\right)$. A posted price $\price\ge0$ results in a sale whenever the buyer's value satisfies $\val\ge\price$. We denote the resulting revenue-price curve and optimal revenue by
\begin{align*}
    \Rprice\left(\price\right)\triangleq\price\Qsale\left(\price\right)=\price\left(1-\prior\left(\price\right)\right)~,
    \quad
    \optrevenue\triangleq\sup\nolimits_{\price\ge0}\Rprice\left(\price\right)~.
\end{align*}
We consider distributions with finite optimal revenue. The case $\optrevenue=0$ is immediate, since every output price satisfies the revenue guarantee below.

The revenue supremum need not be attained. When a maximizer exists, we denote a monopoly price by $\monopolyprice\in\argmax\nolimits_{\price\ge0}\Rprice\left(\price\right)$ and its monopoly quantile by $\monopolyquantile\triangleq\Qsale\left(\monopolyprice\right)$. The price and quantile representations satisfy
\begin{align*}
    \Rprice\left(\price\right)\le\Rquantile\left(\Qsale\left(\price\right)\right)
    \quad\text{whenever }\Qsale\left(\price\right)>0~,
    \quad\optrevenue=\sup\nolimits_{\quantile\in\left(0,1\right]}\Rquantile\left(\quantile\right)~.
\end{align*}

A learning algorithm $\learnalg$ initially knows the distribution family but does not know the distribution $\prior$. The algorithm can acquire information through two forms of access:
\begin{itemize}
    \item \textbf{Sample access.}
    In a sample round $\roundidx$, the algorithm observes a fresh value sample $\sampleval^{\roundidx}\sim\prior$.

    \item \textbf{Pricing-query access.}
    In a query round $\roundidx$, the algorithm selects a price $\price^{\roundidx}\ge0$ and observes only the binary feedback
    \begin{align*}
        \queryfeedback^{\roundidx}\triangleq\indicator\left[\val^{\roundidx}\ge\price^{\roundidx}\right]~,\quad
        \val^{\roundidx}\sim\prior~.
    \end{align*}
    Conditional on the selected price and the preceding history, the feedback satisfies
    \begin{align*}
        \prob{\queryfeedback^{\roundidx}=0\mid\price^{\roundidx},\text{history}}=\prior\left(\price^{\roundidx}\right)~,
        \quad\prob{\queryfeedback^{\roundidx}=1\mid\price^{\roundidx},\text{history}}=\Qsale\left(\price^{\roundidx}\right)~.
    \end{align*}
\end{itemize}

Every sample or query uses an independent draw from the value distribution. The algorithm may choose the access type, query price, and stopping decision adaptively, based on its previous observations and internal randomness. At termination, the algorithm outputs a price $\price^{\learnalg}$.

For an accuracy parameter $\eps\in\left(0,1\right)$ and a failure probability $\failureprob\in\left(0,1\right)$, we call the algorithm $\left(\eps,\failureprob\right)$-PAC if, for every distribution in the specified family, its output satisfies
\begin{align*}
    \prob[\prior,\learnalg]{\Rprice\left(\price^{\learnalg}\right)\ge\left(1-\eps\right)\optrevenue}&\ge1-\failureprob~.
\end{align*}
The probability is over the observations generated from the fixed distribution $\prior$ and the algorithm's internal randomness.

We denote the number of value samples by $\samplecount$ and the number of pricing queries by $\querycount$. Unless stated otherwise, an upper bound on either count holds on every execution. Sample complexity concerns algorithms using no pricing queries. Single-sample query complexity concerns algorithms that first draw one value sample and then use pricing queries. The query-sample tradeoff characterizes the combinations of sample and query budgets that permit PAC learning.

For sample-based algorithms, we use the empirical sale-probability function and empirical revenue-price curve constructed from independent samples $\val_1,\ldots,\val_{\indexnum}\sim\prior$:
\begin{align*}
    \empQsale\left(\price\right)\triangleq\frac{1}{\indexnum}\sum\nolimits_{\coordidx=1}^{\indexnum}\indicator\left[\val_{\coordidx}\ge\price\right]~,
    \quad\empRprice\left(\price\right)\triangleq\price\empQsale\left(\price\right)~,
    \quad
    \price\ge0~.
\end{align*}
These empirical functions use the same sale convention as their population counterparts, including at sample values.

\subsection{Information Theory}

We consider two probability measures $\probmeasure^{*}$ and $\probmeasure$ on the same measurable space. When the measure $\probmeasure^{*}$ is absolutely continuous with respect to $\probmeasure$, their Kullback--Leibler (KL) divergence \cite[Chapter~2.3]{CT06} is
\begin{align*}
    \DKL\left(\probmeasure^{*},\probmeasure\right)&\triangleq\expect[\probmeasure^{*}]{\log\!\left(\frac{d\probmeasure^{*}}{d\probmeasure}\right)}~,
\end{align*}
where the ratio denotes the Radon--Nikodym derivative. Otherwise, we set the divergence to $+\infty$.

We also use the notation $\DKL\left(\bernparamone,\bernparamtwo\right)$ for the divergence between Bernoulli distributions with parameters $\bernparamone,\bernparamtwo\in\left[0,1\right]$:
\begin{align*}
    \DKL\left(\bernparamone,\bernparamtwo\right)&\triangleq\bernparamone\log\frac{\bernparamone}{\bernparamtwo}
    +\left(1-\bernparamone\right)\log\frac{1-\bernparamone}{1-\bernparamtwo}~.
\end{align*}
The expression uses the usual limiting conventions at zero and one. In particular, complementing both Bernoulli parameters leaves the divergence unchanged. For a price $\price$, the query feedback $\indicator\left[\val\ge\price\right]$ has the Bernoulli law with parameter $\Qsale\left(\price\right)=1-\prior\left(\price\right)$. For two value distributions $\prior^{(0)}$ and $\prior^{(\distidx)}$ with sale probabilities $\Qsale^{(0)}$ and $\Qsale^{(\distidx)}$, we denote by $\DKL\left(\prior^{(0)}\left(\price\right),\prior^{(\distidx)}\left(\price\right)\right)$ the divergence between their feedback laws at price $\price$, which equals $\DKL\left(\Qsale^{(0)}\left(\price\right),\Qsale^{(\distidx)}\left(\price\right)\right)$ by the complementation property.

For a fixed learning algorithm and a value distribution $\prior^{(\distidx)}$, we denote the law of the algorithm's transcript by $\probmeasure^{(\distidx)}$ and the corresponding expectation by $\expectop^{(\distidx)}\triangleq\expectop_{\probmeasure^{(\distidx)}}$. The transcript records the access types, observed samples, query prices, query feedback, and final output.

\begin{claim}[Learning Algorithms]
\label{lem:kl-bound}
Two value distributions $\prior^{(0)}$ and $\prior^{(\distidx)}$ agree outside a measurable price set $\priceset^{(\distidx)}\subseteq\left[0,\infty\right)$, so that
\begin{align*}
    \prior^{(0)}\left(\price\right)&=\prior^{(\distidx)}\left(\price\right)
    \qquad\text{for every }\price\notin\priceset^{(\distidx)}~.
\end{align*}
A learning algorithm $\learnalg$ terminates almost surely under both distributions. We denote its sample count by $\samplecount$ and the number of its queries with prices in $\priceset^{(\distidx)}$ by $\querycount^{(\distidx)}$. For every measurable transcript event $\transcriptevent$, the induced transcript laws satisfy
\begin{align*}
    \DKL\left(\prob[\prior^{(0)}]{\transcriptevent},\prob[\prior^{(\distidx)}]{\transcriptevent}\right)
    &\le\expect[\prior^{(0)}]{\samplecount}\DKL\left(\prior^{(0)},\prior^{(\distidx)}\right)
    +\expect[\prior^{(0)}]{\querycount^{(\distidx)}}\sup\nolimits_{\price\in\priceset^{(\distidx)}}\DKL\left(\prior^{(0)}\left(\price\right),\prior^{(\distidx)}\left(\price\right)\right)~.
\end{align*}
\end{claim}

\begin{proof}
We denote the sets of sample rounds and query rounds by $\samplerounds$ and $\queryrounds$, respectively, and the set of queries with prices in the modified region by $\queryrounds^{(\distidx)}\triangleq\left\{\roundidx\in\queryrounds:\price^{\roundidx}\in\priceset^{(\distidx)}\right\}$. Their cardinalities satisfy $\left|\samplerounds\right|=\samplecount$ and $\left|\queryrounds^{(\distidx)}\right|=\querycount^{(\distidx)}$.

We denote the history $\sigma$-algebra before round $\roundidx$ by $\history^{\roundidx}$. Conditional on the same history, the algorithm uses the same rule for selecting its next action under both distributions. These action selections therefore contribute zero conditional KL divergence. A fresh value sample contributes $\DKL\left(\prior^{(0)},\prior^{(\distidx)}\right)$. At a query round, conditioning additionally on the selected price corresponds to the $\sigma$-algebra $\history^{\roundidx}\vee\sigma\left(\price^{\roundidx}\right)$, and the feedback contributes $\DKL\left(\prior^{(0)}\left(\price^{\roundidx}\right),\prior^{(\distidx)}\left(\price^{\roundidx}\right)\right)$. This contribution is zero whenever the query price lies outside $\priceset^{(\distidx)}$.

The data-processing inequality \cite[Chapter~2.8]{CT06} and the chain rule for KL divergence \cite{LS20} therefore give
\begin{align*}
    \DKL\left(\prob[\prior^{(0)}]{\transcriptevent},\prob[\prior^{(\distidx)}]{\transcriptevent}\right)
    &\le\expect[\prior^{(0)}]{\sum\nolimits_{\roundidx\in\samplerounds}\DKL\left(\prior^{(0)},\prior^{(\distidx)}\right)
    +\sum\nolimits_{\roundidx\in\queryrounds^{(\distidx)}}\DKL\left(\prior^{(0)}\left(\price^{\roundidx}\right),\prior^{(\distidx)}\left(\price^{\roundidx}\right)\right)}\\
    &\le\expect[\prior^{(0)}]{\samplecount}\DKL\left(\prior^{(0)},\prior^{(\distidx)}\right)
    +\expect[\prior^{(0)}]{\querycount^{(\distidx)}}\sup\nolimits_{\price\in\priceset^{(\distidx)}}\DKL\left(\prior^{(0)}\left(\price\right),\prior^{(\distidx)}\left(\price\right)\right)~.
\end{align*}
For a random stopping time, the chain-rule identity follows by first considering finite transcript prefixes and then passing to the limit. The final inequality bounds each informative query by the largest possible divergence over the modified price set. This proves the claim.
\end{proof}

%% file: quasi-regular.tex
\providecommand{\pref}[1]{\Cref{#1}}

\newcommand{\inisample}{s^1}

\section{Quasi-Regular Distributions}
\label{sec:quasi-regular}

In this section, we establish the query and sample upper bounds and the query-sample lower bound for quasi-regular distributions. We use the sale probability $Q$ and optimal revenue $r^*$ from \pref{subsec:overview} and \pref{sec:intro}. Throughout the proofs, $r^*$ denotes the revenue supremum, which need not be attained.

\subsection{Overview of the Section}
\label{subsec:quasi-regular:overview}

\begin{figure}[htbp]
    \centering
    \includegraphics[width=.44\textwidth]{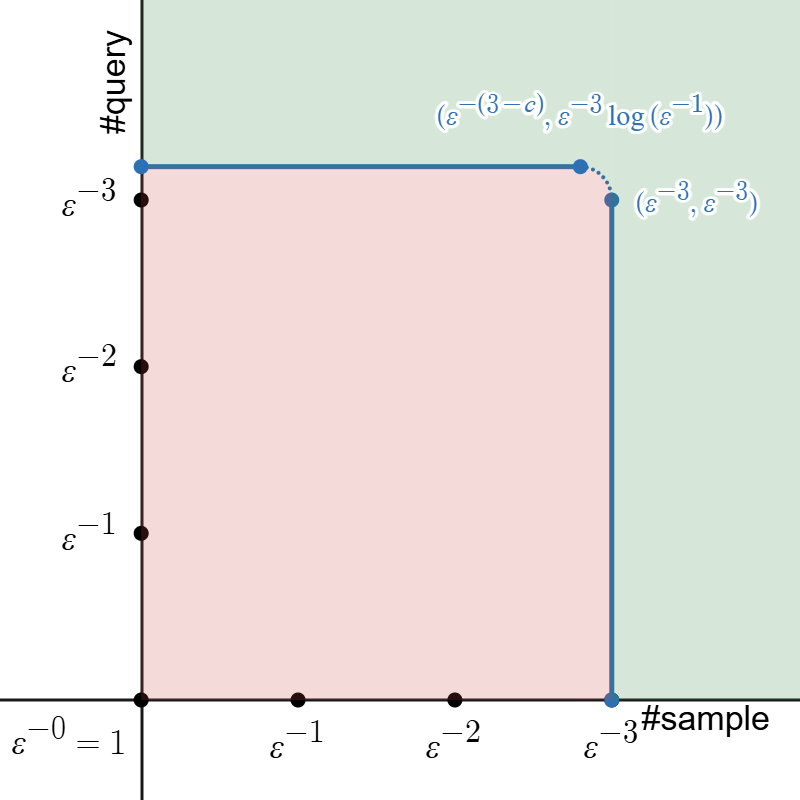}
    \caption{The quasi-regular query-sample tradeoff from \Cref{fig:result-quasi-regular}.
    Both axes are logarithmic; constant factors and $\log(1/\delta)$ are suppressed.
    Green and red denote the learnable and unlearnable regimes.}
    \label{fig:quasi-regular:tradeoff}
\end{figure}

{Our goal is to establish the almost tight tradeoff shown in \Cref{fig:quasi-regular:tradeoff}, reproduced from \Cref{fig:result-quasi-regular}.
Here, $\samplecount$ and $\querycount$ denote the sample and query budgets, respectively.}

{For the upper bounds, it suffices to establish the two endpoints
\begin{align*}
    (\samplecount,\querycount)
    &=\left(1,\ O\bigl(\eps^{-3}\log(1/\eps)\log(1/\delta)\bigr)\right),\\
    (\samplecount,\querycount)
    &=\left(O\bigl(\eps^{-3}\log(1/\delta)\bigr),\ 0\right),
\end{align*}
since an algorithm can simply ignore any additional samples or queries.
We first prove \Cref{thm:quasi-regular:UB-query} in \Cref{subsec:quasi-regular:UB}, giving the single-sample query endpoint.
Next, \Cref{subsubsec:quasi-regular:UB-sample} presents the sample-only endpoint, \Cref{thm:quasi-regular:UB-sample}; the complete proof appears in the quasi-regular subsection of \Cref{sec:sample-complexity}.

For the lower bounds, we need the horizontal family of points and the right-hand corner:
\begin{align*}
    (\samplecount,\querycount)
    &=\left(O\bigl(\eps^{-(3-c)}\log(1/\delta)\bigr),\
      \Omega\bigl(\eps^{-3}\log(1/\eps)\log(1/\delta)\bigr)\right),\\
    (\samplecount,\querycount)
    &=\left(O\bigl(\eps^{-3}\log(1/\delta)\bigr),\
      \Omega\bigl(\eps^{-3}\log(1/\delta)\bigr)\right).
\end{align*}
Here $c>0$ is any fixed constant, the constants in the sample budgets are sufficiently small, and the query lower bounds hold in expectation.
We prove the first family of points in \Cref{subsec:quasi-regular:LB}, as \Cref{thm:quasi-regular:LB-tradeoff}; these match the single-sample query upper bound.
For the second point, we will prove a stronger lower bound that holds even for regular distributions (\Cref{thm:regular:LB-tradeoff} in \Cref{subsec:regular:LB}).
It also gives the matching sample-only lower bound.
The dotted portion indicates the remaining transition near the sample-only threshold.}

\subsection{Query Complexity Upper Bound with a Single Sample}
\label{subsec:quasi-regular:UB}

We first prove the query upper bound for the quasi-regular family.
Our algorithm follows the {four} modules described in \pref{subsec:overview}: we first estimate the monopoly revenue using the single sample and logarithmically many queries, then sparsify the candidate price set and reduce the remaining task to a modified {\BestArmIdentification} problem.

\begin{theorem}[Upper Bound for Quasi-Regular Distributions]
\label{thm:quasi-regular:UB-query}
For the family of quasi-regular distributions, there exists an $(\eps,\delta)$-PAC learning algorithm that first draws a single sample and then makes at most $O\bigl(\frac{1}{\eps^3}\log(\frac{1}{\eps})\log(\frac{1}{\delta})\bigr)$ pricing queries.
\end{theorem}

Throughout the proof, we assume $\eps,\delta\le1/4$ and $0<r^*<\infty$.
Recall that $F(p)=\Pr_{v\sim F}[v<p]$, $Q(p)=1-F(p)$, $\Rprice(p)=pQ(p)$, and $r^*=\sup_{p\ge0}\Rprice(p)$.
We will use the characterization of quasi-regularity from~\cite{FJ25}: the function $B(p)\eqdef pQ(p)/F(p)$ is nonincreasing for $p>0$, with $B(p)=+\infty$ when $F(p)=0$.
All pricing queries below use independent draws from $F$.

{We restate the main steps below.}
\begin{enumerate}
 \item \textbf{Estimate the monopoly revenue} ({Module~\ref{overview:quasi-regular:1}}).
 Using the single sample and $O(\log(1/\delta))$ queries, find an estimate $\Hat{r^{*}}\in[r^*,\frac{11}{10}r^*]$ with probability at least $1-\delta/4$.
 \item \textbf{Sparsify the candidate price set} (Module~\ref{overview:quasi-regular:2}).
 Given $\Hat{r^{*}}$, set
 \[
  [\underline p,\Bar p]\eqdef
  \left[\frac{10}{11}\Hat{r^{*}},\frac{2}{\eps}\Hat{r^{*}}\right],
  \qquad
  \Pcandidate\eqdef\{p_i=\Bar p/i:i\in[\Bar i]\},
  \qquad
  \Bar i\eqdef\left\lceil\frac{\Bar p}{\underline p}\right\rceil.
 \]
 This set has size $\Theta(1/\eps)$ and contains a price with revenue at least $(1-\eps/2)r^*$.
 Write $r_i=\Rprice(p_i)$ and $q_i=Q(p_i)$ for each $i\in[\Bar i]$.
 \item \textbf{Find a threshold index} (Module~\ref{overview:quasi-regular:3}).
 Find $\underline i\in[\Bar i]$ such that $r_{\underline i}\ge0.5\Hat{r^{*}}$ and $r_i\le0.7\Hat{r^{*}}$ for every $i<\underline i$.
 An optimal grid price remains among the indices $\underline i,\ldots,\Bar i$; moreover, every such index satisfies $0.3\Hat{r^{*}}\le r_i\le\Hat{r^{*}}$ and $q_i\ge\eps i/6$.
 Steps~1--3 succeed jointly with probability at least $1-\delta/2$.
 \item \textbf{Solve the modified Best Arm Identification problem} (Module~\ref{overview:quasi-regular:4}).
 Select a nearly optimal price from the remaining candidates using $O(\eps^{-3}\log(1/\eps)\log(1/\delta))$ queries, and return it as $\Hat{p^{*}}$.
\end{enumerate}

\subsubsection{Step 1: Estimating the Monopoly Revenue}

Here, we aim to derive the estimate $\Hat{r^{*}}$ required by Module~\ref{overview:quasi-regular:1}.
{We state the revenue-estimation guarantee first, and then use the {\NoisyPriceLocation} subproblem from Module~\ref{overview:quasi-regular:1} to prove it.}
For Step~1, we apply the lemma with $\alpha=\delta/4$ and denote its output $\rho$ by $\Hat{r^{*}}$.

\begin{lemma}
\label{lem:quasi-regular:scale}
Fix $\alpha\in(0,1/4]$ and a quasi-regular distribution with $0<r^*<\infty$.
One initial sample and at most $O(\log(1/\alpha))$ pricing queries suffice to return $\rho\ge0$ such that $r^*\le\rho\le(11/10)r^*$ with probability at least $1-\alpha$.
\end{lemma}

The main idea is to first find a price $p'$ such that $[Q(1.01p'),Q(p')]\cap[0.01,0.02]\ne\emptyset$.
Quasi-regularity then bounds the revenue of every price above $p'$, while prices below $0.01p'$ cannot outperform $p'$.
Thus, we can concentrate on $[0.01p',p']$, discretize it into a constant-size geometric grid, and estimate the revenue of each grid price.
To find $p'$, we use the following finite-grid formulation of {\NoisyPriceLocation}.

\begin{definition}[{\NoisyPriceLocation}]
\label{def:quasi-regular:noisy-price-location}
Given $\kappa,\eta,\delta'\in(0,1)$, let $G$ be a finite grid of positive prices whose consecutive prices have ratio at most $1+\kappa$.
Suppose that
\[
 Q(\min G)\ge\frac43\eta
 \qquad\text{and}\qquad
 Q(\max G)\le\frac23\eta.
\]
Using pricing queries at prices in $G$, the task is to find consecutive prices $p',p''\in G$ such that
\[
 \frac{p''}{p'}\le1+\kappa,
 \qquad
 [Q(p''),Q(p')]\cap\left[\frac23\eta,\frac43\eta\right]\ne\emptyset,
\]
with probability at least $1-\delta'$.
\end{definition}

Since $Q$ is monotone, this is an instance of noisy binary search.
For the revenue estimate, we only need a fixed target $\eta=0.015$; the following consequence of~\cite[Theorem~1]{GP24} is sufficient.

\begin{lemma}[{\cite[Theorem~1]{GP24}}]
\label{lem:quasi-regular:noisy-price-location}
For any fixed $\eta\in(0,1/2]$, the \textnormal{\NoisyPriceLocation} problem can be solved using at most $O(\log|G|+\log(1/\delta'))$ pricing queries, for every $\kappa\in(0,1)$ and $\delta'\in(0,1/4]$.
The hidden constant depends only on $\eta$.
\end{lemma}

Now, we are ready to prove \Cref{lem:quasi-regular:scale} using \Cref{lem:quasi-regular:noisy-price-location}.

\begin{proof}[Proof of \Cref{lem:quasi-regular:scale}]
Set $\kappa=0.01$, $\eta=0.015$, and $\delta'=\alpha/2$.
We first implement {\NoisyPriceLocation} with total failure probability $\delta'$, including the choice of its initial price range.
Let $\inisample$ be the initial sample and put $\beta=\delta'/4$.
With probability at least $1-2\beta=1-\delta'/2$, the sample satisfies
\[
 Q(\inisample)\ge\beta
 \qquad\text{and}\qquad
 \Pr_{v\sim F}[v\le\inisample]\ge\beta.
\]
Condition on this event.
{By the monotonicity of $B$, we have $B(x)\ge\beta\inisample$ for $0<x<\inisample$ and $B(x)\le\inisample/\beta$ for $x>\inisample$.}
Therefore, the prices $\ell=\beta\inisample/4$ and $h=400\inisample/\beta$ satisfy $Q(\ell)\ge4/5$ and $Q(h)\le1/401$.
The geometric grid
\[
 G=\left\{\ell(1+\kappa)^j:
 0\le j\le\left\lceil\log_{1+\kappa}\frac h\ell\right\rceil\right\}
\]
has size $O(\log(1/\delta'))$ and satisfies the endpoint conditions in \Cref{def:quasi-regular:noisy-price-location}.
Apply \Cref{lem:quasi-regular:noisy-price-location} with failure probability $\delta'/2$.
Using $O(\log|G|+\log(1/\delta'))=O(\log(1/\alpha))$ queries, it returns consecutive prices $p'$ and $p''=(1+\kappa)p'$ such that
\[
 [Q(p''),Q(p')]\cap[0.01,0.02]\ne\emptyset.
\]
The sample event and this call succeed jointly with probability at least $1-\delta'$.

Condition on their success, so that $Q(p')\ge0.01$ and $Q(p'')\le0.02$.
For $p\in[p',p'']$, we have $\Rprice(p)\le(1+\kappa)\Rprice(p')$ by monotonicity of $Q$; for $p\ge p''$, quasi-regularity gives
\[
 \Rprice(p)\le B(p)\le B(p'')
 =\frac{p''Q(p'')}{1-Q(p'')}
 \le\frac{1+\kappa}{0.98}\Rprice(p').
\]
Also, prices below $0.01p'$ have revenue at most $0.01p'\le\Rprice(p')$.
It remains to search the interval $[0.01p',p']$.
For this purpose, define the geometric grid
\[
 G=\left\{p_i=p'(1+\kappa)^{-i}:
 0\le i\le\left\lceil\log_{1+\kappa}100\right\rceil\right\}.
\]
Rounding a price in this interval down to $G$ loses a factor of at most $1+\kappa$ in revenue.
Combining this with the bounds outside the interval, we obtain
\[
 \max_{p_i\in G}\Rprice(p_i)\ge\frac{0.98}{1+\kappa}r^*.
\]

Every price in $G$ has sale probability at least $0.01$.
By the multiplicative Chernoff bound, $O(\log(|G|/\alpha))$ queries at each $p_i\in G$ suffice to obtain an estimate $\Hat r_i\in[0.99\Rprice(p_i),1.01\Rprice(p_i)]$, simultaneously for all prices with probability at least $1-\alpha/2$.
Return $\rho=\max_{p_i\in G}\Hat r_i/0.95$.
Since $\kappa=0.01$, we have
\[
 r^*
 \le\frac{0.99\cdot0.98}{0.95(1+\kappa)}r^*
 \le\rho
 \le\frac{1.01}{0.95}r^*
 \le\frac{11}{10}r^*.
\]
The failure probabilities sum to $\delta'+\alpha/2=\alpha$.
The first search uses $O(\log(1/\alpha))$ queries, and so do the estimates on the constant-size grid $G$.
\end{proof}

\subsubsection{Step 2: Sparsifying the Candidate Price Set}

Given the estimate $\Hat{r^{*}}$, we next aim to construct a size-$\Theta(1/\eps)$ set $\Pcandidate$ containing a nearly optimal price.
The straightforward approach is to place a geometric $(1+\Theta(\eps))$-grid over $[\underline p,\Bar p]$, but this produces $\Theta(\eps^{-1}\log(1/\eps))$ candidates.
To avoid this additional factor, we use the reciprocal grid $p_i=\Bar p/i$ from Module~\ref{overview:quasi-regular:2}.
The key observation is that lowering a price increases inverse revenue by at most the increase in inverse price; since $1/p_i=i/\Bar p$, this gives a uniform bound on the loss from rounding to our grid.

\begin{lemma}
\label{lem:quasi-regular:harmonic}
For a quasi-regular distribution, if $0<s\le p$ and $r=\Rprice(p)>0$, then
\[
 \Rprice(s)\ge\frac{r}{1+r(1/s-1/p)}.
\]
Suppose also that $0<r^*<\infty$, $\eta\in(0,1/4]$, and $\rho\in[(9/10)r^*,(11/10)r^*]$.
For any $M\ge2/\eta$ and integer $n\ge\lceil11M/10\rceil$, the prices $p_i=M\rho/i$, $i\in[n]$, satisfy
\[
 \max_{i\in[n]}\Rprice(p_i)
 \ge\frac{r^*}{1+r^*/(M\rho)}
 \ge(1-\eta)r^*.
\]
In particular, one may take $M=\lceil2/\eta\rceil$ and $n=2M$.
\end{lemma}

\begin{proof}
Quasi-regularity implies
\[
 \frac1{\Rprice(s)}-\frac1s
 =\frac{F(s)}{sQ(s)}
 \le\frac{F(p)}{pQ(p)}
 =\frac1{\Rprice(p)}-\frac1p,
\]
and rearranging gives the first inequality.
{For the second assertion, $p_n=M\rho/n\le(10/11)\rho\le r^*$, so the revenue supremum can be taken over prices $p\ge p_n$.}
For any such price, choose $i=\lceil M\rho/p\rceil\in[n]$.
Then $p_i\le p$ and $1/p_i-1/p\le1/(M\rho)$, so
\[
 \Rprice(p_i)\ge\frac{\Rprice(p)}{1+\Rprice(p)/(M\rho)}.
\]
Taking the supremum over $p\ge p_n$ yields
\[
 \max_{i\in[n]}\Rprice(p_i)
 \ge\frac{r^*}{1+r^*/(M\rho)}
 \ge(1-\eta)r^*.\qedhere
\]
\end{proof}

For our candidate set, apply \Cref{lem:quasi-regular:harmonic} with
$\rho=\Hat{r^{*}}$, $M=2/\eps$, $n=\Bar i=\lceil11/(5\eps)\rceil$, and $\eta=\eps$.
Because $\Hat{r^{*}}\ge r^*$, its guarantee becomes
\[
 \max_{i\in[\Bar i]}r_i
 \ge\frac{r^*}{1+r^*/\Bar p}
 \ge\frac{r^*}{1+\eps/2}
 \ge(1-\eps/2)r^*.
\]
Thus, $\Pcandidate$ has the desired approximation guarantee and contains only $\Theta(1/\eps)$ prices.
Clearly, this step makes no queries.

\subsubsection{Step 3: Finding a Threshold Index}

Next, as in Module~\ref{overview:quasi-regular:3}, we find an index $\underline i$ with $r_{\underline i}\ge0.5\Hat{r^{*}}$ and $r_i\le0.7\Hat{r^{*}}$ for every $i<\underline i$.
The gap between these two thresholds allows us to use constant-accuracy revenue estimates.
Moreover, quasi-regularity extends the revenue lower bound at $\underline i$ to every index $i\ge\underline i$, which also bounds the cost of estimating each remaining price.
The following lemma summarizes Steps~1--3; we state it with general grid parameters so that it can also be used for regular distributions.

\begin{lemma}
\label{lem:quasi-regular:preprocess}
Fix a quasi-regular distribution with $0<r^*<\infty$, parameters $\eta,\alpha\in(0,1/4]$, a real number $M\ge2/\eta$, and an integer $n$ with $\lceil11M/10\rceil\le n\le2M$.
Using one initial sample and at most $O(M\log n\log(n/\alpha))$ pricing queries, an algorithm returns $\rho\ge0$ and $a_0\in[n]$ such that, with probability at least $1-\alpha$, the following hold.
Write
\[
 p_i=M\rho/i,\qquad q_i=Q(p_i),\qquad
 u_i=\Rprice(p_i)/\rho\qquad(i\in[n]).
\]
\begin{enumerate}
 \item We have $\rho\in[r^*,(11/10)r^*]$ and
 \[
  \max_{i\in[n]}\Rprice(p_i)
  \ge\frac{r^*}{1+r^*/(M\rho)}
  \ge(1-\eta)r^*.
 \]
 \item We have $u_{a_0}\ge0.55$ and $u_i\le0.65$ for every $i<a_0$.
 An optimal grid index belongs to $\{a_0,\ldots,n\}$.
 \item For every $i\ge a_0$, we have $1/4\le u_i\le1$ and $q_i\ge i/(4M)$.
\end{enumerate}
The choice $M=\lceil2/\eta\rceil$, $n=2M$ is a special case.
The query bound holds on every execution.
\end{lemma}

The revenue estimate and grid approximation follow from Steps~1 and~2.
To find the threshold, we estimate each candidate revenue to constant accuracy and select the first index whose estimate exceeds a fixed threshold; the estimation error is small enough that no grid maximizer is discarded.
Finally, the inverse-revenue inequality extends this revenue lower bound to all larger indices, giving the revenue and quantile bounds in the last assertion.

\begin{proof}
Apply \Cref{lem:quasi-regular:scale} with failure probability $\alpha/2$ and construct the grid $p_i=M\rho/i$.

\paragraph{Revenue estimate and grid approximation.}
On the success event of scale estimation, we have $r^*\le\rho\le(11/10)r^*$, and \Cref{lem:quasi-regular:harmonic} gives the stated grid approximation.
Below, we condition on this event.

\paragraph{Finding the threshold.}
A pricing query at $p_i$ gives a sale indicator $Y_i\sim\operatorname{Bernoulli}(q_i)$.
The normalized revenue observation $Z_i=(M/i)Y_i$ has mean $u_i\le1$, range $[0,M/i]$, and variance at most $M/i$.
By Bernstein's inequality, $\lceil C(M/i)\log(4n/\alpha)\rceil$ queries suffice to estimate $u_i$ to additive error $0.05$ with failure probability at most $\alpha/(2n)$, for a sufficiently large constant $C$.
Let $\widehat u_i$ be these estimates and choose $a_0$ as the smallest index with $\widehat u_{a_0}\ge0.6$.
By a union bound, all estimates are accurate with probability at least $1-\alpha/2$.
On this event,
\[
 \max_{i\in[n]}u_i\ge\frac{1-\eta}{1.1}>0.65,
\]
so the threshold exists, $u_{a_0}\ge0.55$, and every earlier index satisfies $u_i\le0.65$.
In particular, no grid maximizer is removed.

\paragraph{Revenue and quantile bounds.}
For $i\ge a_0$, the first inequality in \Cref{lem:quasi-regular:harmonic} gives
\[
 u_i\ge\frac{1}{1/u_{a_0}+(i-a_0)/M}
 \ge\frac{1}{2+(i-a_0)/M}
 \ge\frac14,
\]
where the last inequality uses $i\le n\le2M$.
Also, $u_i\le r^*/\rho\le1$ and $q_i=(i/M)u_i\ge i/(4M)$, as required.

\paragraph{Query complexity.}
Denote by $H_n=\sum_{i=1}^n1/i$ the $n$-th harmonic number, with $H_0=0$.
The coarse estimates use
\[
 O\!\left(MH_n\log\frac n\alpha\right)
 =O\!\left(M\log n\log\frac n\alpha\right)
\]
queries, which also absorbs the cost of scale estimation.
These query budgets do not depend on the outcomes, and the two failure probabilities sum to at most $\alpha$.
\end{proof}

In our algorithm, we use \Cref{lem:quasi-regular:preprocess} with $\eta=\eps$, $\alpha=\delta/2$, $M=2/\eps$, and $n=\Bar i$, and write $\Hat{r^{*}}$ and $\underline i$ for its outputs $\rho$ and $a_0$.
The lemma combines Steps~1--3 using the single initial sample, and preserves an optimal candidate price.
It gives $r_{\underline i}\ge0.55\Hat{r^{*}}$ and $r_i\le0.65\Hat{r^{*}}$ for $i<\underline i$.
For the particular grid used here, the lower bound on every remaining revenue can be sharpened:
\[
 \frac{r_i}{\Hat{r^{*}}}
 \ge\frac{1}{20/11+(\eps/2)(i-\underline i)}
 >\frac{1}{20/11+11/10}
 >\frac13,
 \qquad i\in\{\underline i,\ldots,\Bar i\},
\]
where we used $i-\underline i\le\Bar i-1<11/(5\eps)$.
Consequently,
\[
 0.3\Hat{r^{*}}\le r_i\le\Hat{r^{*}},
 \qquad
 q_i=\frac{\eps i}{2}\frac{r_i}{\Hat{r^{*}}}\ge\frac{\eps i}{6},
 \qquad i\ge\underline i.
\]
At the threshold itself, $q_{\underline i}\ge\eps\underline i/4$ follows from $r_{\underline i}\ge0.5\Hat{r^{*}}$.
These are precisely the bounds used in Module~\ref{overview:quasi-regular:3}.

\subsubsection{Step 4: Solving the Modified Best Arm Identification Problem}

Now, we are facing the modified {\BestArmIdentification} problem from Module~\ref{overview:quasi-regular:4}.
Each remaining price $p_i$ is an arm with expected reward $r_i/\Hat{r^{*}}\in[0.3,1]$, so additive and multiplicative accuracies differ only by constant factors.
Given $\eps',\delta'\in(0,1/4]$, an additive-$\eps'$ estimate of this reward costs
\[
 O\!\left(\frac{1}{i\eps\,\eps'^2}\log\frac1{\delta'}\right)
\]
pricing queries.
A union bound over all arms would introduce an additional $\log(1/\eps)$ factor.
To avoid this loss, we adapt the Median Elimination algorithm of~\cite{EMM06}.
We first recall the problem and its standard guarantee.

\begin{definition}[{\BestArmIdentification}]
\label{def:quasi-regular:best-arm-identification}
Consider $N$ arms with unknown reward distributions and means $\mu_1,\ldots,\mu_N$.
Pulling an arm gives an independent draw from its reward distribution.
Given accuracy $t>0$ and failure probability $\alpha\in(0,1)$, the task is to output an arm $\widehat i$ with $\mu_{\widehat i}\ge\max_{i\in[N]}\mu_i-t$, with probability at least $1-\alpha$.
\end{definition}

\begin{lemma}[{\cite[Theorem~10]{EMM06}}]
\label{lem:quasi-regular:median-elimination}
For $N$ arms with rewards in $[0,1]$ and $t,\alpha\in(0,1/4]$, Median Elimination solves the \textnormal{\BestArmIdentification} problem with $O(Nt^{-2}\log(1/\alpha))$ observations.
\end{lemma}

In our problem, the expected rewards are bounded, but the reward from a single pull can be as large as $p_i/\Hat{r^{*}}=2/(\eps i)$.
Applying the bounded-reward guarantee after rescaling would incur a quadratic dependence on this range.
Instead, the bounded means give a variance bound that is linear in the range, so Bernstein's inequality yields the following version of Median Elimination.

\begin{lemma}
\label{lem:quasi-regular:bernstein-elimination}
Consider $N$ arms with independent rewards in $[0,b]$, means $\mu_i\in[0,2]$, and $b\ge1/2$.
For $t,\alpha\in(0,1/4]$, a variant of Median Elimination solves the \textnormal{\BestArmIdentification} problem using at most $O(Nbt^{-2}\log(1/\alpha))$ observations.
\end{lemma}

\begin{proof}
For each reward $X\in[0,b]$, we have $\operatorname{Var}(X)\le\mathbb E[X^2]\le b\mathbb E[X]\le2b$.
We run Median Elimination with sample sizes chosen using this variance bound.
Let $A_s$ be the surviving arms in round $s$, starting with all $N$ arms, and set
\[
 t_s=\frac t4\left(\frac34\right)^{s-1},
 \qquad
 \alpha_s=\frac{\alpha}{2^s}.
\]
In each round with $|A_s|>1$, estimate every surviving mean using $\lceil Cbt_s^{-2}\log(16/\alpha_s)\rceil$ observations and retain the $\lceil|A_s|/2\rceil$ arms with the largest estimates.
Return the final surviving arm.

Conditional on the preceding rounds, Bernstein's inequality bounds the probability that any specified estimate has error greater than $t_s/2$ by $\alpha_s/8$.
Thus, an arm of maximum mean in $A_s$ has an accurate estimate with probability at least $1-\alpha_s/8$.
Also, the expected number of inaccurate estimates is at most $|A_s|\alpha_s/8$; by Markov's inequality, fewer than $|A_s|/4$ estimates are inaccurate with probability at least $1-\alpha_s/2$.
When both events hold, either a maximizing arm survives, or an accurately estimated surviving arm has an estimate at least as large as the maximizing arm's estimate.
In either case,
\[
 \max_{i\in A_{s+1}}\mu_i\ge\max_{i\in A_s}\mu_i-t_s.
\]
Since $\sum_s t_s=t$ and $\sum_s\alpha_s=\alpha$, the returned arm satisfies the required guarantee.

Moreover, $|A_s|\le2N/2^{s-1}$ whenever at least two arms remain.
Hence, the total number of observations is
\[
 O\!\left(Nbt^{-2}\sum_{s\ge1}
 \left(\frac89\right)^{s-1}
 \left(\log\frac1\alpha+s\right)\right)
 =O\!\left(Nbt^{-2}\log\frac1\alpha\right).\qedhere
\]
\end{proof}

To apply this lemma, we group the remaining indices between consecutive powers of two.
A group with indices $2^j\le i<2^{j+1}$ contains at most $2^j$ arms, each with reward range at most $2/(\eps2^j)$.
Thus, the product of the group size and its reward range is $O(1/\eps)$, and each group can be searched with the same query budget.
We then compare the prices returned by these searches.

\subsubsection{Putting Everything Together}

Finally, we finish the proof of \Cref{thm:quasi-regular:UB-query} by combining the above lemmas together.

\begin{proof}[Proof of \Cref{thm:quasi-regular:UB-query}]
Run Steps~1--3 as specified above.
By \Cref{lem:quasi-regular:preprocess} and its application to our grid, with probability at least $1-\delta/2$ we have
\[
 \max_{\underline i\le i\le\Bar i}r_i\ge(1-\eps/2)r^*,
 \qquad
 0.3\Hat{r^{*}}\le r_i\le\Hat{r^{*}}
 \quad(i\ge\underline i).
\]
For the accuracy analysis, fix a successful transcript of these steps.

For each $0\le j\le\lfloor\log_2\Bar i\rfloor$, let
\[
 A_j=\{i:\underline i\le i\le\Bar i,\ 2^j\le i<2^{j+1}\},
\]
and consider the nonempty groups.
A pull of arm $i\in A_j$ consists of querying price $p_i$ and receiving reward $p_i/\Hat{r^{*}}$ on a sale and zero otherwise.
Its mean is $r_i/\Hat{r^{*}}\le1$, and its range is bounded by
$b=\Bar p/(2^j\Hat{r^{*}})=2/(\eps2^j)\ge1/2$.
Apply \Cref{lem:quasi-regular:bernstein-elimination} with accuracy $t=\eps/16$ and failure probability $\alpha=\delta/4$, and denote the returned price by $\Hat p_j$.
Because $|A_j|\le2^j$, each call uses $O(\eps^{-3}\log(1/\delta))$ queries.

Let $j^*$ be a group containing a revenue-maximizing candidate price.
This group is fixed by the transcript of Steps~1--3.
Its call succeeds with probability at least $1-\delta/4$, giving
\[
 \Rprice(\Hat p_{j^*})
 \ge\max_{\underline i\le i\le\Bar i}r_i
 -\frac{\eps}{16}\Hat{r^{*}}.
\]
We only need this guarantee for the group containing an optimal candidate; the other groups may return arbitrary prices from their respective sets.

Finally, estimate the revenue of each $\Hat p_j$ to additive error $(\eps/16)\Hat{r^{*}}$ with failure probability $\delta/2^{j+3}$, and return the price with the largest estimate as $\Hat{p^{*}}$.
Conditional on the prices returned by the group searches, Bernstein's inequality gives the required estimates using
\[
 O\!\left(\frac{1}{\eps^3 2^j}
 \log\frac{2^{j+4}}{\delta}\right)
\]
queries for $\Hat p_j$.
Their failure probabilities sum to at most $\delta/4$.
When these estimates and the search in group $j^*$ succeed,
\begin{align*}
 \Rprice(\Hat{p^{*}})
 &\ge\Rprice(\Hat p_{j^*})-\frac{\eps}{8}\Hat{r^{*}}\\
 &\ge\max_{\underline i\le i\le\Bar i}r_i
      -\frac{3\eps}{16}\Hat{r^{*}}\\
 &\ge(1-\eps/2)r^*-\frac{3\eps}{16}\cdot\frac{11}{10}r^*
 \ge(1-\eps)r^*.
\end{align*}
Together with Steps~1--3, the total failure probability is at most $\delta/2+\delta/4+\delta/4=\delta$.

There are $O(\log\Bar i)=O(\log(1/\eps))$ groups, so the group searches use $O(\eps^{-3}\log(1/\eps)\log(1/\delta))$ queries in total.
The final comparison uses
\[
 O\!\left(\frac1{\eps^3}\sum_{j\ge0}2^{-j}
 \left(\log\frac1\delta+j+1\right)\right)
 =O\!\left(\frac1{\eps^3}\log\frac1\delta\right)
\]
queries, and the cost of Steps~1--3 is also absorbed by the claimed bound.
All subroutines use prescribed query budgets, so this bound holds on every execution.
Only Step~1 draws a value sample.
\end{proof}

\subsection{Sample Complexity Upper Bound}
\label{subsubsec:quasi-regular:UB-sample}

Next, we consider learning from value samples without pricing queries. To be specific, we prove the following: 

\begin{restate}[\Cref{thm:quasi-regular:UB-sample}]
Fix $0<\eps,\delta\le1/4$.
For the family of quasi-regular distributions, there exists an $(\eps, \delta)$-PAC learning algorithm that draws at most {$O\bigl(\frac{1}{\eps^{3}}\log(\frac{1}{\delta})\bigr)$} many samples and makes no pricing queries.
\end{restate}

We first briefly introduce the main ideas of the proof. 
Quasi-regularity guarantees that restricting to prices with sale probability at least $\Theta(\eps)$ loses only an $O(\eps)$ fraction of the optimal revenue; see \pref{eq:quasi-regular:sample:quantile-cutoff} in \pref{sec:sample-complexity} for the complete argument and the proof. 
We therefore maximize empirical revenue subject to an empirical sale-probability cutoff.
To compare the selected price with a nearly optimal feasible price, we need a relative-error guarantee that holds simultaneously for all prices above the cutoff, and an upper bound on empirical sale probabilities below it. The sale-probability form of the multiplicative-accuracy DKW inequality, \pref{thm:multiplicative-DKW}, and its corresponding corollary, \pref{cor:multiplicative-DKW-sale-probability}, provide both guarantees. We prove the two theorems in \pref{sec:DKW}.

For $\eta=\eps/8$, we apply \pref{cor:multiplicative-DKW-sale-probability} with quantile threshold $q=\eta$ and relative-error tolerance $\eps'=\eta$, using $O(\eps^{-3}\log(1/\delta))$ samples.
We choose an empirical revenue maximizer among prices whose empirical sale probabilities are at least $2\eta$.
On the uniform concentration event, every price with true sale probability at least $4\eta$ is feasible, and every feasible price has true sale probability at least $\eta$.
Thus the relative-error bounds apply to both the returned price and a nearly optimal feasible price.
The uniform guarantee avoids a union bound over a price grid and the resulting extra $\log(1/\eps)$ factor.
We defer the complete cutoff and revenue-comparison proof to the quasi-regular proof in \pref{sec:sample-complexity}.

\input{quasi-regular-LB}

%% file: quasi-regular-LB.tex
\subsection{Lower Bounds}
\label{subsec:quasi-regular:LB}

Having established the query and sample upper bounds, we now prove lower bounds that account for both value samples and pricing queries.
We construct hard distributions whose $(1-\eps)$-optimal prices identify them, so that a PAC algorithm must collect enough information to distinguish each of them from a base distribution.
Specifically, the $i$-th hard distribution $F^{(i)}$ agrees with a base distribution $F^{(0)}$ outside a price interval $P^{(i)}$, and these intervals are pairwise disjoint.
A value sample has positive KL divergence between $F^{(0)}$ and every $F^{(i)}$.
A pricing query at price $p$ observes only whether the value is at least $p$, and this observation has the same law under $F^{(0)}$ and $F^{(i)}$ unless $p\in P^{(i)}$, because the two distributions agree outside $P^{(i)}$.
Hence a pricing query at price $p$ has positive KL divergence only between $F^{(0)}$ and the single $F^{(i)}$ with $p\in P^{(i)}$.
The following lemma sums the bound of \pref{lem:kl-bound} over the hard distributions, with weights that balance the query bounds of different intervals.
It shows that an algorithm whose output identifies each hard distribution must collect, on weighted average over the pairs $(F^{(0)},F^{(i)})$, KL divergence at least $\frac1{16}\log\frac1\delta$ from its value samples and pricing queries.

\begin{lemma}
\label{lem:LB:weighted-information}
Consider a base distribution $F^{(0)}$ and alternative distributions $F^{(i)}$ indexed by a finite set $I$, where each $F^{(i)}$ agrees with $F^{(0)}$ outside a price interval $P^{(i)}$.
Suppose that $\DKL(F^{(0)},F^{(i)})\le d_i$ and that $\DKL(F^{(0)}(p),F^{(i)}(p))\le b_i$ for every $p\in P^{(i)}$.
For a learning algorithm, denote by $S$ its number of value samples, by $T^{(i)}$ its number of pricing queries with prices in $P^{(i)}$, and by $\prob[F^{(j)}]{\cdot}$ and $\expect[F^{(j)}]{\cdot}$ the probability and expectation over its transcript under $F^{(j)}$.
Suppose that output events $\scrE^{(i)}$ satisfy $\prob[F^{(i)}]{\scrE^{(i)}}\ge1-\delta$ for every $i\in I$, where $0<\delta\le1/4$.
Let $\{w_i\}_{i\in I}$ be positive weights with sum $W\triangleq\sum_{i\in I}w_i$.
If $\sum_{i\in I}w_i\prob[F^{(0)}]{\scrE^{(i)}}\le W/2$, then
\begin{equation}
\label{eq:LB:weighted-information}
 \sum_{i\in I}w_i\bigl(d_i\,\expect[F^{(0)}]{S}+b_i\,\expect[F^{(0)}]{T^{(i)}}\bigr)
 \ge\frac{W}{16}\log\frac1\delta.
\end{equation}
\end{lemma}

\begin{proof}
Denote $a_i\triangleq\prob[F^{(0)}]{\scrE^{(i)}}$ and $v_i\triangleq\prob[F^{(i)}]{\scrE^{(i)}}$.
We obtain that
\begin{align*}
 &\sum_{i\in I}w_i\bigl(d_i\,\expect[F^{(0)}]{S}+b_i\,\expect[F^{(0)}]{T^{(i)}}\bigr)\\
 &\ge\sum_{i\in I}w_i\DKL(a_i,v_i)\ge W\,\DKL\Bigl(\frac{\sum_iw_ia_i}{W},\frac{\sum_iw_iv_i}{W}\Bigr)
 \ge W\,\DKL\Bigl(\frac12,1-\delta\Bigr)
 \ge\frac W{16}\log\frac1\delta,
\end{align*}
where the first inequality applies \pref{lem:kl-bound} with the price set $P^{(i)}$ to each pair $(F^{(0)},F^{(i)})$, and the second uses the joint convexity of binary KL divergence.
The third inequality uses the monotonicity of binary KL divergence, together with $\sum_iw_iv_i/W\ge1-\delta$ and $\sum_iw_ia_i/W\le1/2$.
The last inequality uses $\DKL(\frac12,1-\delta)=\frac12\log\frac1{4\delta(1-\delta)}$ and $\delta\le1/4$.
\end{proof}

\subsubsection{Query-Sample Tradeoff}

The $\log\frac1\eps$ factor in the query bound of \pref{thm:quasi-regular:UB-query} cannot be removed whenever the algorithm draws at most $\frac{C}{\eps^{3-c}}\log\frac1\delta$ value samples for a constant $c>0$, which is polynomially fewer than the $\Theta(\eps^{-3}\log\frac1\delta)$ value samples that suffice without pricing queries.
Starting from the equal-revenue distribution, we construct $L=\Theta(\eps^{-\min\{c,1\}/2})$ hard distributions, each of which raises the revenue on its own price interval, so that a $(1-\eps)$-optimal price identifies the hard distribution.
A value sample is informative about every hard distribution, whereas a pricing query is informative only about the one whose interval contains its price.
Since the KL divergence of a query grows linearly with the index of its interval, we apply \pref{lem:LB:weighted-information} with weights $1/i$, under which every query has the same weighted contribution.
These weights sum to $\Theta(\log L)$, which produces the $\log\frac1\eps$ factor.
Under the sample budget, the value samples supply at most half of the KL divergence required by \pref{lem:LB:weighted-information}, so the pricing queries must supply the rest.
The following theorem formally states this lower bound.

\begin{theorem}[Lower Bound for Quasi-Regular Distributions]
\label{thm:quasi-regular:LB-tradeoff}
There exists an absolute constant $C>0$ such that the following holds.
Fix a constant $c>0$, $0<\delta\le1/4$, and sufficiently small $\eps>0$.
If an $(\eps,\delta)$-PAC learning algorithm for quasi-regular distributions draws at most $\frac{C}{\eps^{3-c}}\log\frac1\delta$ value samples, then there exists a quasi-regular distribution on which it makes at least $\frac{C\min\{c,1\}}{\eps^3}\log\frac1\eps\log\frac1\delta$ pricing queries in expectation.
\end{theorem}

To prove \Cref{thm:quasi-regular:LB-tradeoff}, we first construct the base and hard distributions.
Fix a constant $c>0$ and a sufficiently small $\eps>0$, and set $\bar c\triangleq\min\{c,1\}$, $K\triangleq\lfloor1/(8\eps)\rfloor$, and $L\triangleq\lfloor\eps^{-\bar c/2}\rfloor$.
These choices satisfy $1/(16\eps)\le K\le1/(8\eps)$, $L\ge4$, and $L+1\le K/4$, so every $i\in[2:L+1]$ satisfies $i/K\le1/4$.
For $i\in[2:L+1]$, denote by $P^{(i)}\triangleq(\frac Ki,\frac K{i-1})$ the price interval on which the $i$-th hard distribution $F^{(i)}$ differs from $F^{(0)}$.

We define the distributions through their revenue-quantile curves.
The base distribution $F^{(0)}$ is the equal-revenue distribution, with $\Rquantile^{(0)}(q)\triangleq1$ for $q\in[0,1]$.
For $i\in[2:L+1]$, the $i$-th hard distribution $F^{(i)}$ has the revenue-quantile curve
\begin{equation*}
 \Rquantile^{(i)}(q)\triangleq
 \begin{cases}
  1+\frac12\bigl(q-\frac{i-1}K\bigr),&\frac{i-1}K\le q<\frac{2i-1}{2K},\\
  1+\frac12\bigl(\frac iK-q\bigr),&\frac{2i-1}{2K}\le q<\frac iK,\\
  1,&\text{otherwise},
 \end{cases}
\end{equation*}
whose two affine sides of slopes $\pm\frac12$ meet at the revenue $1+\frac1{4K}$ at the price $p^{i,*}\triangleq\frac{4K+1}{2(2i-1)}$.
Inverting the price $\Rquantile^{(i)}(q)/q$ gives $F^{(0)}(p)=\max\{1-1/p,0\}$ and
\begin{equation*}
 F^{(i)}(p)=
 \begin{cases}
  1-\frac{1+i/(2K)}{p+1/2},&\frac Ki<p\le p^{i,*},\\
  1-\frac{1-(i-1)/(2K)}{p-1/2},&p^{i,*}<p<\frac K{i-1},\\
  F^{(0)}(p),&p\notin P^{(i)}.
 \end{cases}
\end{equation*}
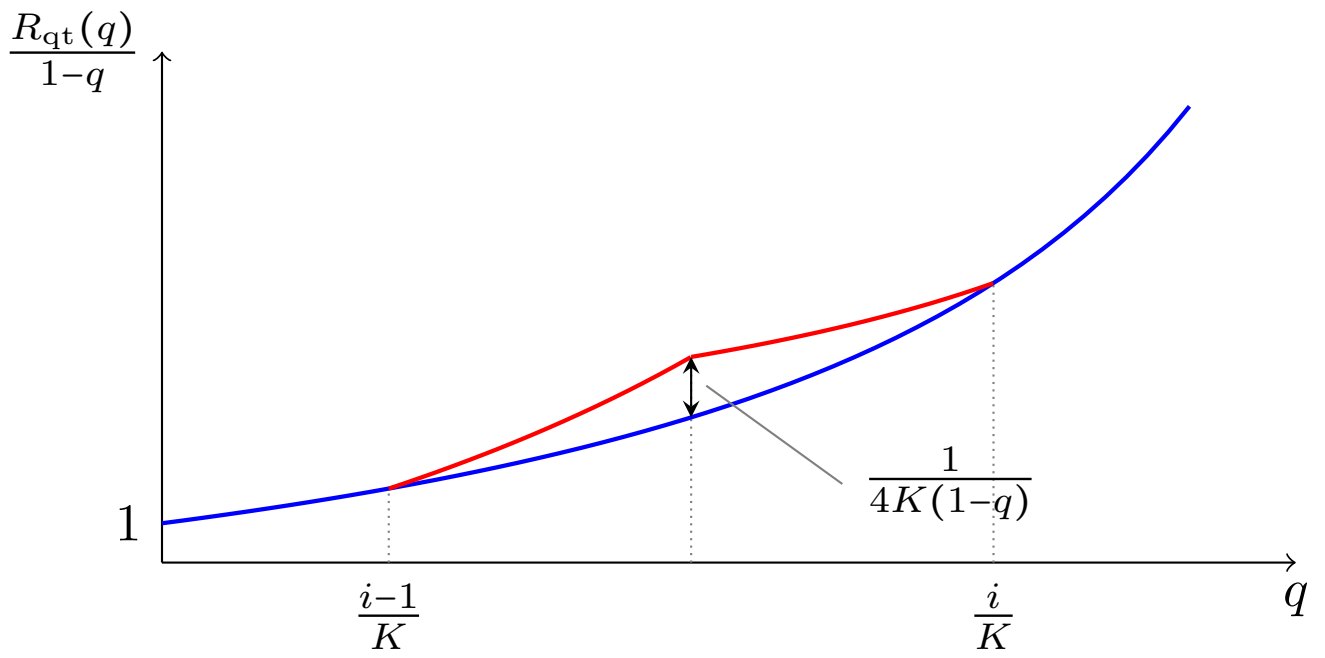
\begin{figure}[htbp]
\centering
\begin{tikzpicture}[x=10cm,y=1.3cm,font=\small]
 \draw[->] (0,0) -- (0.75,0) node[below]{$q$};
 \draw[->] (0,0) -- (0,2.6) node[left]{$\frac{\Rquantile(q)}{1-q}$};
 \draw[gray,densely dotted] (0.15,0) -- (0.15,{1/0.85-0.8}) (0.35,0) -- (0.35,{1.2/0.65-0.8}) (0.55,0) -- (0.55,{1/0.45-0.8});
 \draw[blue,thick,domain=0:0.68,samples=50] plot (\x,{1/(1-\x)-0.8});
 \draw[red,thick,domain=0.15:0.35,samples=20] plot (\x,{(1+(\x-0.15))/(1-\x)-0.8});
 \draw[red,thick,domain=0.35:0.55,samples=20] plot (\x,{(1+(0.55-\x))/(1-\x)-0.8});
 \draw[<->,>=stealth] (0.35,{1/0.65-0.8}) -- (0.35,{1.2/0.65-0.8});
 \draw[gray] (0.36,0.9) -- (0.45,0.4) node[right,black]{$\frac{1}{4K(1-q)}$};
 \node[left] at (0,0.2) {$1$};
 \node[below] at (0.15,0) {$\frac{i-1}K$};
 \node[below] at (0.55,0) {$\frac iK$};
\end{tikzpicture}
\caption{The ratio $\Rquantile(q)/(1-q)$ under $F^{(0)}$ (blue) and $F^{(i)}$ (red), not to scale. The tent of $\Rquantile^{(i)}$ raises the ratio by at most $\frac{1}{4K(1-q)}$, and the ratio stays nondecreasing, which is the quasi-regularity condition.}
\label{fig:quasi-regular:LB}
\end{figure}

\noindent The two distributions agree at both endpoints of $P^{(i)}$ and have no atoms, so they assign the same probability to $P^{(i)}$.
The affine sides have nonzero intercepts to avoid an atom.
If the increasing side passed through the origin, the price would be constant on a quantile range of length $\frac{i-1}{K(K-1)}$, which creates an atom of this mass at $\frac K{i-1}$.
Since $F^{(0)}$ has no atom there, the data-processing inequality would then give a sample KL divergence of at least $\frac{i-1}{K(K-1)}$, much larger than the bound $O(i^2/K^3)$ needed for the tradeoff.

The following lemma verifies that these distributions are quasi-regular.

\begin{lemma}[Quasi-Regularity]
\label{lem:quasi-regular:quasi-regular}
The base distribution $F^{(0)}$ and the hard distributions $\{F^{(i)}\}_{i\in[2:L+1]}$ are quasi-regular distributions.
\end{lemma}

\begin{proof}
By the equivalent form of \Cref{def:lambda-quasi-regular} with $\regularity=1$ stated after it, $F^{(i)}$ is quasi-regular if the function $pQ_i(p)/F^{(i)}(p)$ is nonincreasing in $p$, where $Q_i\triangleq1-F^{(i)}$ is the sale probability.
At the price with sale probability $q$, this function equals $\Rquantile^{(i)}(q)/(1-q)$, and $q$ decreases as $p$ increases, so it suffices that $\Rquantile^{(i)}(q)/(1-q)$ is nondecreasing in $q$ (see \Cref{fig:quasi-regular:LB}).
Writing each affine piece of $\Rquantile^{(i)}$ in terms of $1-q$, we obtain
\begin{equation*}
 \frac{\Rquantile^{(i)}(q)}{1-q}=
 \begin{cases}
  -\frac12+\bigl(\frac32-\frac{i-1}{2K}\bigr)\frac1{1-q},&\frac{i-1}K\le q<\frac{2i-1}{2K},\\
  \frac12+\bigl(\frac12+\frac i{2K}\bigr)\frac1{1-q},&\frac{2i-1}{2K}\le q<\frac iK,\\
  \frac1{1-q},&\text{otherwise}.
 \end{cases}
\end{equation*}
Each piece is a constant plus a positive multiple of $\frac1{1-q}$, and the pieces join continuously, so $\Rquantile^{(i)}(q)/(1-q)$ is nondecreasing.
The third piece alone proves the claim for $F^{(0)}$.
\end{proof}

We next show that a $(1-\eps)$-optimal price under $F^{(i)}$ identifies $F^{(i)}$.
Under $F^{(i)}$, the optimal revenue is $1+\frac1{4K}\ge1+2\eps$, whereas the revenue at prices outside $P^{(i)}$ is at most one.
Since $(1-\eps)(1+2\eps)>1$, every $(1-\eps)$-optimal price under $F^{(i)}$ lies in $P^{(i)}$.
The following lemma bounds the KL divergence of a value sample between $F^{(0)}$ and $F^{(i)}$.

\begin{lemma}[KL Divergence of Samples]
\label{lem:quasi-regular:LB:sample-KL}
For every $i\in[2:L+1]$, we have $\DKL(F^{(0)},F^{(i)})\le4i^2/K^3$.
\end{lemma}

\begin{proof}
The density $f_0(p)=1/p^2$ for $p>1$ of $F^{(0)}$ and the density $f_i$ of $F^{(i)}$ agree outside $P^{(i)}$.
On $P^{(i)}$, differentiating the displayed CDF gives $f_i(p)=\frac{1+i/(2K)}{(p+1/2)^2}$ for $p<p^{i,*}$ and $f_i(p)=\frac{1-(i-1)/(2K)}{(p-1/2)^2}$ for $p>p^{i,*}$.
Denote by $z(p)\triangleq f_i(p)/f_0(p)-1$ the relative density change on $P^{(i)}$.
Since $\frac1{2p}\le\frac i{2K}\le\frac18$ on $P^{(i)}$, the ratio $\frac{f_i(p)}{f_0(p)}=\frac{1+i/(2K)}{(1+1/(2p))^2}$ lies in $[(1+\frac i{2K})^{-1},1+\frac i{2K}]$ for $p<p^{i,*}$, and the ratio $\frac{f_i(p)}{f_0(p)}=\frac{1-(i-1)/(2K)}{(1-1/(2p))^2}$ lies in $[1-\frac i{2K},(1-\frac i{2K})^{-2}]$ for $p>p^{i,*}$.
Since $(1-x)^{-2}-1\le3x$ for $0\le x\le\frac18$, both cases give $|z(p)|\le\frac{3i}{2K}\le\frac{2i}K\le\frac12$.
Since the two distributions assign the same probability to $P^{(i)}$, $\int_{P^{(i)}}f_0(p)z(p)\,\d p=0$, and the mass of $P^{(i)}$ under $F^{(0)}$ is $\frac iK-\frac{i-1}K=\frac1K$.
As in~\cite[Lemma~4.4]{HMR18}, only the changed interval contributes to the divergence.
Expanding the KL divergence by its definition and substituting $f_i=(1+z)f_0$ on $P^{(i)}$, we obtain that
\begin{align*}
 &\DKL(F^{(0)},F^{(i)})\\
 &=\int_{P^{(i)}}f_0(p)\log\frac{f_0(p)}{f_i(p)}\,\d p
 =\int_{P^{(i)}}f_0(p)\bigl[-\log(1+z(p))+z(p)\bigr]\,\d p\\
 &\le\int_{P^{(i)}}f_0(p)z(p)^2\,\d p
 \le\frac1K\Bigl(\frac{2i}K\Bigr)^2
 =\frac{4i^2}{K^3},
\end{align*}
where the second equality uses $\int_{P^{(i)}}f_0(p)z(p)\,\d p=0$, the first inequality uses $-\log(1+z)+z\le z^2$ for $|z|\le\frac12$, and the second inequality uses $|z(p)|\le\frac{2i}K$ and the mass $\frac1K$ of $P^{(i)}$.
\end{proof}

The following lemma bounds the KL divergence of a pricing query at a price in $P^{(i)}$.

\begin{lemma}[KL Divergence of Queries]
\label{lem:quasi-regular:LB:query-KL}
For every $i\in[2:L+1]$ and every query price $p\in P^{(i)}$, we have $\DKL(F^{(0)}(p),F^{(i)}(p))\le i/(8K^3)$.
\end{lemma}

\begin{proof}
For $p\in P^{(i)}$, the base sale probability is $Q_0(p)=1/p\le i/K\le1/4$.
Since the revenue under $F^{(i)}$ lies in $[1,1+\frac1{4K}]$, we have $Q_0(p)\le Q_i(p)\le(1+\frac1{4K})Q_0(p)\le1/2$.
Expanding the Bernoulli KL divergence by its definition, we obtain that
\begin{align*}
 &\DKL(F^{(0)}(p),F^{(i)}(p))\\
 &=Q_0(p)\log\frac{Q_0(p)}{Q_i(p)}+(1-Q_0(p))\log\frac{1-Q_0(p)}{1-Q_i(p)}\\
 &\le\frac{(Q_i(p)-Q_0(p))^2}{Q_i(p)(1-Q_i(p))}
 \le\frac{(Q_0(p)/(4K))^2}{Q_0(p)/2}
 =\frac{Q_0(p)}{8K^2}
 \le\frac{i}{8K^3},
\end{align*}
where the first inequality is the chi-squared bound, obtained by applying $\log x\le x-1$ to both logarithms, the second inequality uses the two bounds on $Q_i(p)$ above, and the last inequality uses $Q_0(p)\le i/K$.
\end{proof}

We now prove \Cref{thm:quasi-regular:LB-tradeoff} by combining the good-price separation with \Cref{lem:quasi-regular:LB:sample-KL,lem:quasi-regular:LB:query-KL} through \pref{lem:LB:weighted-information}.

\begin{proof}[Proof of \Cref{thm:quasi-regular:LB-tradeoff}]
Consider an $(\eps,\delta)$-PAC learning algorithm $\calA$, and denote by $p^\calA$ its output price, by $S$ its number of value samples, by $T$ its number of pricing queries, and by $T^{(i)}$ the number of its queries with prices in $P^{(i)}$.
Denote by $\prob[F^{(j)}]{\cdot}$ and $\expect[F^{(j)}]{\cdot}$ the probability and expectation over its transcript under $F^{(j)}$.
The events $\scrE^{(i)}\triangleq\{p^\calA\in P^{(i)}\}$ for $i\in[2:L+1]$ are pairwise disjoint.
Since every $(1-\eps)$-optimal price under $F^{(i)}$ lies in $P^{(i)}$, the PAC guarantee gives $\prob[F^{(i)}]{\scrE^{(i)}}\ge1-\delta$.
We apply \pref{lem:LB:weighted-information} with the index set $[2:L+1]$, the weights $w_i=1/i$, and the bounds $d_i=4i^2/K^3$ and $b_i=i/(8K^3)$ from \Cref{lem:quasi-regular:LB:sample-KL,lem:quasi-regular:LB:query-KL}.
The weights sum to $W\triangleq\sum_{i=2}^{L+1}1/i\ge1$ since $L\ge4$.
Since the events $\scrE^{(i)}$ are pairwise disjoint, they satisfy $\sum_iw_i\prob[F^{(0)}]{\scrE^{(i)}}\le\max_iw_i=1/2\le W/2$, and we obtain
\begin{equation*}
 \sum_{i=2}^{L+1}\Bigl(\frac{4i}{K^3}\,\expect[F^{(0)}]{S}+\frac{1}{8K^3}\,\expect[F^{(0)}]{T^{(i)}}\Bigr)
 \ge\frac W{16}\log\frac1\delta.
\end{equation*}
The weight $1/i$ cancels the factor $i$ in the query bound, so every pricing query has the same weighted contribution $1/(8K^3)$ whichever interval contains its price.
Each value sample appears in all terms on the left, whereas each pricing query appears only in the term of the interval containing its price, so $\sum_{i=2}^{L+1}T^{(i)}\le T$.
Since also $\sum_{i=2}^{L+1}i\le(L+1)^2$, multiplying by $8K^3$ gives
\begin{equation*}
 32(L+1)^2\,\expect[F^{(0)}]{S}+\expect[F^{(0)}]{T}
 \ge\frac{K^3W}2\log\frac1\delta.
\end{equation*}
Since $K\ge1/(16\eps)$, $(L+1)^2\le4\eps^{-\bar c}$ with $\bar c\le c$, and $W\ge\int_2^{L+2}\frac{\d x}x=\log\frac{L+2}2\ge\frac{\bar c}4\log\frac1\eps$ for sufficiently small $\eps$, there is an absolute constant $C>0$ such that $\frac{C}{\eps^{3-c}}\le\frac{K^3}{128(L+1)^2}$ and $\frac{C\bar c}{\eps^3}\log\frac1\eps\le\frac{K^3W}4$.
By the first inequality, the sample budget $\expect[F^{(0)}]{S}\le\frac{C}{\eps^{3-c}}\log\frac1\delta$ gives $32(L+1)^2\,\expect[F^{(0)}]{S}\le\frac{K^3}4\log\frac1\delta\le\frac{K^3W}4\log\frac1\delta$, where the last step uses $W\ge1$.
The last display and the second inequality therefore give
\begin{equation*}
 \expect[F^{(0)}]{T}
 \ge\frac{K^3W}4\log\frac1\delta
 \ge\frac{C\min\{c,1\}}{\eps^3}\log\frac1\eps\log\frac1\delta.
\end{equation*}
\end{proof}

%% file: regular.tex
\section{Regular Distributions}
\label{sec:regular}


In this section, we establish the query and sample upper bounds and the query-sample lower bound for regular distributions.
We use the same notation and conventions as in \Cref{sec:quasi-regular}.

\subsection{Overview of the Section}
\label{subsec:regular:overview}

\begin{figure}[htbp]
    \centering
    \includegraphics[width=.44\textwidth]{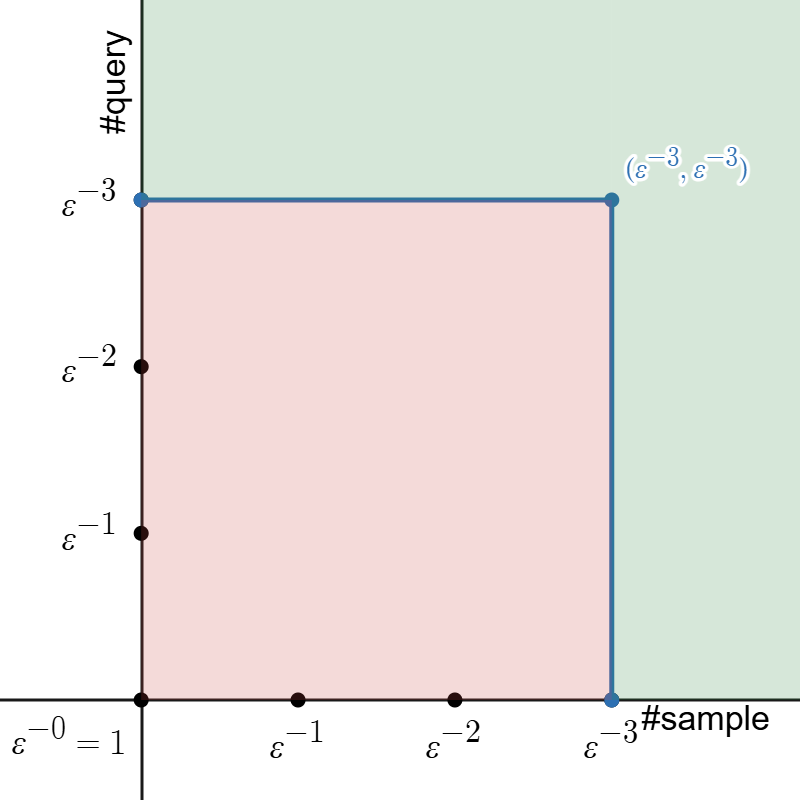}
    \caption{The regular query-sample tradeoff from \Cref{fig:result-regular}.
    Both axes are logarithmic; constant factors and $\log(1/\delta)$ are suppressed.
    Green and red denote the learnable and unlearnable regimes.}
    \label{fig:regular:tradeoff}
\end{figure}

Our goal is to establish the tight tradeoff shown in \Cref{fig:regular:tradeoff}, reproduced from \Cref{fig:result-regular}.
As before, $\samplecount$ and $\querycount$ denote the sample and query budgets, respectively.

For the upper bounds, it suffices to establish the two endpoints
\begin{align*}
    (\samplecount,\querycount)
    &=\left(1,\ O\bigl(\eps^{-3}\log(1/\delta)\bigr)\right),\\
    (\samplecount,\querycount)
    &=\left(O\bigl(\eps^{-3}\log(1/\delta)\bigr),\ 0\right),
\end{align*}
since an algorithm can simply ignore any additional samples or queries.
We first prove \Cref{thm:regular:UB-query} in \Cref{subsec:regular:UB}, giving the single-sample query endpoint.
Next, \Cref{subsec:regular:UB-sample} presents the sample-only endpoint, \Cref{thm:regular:UB-sample}; the complete proof appears in the regular-distribution subsection of \Cref{sec:sample-complexity}.

For the lower bound, it suffices to establish the corner
\[
    (\samplecount,\querycount)
    =\left(O\bigl(\eps^{-3}\log(1/\delta)\bigr),\
      \Omega\bigl(\eps^{-3}\log(1/\delta)\bigr)\right),
\]
where the constant in the sample budget is sufficiently small and the query lower bound holds in expectation.
We prove this bound in \Cref{subsec:regular:LB}, as \Cref{thm:regular:LB-tradeoff}.
It matches the single-sample query upper bound and also gives the matching sample-only lower bound.

\subsection{Query Complexity Upper Bound with a Single Sample}
\label{subsec:regular:UB}

We first prove the query upper bound for the regular family.
Since regularity implies quasi-regularity, we can reuse Steps~1--3 of the algorithm in \Cref{subsec:quasi-regular:UB}.
The additional concavity of the revenue-quantile curve allows us to replace the final {\BestArmIdentification} step by the {\NoisyConcaveOptimization} procedure from Module~\ref{overview:regular:4}.
This removes the $\log(1/\eps)$ factor from the quasi-regular query bound.

\begin{theorem}[Upper Bound for Regular Distributions]
\label{thm:regular:UB-query}
Fix $0<\eps,\delta\le1/4$.
For the family of regular distributions, there exists an $(\eps,\delta)$-PAC learning algorithm that first draws a single sample and then makes at most $O\bigl(\frac{1}{\eps^3}\log(\frac{1}{\delta})\bigr)$ pricing queries.
\end{theorem}

Our algorithm consists of the following two steps.
\begin{enumerate}[label=\textbf{Step \arabic*:},leftmargin=*]
    \item \textbf{Run Steps~1--3 of the quasi-regular algorithm.}
    Use accuracy $\eps/16$ and failure probability $\delta/2$ to estimate the monopoly revenue, construct the reciprocal grid, and find a threshold index that preserves an optimal grid price.
    Denote the resulting estimate and threshold by $\Hat{r^{*}}$ and $\underline i$, and write
    \[
        M\eqdef32/\eps,\qquad
        \Bar p\eqdef M\Hat{r^{*}},\qquad
        \Bar i\eqdef\lceil11M/10\rceil,\qquad
        \Pcandidate\eqdef\{p_i\eqdef\Bar p/i:i\in[\Bar i]\}.
    \]
    As before, let $r_i=\Rprice(p_i)$ and $q_i=Q(p_i)$.
    \item \textbf{Solve the noisy concave optimization problem.}
    Run a noise-tolerant ternary search on the remaining indices $[\underline i:\Bar i]$.
    At each iteration, compare revenue estimates at the two trisection indices and discard one third of the interval.
    Estimate the remaining prices when the interval has constant size, and return the price with the largest estimate among all prices estimated during the search, including those outside the final interval.
\end{enumerate}

The guarantee for Step~1 follows directly from \Cref{lem:quasi-regular:preprocess}.
Apply it with $\eta=\eps/16$, $\alpha=\delta/2$, $M=32/\eps$, and $n=\Bar i$, and write $\Hat{r^{*}}$ and $\underline i$ for its outputs $\rho$ and $a_0$.
With probability at least $1-\delta/2$, we have $r^*\le\Hat{r^{*}}\le(11/10)r^*$ and
\begin{equation}
    \max_{i\in[\underline i:\Bar i]}r_i
    =\max_{i\in[\Bar i]}r_i
    \ge(1-\eps/16)r^*.
    \label{eq:regular:preprocess-grid}
\end{equation}
Moreover, every remaining index satisfies
\[
    \frac14\Hat{r^{*}}\le r_i\le\Hat{r^{*}},
    \qquad q_i\ge\frac{i}{4M},
    \qquad i\in[\underline i:\Bar i].
\]
This step uses the single initial sample and $O\bigl(\eps^{-1}\log(1/\eps)\log(1/(\eps\delta))\bigr)$ queries.
Next, we analyze the performance of Step 2 via the above guarantees and prove \Cref{thm:regular:UB-query}.

\begin{proof}[Proof of \Cref{thm:regular:UB-query}]
Set $\zeta=\eps/10^3$ and $\delta_i=\delta/(8i^2)$.
A query at $p_i$ gives a sale indicator $Y_i\sim\operatorname{Bernoulli}(q_i)$, so $p_iY_i$ has mean $r_i$, range $[0,p_i]$, and variance at most
\[
    p_i r_i\le\frac{M}{i}\bigl(\Hat{r^{*}}\bigr)^2.
\]
By Bernstein's inequality, $O\bigl(\frac{M}{i\zeta^2}\log(1/\delta_i)\bigr)$ queries suffice to obtain an estimate $\Hat r_i$ with additive error at most $\zeta\Hat{r^{*}}$ and failure probability at most $\delta_i$.
We estimate each index only when it is first requested and reuse that estimate thereafter.
Since each estimate uses fresh queries, its failure probability is at most $\delta_i$ even conditional on the history before its first request.
A union bound shows that all requested estimates are accurate with probability at least $1-\sum_{i\ge1}\delta_i\ge1-\delta/2$.

Initialize the search interval as $[a,b]=[\underline i,\Bar i]$.
While $L\eqdef b-a>12$, set
\[
    \ell\eqdef a+\lfloor L/3\rfloor,
    \qquad h\eqdef a+\lfloor2L/3\rfloor.
\]
If $\Hat r_\ell\le\Hat r_h$, keep $[\ell,b]$; otherwise, keep $[a,h]$.
At termination, estimate all remaining indices and return the price $\Hat{p^{*}}$ with the largest estimate among all indices estimated during the search.

To analyze the revenue guarantee, condition also on the accuracy of all these estimates.
Fix an index $k$ maximizing $r_i$ over the initial interval $[\underline i:\Bar i]$.
We consider the first iteration that discards $k$, if such an iteration occurs.
Write $x_i\eqdef i/\Bar p$, so $q_i=x_i r_i$. Suppose the first such iteration keeps $[\ell,b]$, so $k<\ell$.
Unimodality gives $r_k\ge r_\ell\ge r_h$.
Let $D=r_k-r_\ell$ and $d=r_\ell-r_h$.
Since $\Hat r_\ell\le\Hat r_h$, accuracy of the estimates gives $0\le d\le2\zeta\Hat{r^{*}}$.
Concavity implies
\[
    D(q_h-q_\ell)\le d(q_\ell-q_k).
\]
Substituting $q_i=x_i r_i$ and collecting the terms containing $D$, we obtain
\[
    D\bigl[r_\ell(x_h-x_\ell)-d(x_h-x_k)\bigr]
    \le d r_\ell(x_\ell-x_k).
\]
The trisection points satisfy
\[
    \frac{x_\ell-x_k}{x_h-x_\ell}\le2,
    \qquad
    \frac{x_h-x_k}{x_h-x_\ell}\le4.
\]
Since $r_\ell\ge\Hat{r^{*}}/4$ and $d\le2\zeta\Hat{r^{*}}$, the expression in brackets is at least $\frac12 r_\ell(x_h-x_\ell)$.
It follows that $D\le4d\le8\zeta\Hat{r^{*}}$, so the estimated price $p_\ell$ is already close to the original grid optimum.

For the other case, the first such iteration keeps $[a,h]$ and $k>h$.
Let $D=r_k-r_h$ and $d=r_h-r_\ell\in[0,2\zeta\Hat{r^{*}}]$.
Concavity and the same substitution give
\[
    D\bigl[r_h(x_h-x_\ell)-d(x_k-x_\ell)\bigr]
    \le d r_h(x_k-x_h).
\]
The corresponding trisection ratios are again at most $2$ and $4$, so $D\le8\zeta\Hat{r^{*}}$ also holds in this case.
Thus, either $k$ survives to the final enumeration, or a price within $8\zeta\Hat{r^{*}}$ of its revenue has already been estimated.
Choosing the largest estimate loses at most another $2\zeta\Hat{r^{*}}$.
Together with \eqref{eq:regular:preprocess-grid}, this gives
\begin{align*}
    \Rprice(\Hat{p^{*}})
    &\ge\max_{i\in[\Bar i]}r_i-10\zeta\Hat{r^{*}}\\
    &\ge(1-\eps/16)r^*-11\zeta r^*
    \ge(1-\eps)r^*.
\end{align*}
Only the first iteration that discards $k$ enters this argument, so the approximation losses do not accumulate over the search.

It remains to bound the number of queries.
Let $L_1,\ldots,L_K>12$ be the interval lengths before the ternary cuts.
They satisfy $L_{t+1}\le(3/4)L_t$, and both indices requested at iteration $t$ are at least $L_t/4$.
Consequently, $L_t>12(4/3)^{K-t}$.
Both $1/i$ and $\log(ei)/i$ are nonincreasing for $i\ge1$, so the sums over all distinct estimated indices, including the at most $13$ final indices, satisfy
\begin{align*}
    \sum_{i\text{ estimated}}\frac1i
    &\le13+\frac23\sum_{j\ge0}(3/4)^j=O(1),\\*
    \sum_{i\text{ estimated}}\frac{\log(ei)}i
    &\le13+\frac23\sum_{j\ge0}(3/4)^j
      \bigl(\log(3e)+j\log(4/3)\bigr)=O(1).
\end{align*}
Each index is estimated only once, so the search uses
\[
    O\!\left(\frac{M}{\zeta^2}
    \sum_{i\text{ estimated}}\frac{\log(8i^2/\delta)}i\right)
    =O\!\left(\eps^{-3}\log\frac1\delta\right)
\]
queries.
The cost of Step~1 is absorbed by this bound, and the query budgets hold on every execution.
Combining the two failure probabilities gives a total failure probability of at most $\delta$.
\end{proof}

\subsection{Sample Complexity Upper Bound}
\label{subsec:regular:UB-sample}

For learning from value samples alone, the quasi-regular algorithm applies directly because every regular distribution is quasi-regular.

\begin{restate}[\Cref{thm:regular:UB-sample}]
Fix $0<\eps,\delta\le1/4$.
For the family of regular distributions, there exists an $(\eps,\delta)$-PAC learning algorithm that draws at most $O\bigl(\frac{1}{\eps^3}\log(\frac{1}{\delta})\bigr)$ samples and makes no pricing queries.
\end{restate}

We use the same empirical revenue maximizer and sale-probability cutoff as in \Cref{thm:quasi-regular:UB-sample}.
The complete argument appears in the regular-distribution subsection of \Cref{sec:sample-complexity}.

\subsection{Lower Bounds}
\label{subsec:regular:LB}

For regular distributions, the query bound in \pref{thm:regular:UB-query} cannot be improved even if the algorithm also draws up to $\frac{C}{\eps^3}\log\frac1\delta$ value samples for a small absolute constant $C$.
We use the two exact regular distributions $F^{(0)}$ and $F^{(1)}$ from the sample lower bound in~\cite[Theorem~4.8]{HMR18}, whose sets of $(1-\eps)$-optimal prices are disjoint.
They agree below a price $u=\Theta(1/\eps)$, and above $u$ both assign probability $\Theta(\eps)$ and their densities differ by a factor $1\pm O(\eps)$.
As a result, a value sample has KL divergence $O(\eps^3)$ between them, as shown in~\cite[Lemma~4.9]{HMR18}.
As explained in \Cref{subsec:quasi-regular:LB}, a pricing query at a price $p\le u$ has the same law under the two distributions.
A pricing query at a price $p>u$ also has KL divergence $O(\eps^3)$, because both sale probabilities are $O(\eps)$ and differ by a factor $1-O(\eps)$.
By \pref{lem:LB:weighted-information} with a single alternative distribution, a PAC algorithm must collect KL divergence at least $\frac1{16}\log\frac1\delta$ from its value samples and pricing queries.
Under the sample budget, the value samples supply at most half of it, so the pricing queries must supply the rest, which requires $\Omega(\eps^{-3}\log\frac1\delta)$ queries.
The following theorem formally states this lower bound.

\begin{theorem}[Lower Bound for Regular Distributions]
\label{thm:regular:LB-tradeoff}
There exists an absolute constant $C>0$ such that the following holds.
Fix $0<\delta\le1/4$ and $0<\eps\le1/48$.
If an $(\eps,\delta)$-PAC learning algorithm for regular distributions draws at most $\frac{C}{\eps^3}\log\frac1\delta$ value samples, then there exists a regular distribution on which it makes at least $\frac{C}{\eps^3}\log\frac1\delta$ pricing queries in expectation.
\end{theorem}

To prove \Cref{thm:regular:LB-tradeoff}, we first construct the two distributions.
We use the two regular distributions constructed in~\cite[proof of Theorem~4.8]{HMR18}, with the same parameter $\eps_0=3\eps$ as there.
Set $B\triangleq1-6\eps$ and $u\triangleq B/(6\eps)$, and define $F^{(j)}(p)\triangleq1-Q_j(p)$ on $p\ge0$ for $j\in\{0,1\}$, where $Q_0(p)\triangleq1/(p+1)$ and
\begin{equation*}
 Q_1(p)\triangleq
 \begin{cases}
  1/(p+1),&p\le u,\\
  B^2/(p-B),&p>u.
 \end{cases}
\end{equation*}
The two branches of $Q_1$ both equal $6\eps$ at $u$, so $Q_0$ and $Q_1$ decrease continuously from one to zero and are the sale probabilities of $F^{(0)}$ and $F^{(1)}$.
The two distributions agree on $[0,u]$ and both assign probability $6\eps$ to $(u,\infty)$.
\cite[proof of Theorem~4.8]{HMR18} shows that the sets of $(1-3\eps)$-optimal prices under $F^{(0)}$ and $F^{(1)}$ are disjoint, so their sets of $(1-\eps)$-optimal prices are also disjoint.
\cite[Lemma~4.9]{HMR18} bounds the sum of the KL divergences between $F^{(0)}$ and $F^{(1)}$ in both directions by $8(3\eps)^3/(1-6\eps)^2$.
Since $1-6\eps\ge7/8$, a value sample therefore satisfies
\begin{equation*}
 \DKL(F^{(0)},F^{(1)})
 \le\frac{8(3\eps)^3}{(1-6\eps)^2}
 \le2^9\eps^3.
\end{equation*}

The following lemma bounds the KL divergence of a pricing query at a price above $u$.

\begin{lemma}[KL Divergence of Queries]
\label{lem:regular:LB:query-KL}
For every query price $p>u$, we have $\DKL(F^{(0)}(p),F^{(1)}(p))\le2^{12}\eps^3$.
\end{lemma}

\begin{proof}
For $p>u$, the base sale probability is $Q_0(p)\le Q_0(u)=6\eps$.
Since $(p+1)/(p-B)$ decreases from $(u+1)/(u-B)=B^{-2}$ to one on $(u,\infty)$, the ratio $Q_1(p)/Q_0(p)=B^2(p+1)/(p-B)$ lies in $[B^2,1]$.
Hence $Q_0(p)-Q_1(p)\le(1-B^2)Q_0(p)\le12\eps Q_0(p)$, $Q_1(p)\ge B^2Q_0(p)\ge Q_0(p)/2$, and $1-Q_1(p)\ge1-6\eps\ge\frac12$.
Expanding the Bernoulli KL divergence by its definition, we obtain that
\begin{align*}
 &\DKL(F^{(0)}(p),F^{(1)}(p))\\
 &=Q_0(p)\log\frac{Q_0(p)}{Q_1(p)}+(1-Q_0(p))\log\frac{1-Q_0(p)}{1-Q_1(p)}\\
 &\le\frac{(Q_0(p)-Q_1(p))^2}{Q_1(p)(1-Q_1(p))}
 \le\frac{(12\eps Q_0(p))^2}{Q_0(p)/4}
 =576\eps^2Q_0(p)
 \le2^{12}\eps^3,
\end{align*}
where the first inequality is the chi-squared bound from the proof of \Cref{lem:quasi-regular:LB:query-KL}, the second inequality uses the three bounds above, and the last inequality uses $Q_0(p)\le6\eps$.
\end{proof}

We now prove \Cref{thm:regular:LB-tradeoff} by combining the disjointness of the good-price sets with the two divergence bounds through \pref{lem:LB:weighted-information}.

\begin{proof}[Proof of \Cref{thm:regular:LB-tradeoff}]
Consider an $(\eps,\delta)$-PAC learning algorithm, with $p^\calA$, $S$, $T$, $\prob[F^{(j)}]{\cdot}$, and $\expect[F^{(j)}]{\cdot}$ as in the proof of \Cref{thm:quasi-regular:LB-tradeoff}.
Denote by $\scrE^{(1)}$ the event that $p^\calA$ is $(1-\eps)$-optimal under $F^{(1)}$, and by $T^{(1)}$ the number of pricing queries with prices in $(u,\infty)$.
The PAC guarantee under $F^{(1)}$ gives $\prob[F^{(1)}]{\scrE^{(1)}}\ge1-\delta$.
Since the two sets of $(1-\eps)$-optimal prices are disjoint, the PAC guarantee under $F^{(0)}$ gives $\prob[F^{(0)}]{\scrE^{(1)}}\le\delta\le1/2$.
We apply \pref{lem:LB:weighted-information} with $I=\{1\}$, the price interval $P^{(1)}=(u,\infty)$, the weight $w_1=1$, and the bounds $d_1=2^9\eps^3$ from~\cite[Lemma~4.9]{HMR18} and $b_1=2^{12}\eps^3$ from \Cref{lem:regular:LB:query-KL}.
Since $T^{(1)}\le T$, we obtain
\begin{equation*}
 2^9\eps^3\,\expect[F^{(0)}]{S}+2^{12}\eps^3\,\expect[F^{(0)}]{T}
 \ge\frac1{16}\log\frac1\delta.
\end{equation*}
Set $C\triangleq2^{-17}$.
The sample budget $\expect[F^{(0)}]{S}\le\frac{C}{\eps^3}\log\frac1\delta$ gives $2^9\eps^3\,\expect[F^{(0)}]{S}\le2^9C\log\frac1\delta\le\frac1{32}\log\frac1\delta$.
The last display therefore gives
\begin{equation*}
 \expect[F^{(0)}]{T}
 \ge\frac1{2^{17}\eps^3}\log\frac1\delta
 =\frac{C}{\eps^3}\log\frac1\delta,
\end{equation*}
so the algorithm makes at least $\frac{C}{\eps^3}\log\frac1\delta$ pricing queries in expectation on the regular distribution $F^{(0)}$.
\end{proof}

%% file: quasi-MHR.tex
\section{Quasi-MHR Distributions}
\label{sec:quasi-MHR}

{In this section, we establish the query and sample upper bounds and the tight query-sample tradeoff for quasi-MHR distributions.
We use the same notation and conventions as in \Cref{sec:quasi-regular}, and write $H(p)=-\log Q(p)$ for the cumulative hazard.}

{\subsection{Overview of the Section}
\label{subsec:quasi-MHR:overview}}

\begin{figure}[htbp]
    \centering
    \includegraphics[width=.44\textwidth]{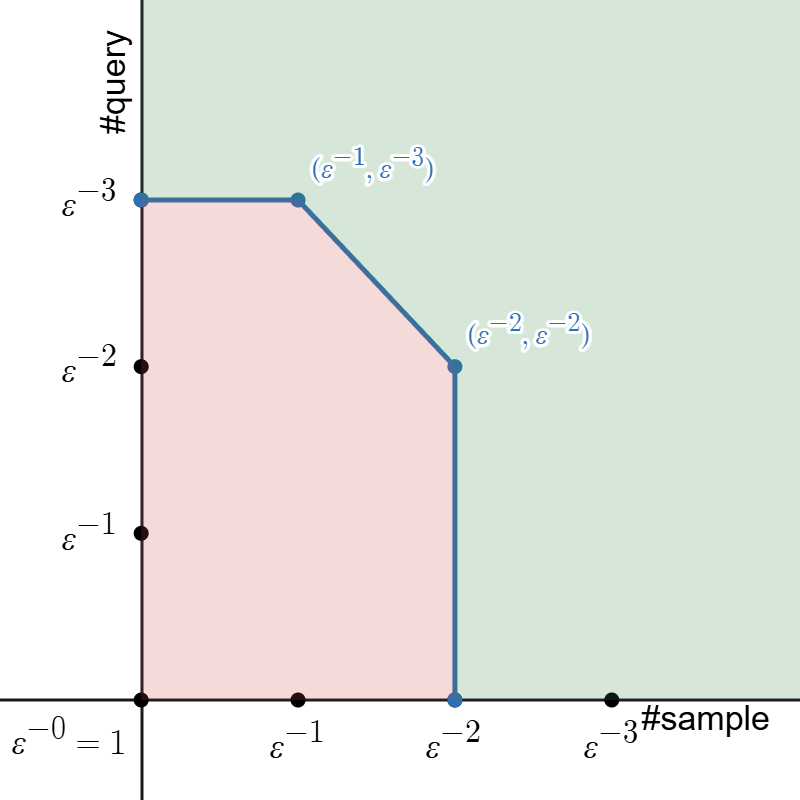}
    \caption{The quasi-MHR query-sample tradeoff from \Cref{fig:result-quasi-MHR}.
    Both axes are logarithmic; constant factors and $\log(1/\delta)$ are suppressed.
    Green and red denote the learnable and unlearnable regimes.}
    \label{fig:quasi-MHR:tradeoff}
\end{figure}

{Our goal is to establish the tight tradeoff shown in \Cref{fig:quasi-MHR:tradeoff}, reproduced from \Cref{fig:result-quasi-MHR}.
As before, $\samplecount$ and $\querycount$ denote the sample and query budgets, respectively.}

{For the upper bounds, we establish the two endpoints
\begin{align*}
    (\samplecount,\querycount)
    &=\left(1,\ O\bigl(\eps^{-3}\log(1/\delta)\bigr)\right),\\
    (\samplecount,\querycount)
    &=\left(O\bigl(\eps^{-2}\log(1/\delta)\bigr),\ 0\right),
\end{align*}
and the family of intermediate slope
\[
    (\samplecount,\querycount)
    =\left(O\bigl(\eps^{-(2-c)}\log(1/\delta)\bigr),\
      O\bigl(\eps^{-(2+c)}\log(1/\delta)\bigr)\right),
    \qquad c\in[0,1].
\]
We first prove \Cref{thm:quasi-MHR:UB-query} in \Cref{subsec:quasi-MHR:UB}, giving the single-sample query endpoint.
Next, \Cref{subsec:quasi-MHR:UB-sample} presents the sample-only endpoint, \Cref{thm:quasi-MHR:UB-sample}; the complete proof appears in the quasi-MHR subsection of \Cref{sec:sample-complexity}.
Finally, \Cref{subsec:quasi-MHR:UB-tradeoff} proves \Cref{thm:quasi-MHR:UB-tradeoff}, giving the intermediate slope by using samples to reduce the number of prices compared by queries.}

{For the lower bounds, we need the matching slope lower bound
\[
    (\samplecount,\querycount)
    =\left(O\bigl(\eps^{-(2-c)}\log(1/\delta)\bigr),\
      \Omega\bigl(\eps^{-(2+c)}\log(1/\delta)\bigr)\right),
    \qquad c\in[0,1],
\]
where the constant in the sample budget is sufficiently small and the query lower bound holds in expectation.
We prove this slope lower bound in \Cref{subsec:quasi-MHR:LB}, as \Cref{thm:quasi-MHR:LB-tradeoff}.
The case $c=1$ also gives the matching single-sample query lower bound, while $c=0$ gives the matching sample-only lower bound stated in \Cref{thm:quasi-MHR:LB-sample}.}

{\subsection{Query Complexity Upper Bound with a Single Sample}
\label{subsec:quasi-MHR:UB}}

{We first prove the query upper bound for the quasi-MHR family.
As explained in \Cref{rmk:overview:quasi-MHR}, the constant monopoly-quantile bound simplifies the quasi-regular algorithm: a constant-factor price interval suffices, and the final comparison uses the original bounded-reward {\BestArmIdentification} problem.}

\begin{theorem}[Upper Bound for Quasi-MHR Distributions]
\label{thm:quasi-MHR:UB-query}
Fix $0<\eps,\delta\le1/4$.
For the family of quasi-MHR distributions, there exists an $(\eps,\delta)$-PAC learning algorithm that first draws one value sample and then makes at most {$O\bigl(\frac{1}{\eps^{3}}\log(\frac{1}{\delta})\bigr)$} pricing queries.
\end{theorem}

{As before, we assume $0<r^*<\infty$.
Our algorithm consists of the following three steps.
\begin{enumerate}[label=\textbf{Step \arabic*:},leftmargin=*]
    \item \textbf{Localize the search interval.}
    Use the revenue estimate from Module~\ref{overview:quasi-regular:1} and quasi-MHR to find an interval $I=[a/B,a]$, where $B=3e$, that contains a nearly optimal price and satisfies $Q(p)\ge1/B$ throughout.
    This is the quasi-MHR version of the scale estimation and threshold steps; \Cref{lem:quasi-MHR:localization} below gives its guarantee.
    \item \textbf{Construct a candidate price set.}
    Take a downward geometric grid $\mathcal P$ on $I$ with successive ratios at most $1+\eps/16$, including both endpoints.
    Since $B$ is constant, the grid has $O(\eps^{-1})$ prices.
    \item \textbf{Select a nearly optimal price.}
    Treat each $p\in\mathcal P$ as an arm with reward $(p/a)Y\in[0,1]$, where $Y$ is the sale indicator from a fresh pricing query.
    Apply Median Elimination with additive accuracy $\eps/(8B)$ and failure probability $\delta/4$, and return its selected price.
\end{enumerate}

The following lemma supplies Step~1.
It combines the revenue estimate from \Cref{lem:quasi-regular:scale} with the quasi-MHR monopoly-quantile bound, and will also be used for the tradeoff and MHR upper bounds.}

\begin{lemma}
\label{lem:quasi-MHR:localization}
Fix $0<\eps,\delta\le1/4$.
For a quasi-MHR distribution with $0<r^*<\infty$, a procedure using one value sample and at most $O(\eps^{-2}\log(1/\delta))$ pricing queries returns $a>0$ such that, with probability at least $1-\delta/4$, the interval $I\triangleq[a/B,a]$, where $B=3e$, satisfies $Q(p)\ge1/B$ for every $p\in I$, $r^*\ge\Rprice(a)\ge a/B$, and
\begin{equation}\label{eq:quasi-MHR:localization}
 \max_{p\in I}\Rprice(p)\ge(1-\eps/16)r^*.
\end{equation}
The query bound holds on every run, including runs on which the guarantee fails.
\end{lemma}

\begin{proof}
{We first recall the monopoly-quantile bound in~\cite{FJ25}. For a price $p$ with $1<H(p)<\infty$, set $t=1/H(p)$.
Quasi-MHR gives $H(tp)\le tH(p)=1$, and hence $\Rprice(tp)/\Rprice(p)\ge e^{H(p)-1}/H(p)>1$.
Thus, restricting to prices with $Q(p)\ge1/e$ does not decrease the revenue supremum.
These prices satisfy $p\le er^*$, so upper semicontinuity of $\Rprice$ gives a monopoly price $p^*$ with $Q(p^*)\ge1/e$ and $r^*\le p^*\le er^*$.}

{Quasi-MHR implies quasi-regularity: both factors in the identity $(Q(p)^{-1}-1)/p=(H(p)/p)(e^{H(p)}-1)/H(p)$ are nonnegative and nondecreasing.}
Apply \Cref{lem:quasi-regular:scale} with failure probability $\delta/8$ to obtain $\rho\in[r^*,1.1r^*]$ using one sample and $O(\log(1/\delta))$ pricing queries.
{Set $\kappa=\eps/16$ and form a geometric grid $G$ of ratio $1+\kappa$ on $[\rho/(2e),2e\rho]$, including both endpoints.}
At each $p\in G$, use $\lceil18e^2\log(16|G|/\delta)\rceil$ fresh queries to estimate $Q(p)$, and select $p$ when its empirical sale probability is at least $2/(3e)$.
{Denote by $a$ the largest selected price.}

Hoeffding's inequality and a union bound give additive error at most $1/(6e)$ simultaneously on $G$ with probability at least $1-\delta/8$.
Together with successful scale estimation, this event has probability at least $1-\delta/4$.
On this event, every grid price with $Q(p)\ge1/e$ is selected, and every selected price has $Q(p)\ge1/(2e)$.

{Since $r^*\le p^*\le er^*$, the monopoly price lies in $[\rho/(2e),2e\rho]$.
Denote by $g$ the largest grid price no greater than $p^*$.
Then $g\ge p^*/(1+\kappa)$ and $Q(g)\ge1/e$, so $a\ge g$.
If $a\le p^*$, monotonicity of $Q$ gives $\Rprice(a)\ge r^*/(1+\kappa)$.
If $a>p^*$, the selected-price guarantee gives $a/B\le(2/3)\Rprice(a)\le(2/3)r^*<p^*$, so $p^*\in I$.
In both cases, $\max_{p\in I}\Rprice(p)\ge r^*/(1+\kappa)\ge(1-\eps/16)r^*$, proving the revenue guarantee in \eqref{eq:quasi-MHR:localization}.}
The other two guarantees follow from $Q(p)\ge Q(a)\ge1/(2e)\ge1/B$ on $I$ and $\Rprice(a)\le r^*$.

The grid has $O(\eps^{-1})$ points.
The grid is formed after estimating the revenue scale, so its size is independent of the initial sample's magnitude.
Thus the procedure uses $O(\eps^{-1}\log(1/(\eps\delta)))=O(\eps^{-2}\log(1/\delta))$ pricing queries, with a deterministic bound on every branch.
Thus an additional logarithmic factor in $1/\delta$ is unnecessary.

We can also use the sharper noisy-binary-search implementation, although the grid procedure already suffices for the theorem's query bound.
On the same geometric grid, apply \cite[Theorem~1]{GP24} with target $\tau=1-3/(4e)$, tolerance $\omega=1/(16e)$, and failure probability $\delta/8$.
The endpoint sale probabilities bracket this target, and the output adjacent prices $a,b$ satisfy $Q(a)>11/(16e)$, $Q(b)<13/(16e)$, and $b\le(1+\kappa)a$.
Every price approaching $r^*$ with sale probability at least $1/e$ is below $b$, so $a$ is at least that price divided by $1+\kappa$; moreover, $a/B\le(16/33)\Rprice(a)\le(16/33)r^*$.
The same interval argument therefore proves \eqref{eq:quasi-MHR:localization}.
The noisy search uses $O(\log(1/\eps)+\log(1/\delta))$ queries apart from the scale estimate; a predetermined query cap enforces this bound on unsuccessful scale-estimation runs as well.
\end{proof}

{With this interval in hand, the remaining two steps are a geometric discretization and a bounded-reward arm comparison.
We now complete the proof of \Cref{thm:quasi-MHR:UB-query}.}

\begin{proof}[{Proof of \Cref{thm:quasi-MHR:UB-query}}]
We apply \Cref{lem:quasi-MHR:localization} and condition on successful localization.
{For Step~2,} we take a downward geometric grid $\mathcal P$ on $I$ with successive ratios at most $1+\eps/16$, including both endpoints.
Rounding down cannot decrease the sale probability, so $\max_{p\in\mathcal P}\Rprice(p)\ge(1-\eps/16)\max_{p\in I}\Rprice(p)\ge(1-\eps/8)r^*$.

{For Step~3,} as in \pref{thm:quasi-regular:UB-query}, we normalize revenues and compare prices using fresh pricing queries. Here a query at $p\in\mathcal P$ gives the bounded reward $(p/a)Y\in[0,1]$, where $Y$ is its sale indicator; its mean is $\Rprice(p)/a$.
We apply Median Elimination~\cite[Theorem~10]{EMM06} with additive accuracy $\eps/(8B)$ and failure probability $\delta/4$, using $O(|\mathcal P|\eps^{-2}\log(1/\delta))$ queries.
Since $|\mathcal P|=O(\eps^{-1})$, this is $O(\eps^{-3}\log(1/\delta))$ queries.
Its output loses at most $\eps a/(8B)\le\eps r^*/8$ relative to the best grid price and therefore has revenue at least $(1-\eps)r^*$.
The localization cost is absorbed by this bound, the two failure probabilities sum to at most $\delta/2$, and no further value samples are used.
\end{proof}

{\subsection{Sample Complexity Upper Bound}
\label{subsec:quasi-MHR:UB-sample}}

{We next consider learning from value samples alone.}
The proof of \pref{lem:quasi-MHR:localization} shows that restricting to prices with sale probability at least $1/e$ does not decrease the revenue supremum.
Consequently, the empirical-cutoff argument from \pref{thm:quasi-regular:UB-sample} applies with a constant cutoff.
The ordinary additive-accuracy DKW inequality~\cite{M90} then gives the required $O(\eps)$ relative accuracy throughout the relevant range.

\begin{restate}[\Cref{thm:quasi-MHR:UB-sample}]
Fix $0<\eps,\delta\le1/4$.
For the family of quasi-MHR distributions, there exists an $(\eps, \delta)$-PAC learning algorithm that draws at most {$O\bigl(\frac{1}{\eps^{2}}\log(\frac{1}{\delta})\bigr)$} many samples and makes no pricing queries.
\end{restate}

We maximize empirical revenue among prices with empirical sale probability at least $2/(3e)$.
With $O(\eps^{-2}\log(1/\delta))$ samples, the uniform additive error is at most $\eps/(24e)$: every price with true sale probability at least $1/e$ is feasible, and every feasible price has true sale probability at least $1/(3e)$.
Thus we can convert additive accuracy into relative accuracy and use the same empirical revenue comparison as in \pref{thm:quasi-regular:UB-sample}.
The cutoff guarantee and complete proof appear in the quasi-MHR proof in \pref{sec:sample-complexity}.

{\subsection{Query-Sample Tradeoff Upper Bound}
\label{subsec:quasi-MHR:UB-tradeoff}}

{We now use additional value samples to reduce the number of candidates compared by queries.
As in Module~\ref{overview:quasi-MHR:2}, we choose one empirical revenue maximizer from each of a collection of short price intervals.
We first apply \Cref{lem:quasi-MHR:localization}, so the sample-based construction and the final query comparison use the same interval.
After scaling this interval to constant length, we divide it into $O(\eps^{-c})$ bins of width $O(\eps^c)$ and use independent value samples to choose a candidate from each bin.}
Within a bin of width $d$, the probability mass between two prices with revenue gap $\Delta$ is at most $O(d+\Delta)$; this limits the fluctuation in their empirical revenue comparison. Consequently, with $d=O(\eps^c)$, only $O(\eps^{-(2-c)}\log(1/\delta))$ samples are needed to find an $O(\eps)$-optimal candidate in the bin containing an optimal price.
We then compare the $O(\eps^{-c})$ candidates using $O(\eps^{-2}\log(1/\delta))$ queries per candidate, giving $O(\eps^{-(2+c)}\log(1/\delta))$ queries in total. Only the bin containing a fixed approximately optimal price needs to produce a good candidate, so no union bound over all bins or grid prices is needed.

\begin{theorem}[Upper Bound for Quasi-MHR Distributions]
\label{thm:quasi-MHR:UB-tradeoff}
Fix $0<\eps,\delta\le1/4$ and a constant $c\in[0,1]$.
For the family of quasi-MHR distributions, there exists an $(\eps,\delta)$-PAC learning algorithm that first draws {$\Theta\bigl(\frac{1}{\eps^{2-c}}\log(\frac{1}{\delta})\bigr)$} value samples and then makes at most {$O\bigl(\frac{1}{\eps^{2+c}}\log(\frac{1}{\delta})\bigr)$} pricing queries.
\end{theorem}

\begin{proof}
Set $B=3e$.
Draw one sample for localization and, independently, $m=\lceil C\eps^{-(2-c)}\log(1/\delta)\rceil$ samples for candidate construction, where $C$ is a sufficiently large absolute constant.
All these samples are drawn before any pricing query.
{Apply \Cref{lem:quasi-MHR:localization} using only the first sample and fresh queries.}
Conditional on its successful event and its output $a$, the $m$ unused samples are still independent observations from $F$, because the localization procedure does not inspect them.

Scale values and prices by $a/B$.
In these scaled coordinates, $Q$ denotes the sale probability of the scaled distribution, and we denote by $R(x)\triangleq xQ(x)$ its revenue.
The localized interval is $[1,B]$, with $Q(B)\ge1/B$ and $R_I^*\triangleq\max_{x\in[1,B]}R(x)\ge1$.
Denote by $P$ the scaled value distribution and by $P_m$ its empirical measure from the $m$ reserved samples; $\widehat Q$ and $\widehat R(x)\triangleq x\widehat Q(x)$ denote the corresponding empirical sale probability and revenue.

We establish the concentration estimate that avoids a union bound over all candidate prices.
{For a fixed $v\in[1,B]$ and an interval $J\subseteq[1,B]$ containing $v$ and of length at most $d\in(0,B]$, there are absolute constants $C_0,c_0>0$ such that}
\begin{equation}\label{eq:quasi-MHR:local-comparison}
 \Pr\left[\exists x\in J:
 R(v)-R(x)\ge r,\ \widehat R(x)\ge\widehat R(v)\right]
 \le C_0\exp\left(-\frac{c_0mr^2}{B^2(d+r)}\right),
\end{equation}
for every $r>0$.

We first control the empirical mass between two prices using a nested family $\mathcal A$ of intervals with a common endpoint, each having probability at most $V$.
Its empirical measure satisfies
\begin{equation}\label{eq:quasi-MHR:nested-concentration}
 \Pr\left[\sup_{A\in\mathcal A}|P_m(A)-P(A)|>\eta\right]
 \le C_1\exp\left(-\frac{c_1m\eta^2}{V+\eta}\right),
\end{equation}
for every $\eta>0$, where $C_1,c_1>0$ are universal constants.
Indeed, denote by $U$ the increasing union of these intervals and by $K$ the number of samples in $U$.
Then $P(U)\le V$, and Bernstein's inequality controls $|K/m-P(U)|$ by $\eta/2$ except with probability at most $2\exp(-cm\eta^2/(V+\eta))$.
{Conditional on $K$, the observations in $U$ are independent draws from the law conditional on $U$.}
The standard additive-accuracy DKW inequality~\cite{M90} controls their conditional empirical probabilities for all anchored intervals simultaneously.
{On $K/m\le V+\eta/2$, the conditional deviation $m\eta/(2K)$ has an exponent at least $cm\eta^2/(V+\eta)$.}
Decomposing $P_m(A)-P(A)$ into the conditional empirical error multiplied by $K/m$ and the count error proves \eqref{eq:quasi-MHR:nested-concentration}.

For \eqref{eq:quasi-MHR:local-comparison}, group prices by their true revenue gap $\Delta(x)\triangleq R(v)-R(x)$ in a dyadic range $[s,2s)$.
If $x=v+t\ge v$, the identity $\Delta(x)=xP([v,x))-tQ(v)$ gives $P([v,x))=(\Delta(x)+tQ(v))/x\le2s+d$.
If $x=v-t\le v$, the identity $\Delta(x)=tQ(v)-xP([x,v))$ gives $P([x,v))\le d$.
{In either case, an empirical reversal requires $|\widehat Q(v)-Q(v)|\ge s/(2d)$ or $|P_m(A_x)-P(A_x)|\ge s/(2B)$, where $A_x$ is the corresponding anchored interval.}
Hoeffding's inequality controls the first event.
Applying \eqref{eq:quasi-MHR:nested-concentration} separately to the anchored intervals on each side of $v$, whose masses are at most $d+2s$, controls the second event.
Since $d\le B$, the combined probability is at most $C\exp(-cms^2/(B^2(d+s)))$.
Summing over $s=2^jr$, for integers $j\ge0$, proves \eqref{eq:quasi-MHR:local-comparison}: the exponents grow at least geometrically because $(2^jr)^2/(d+2^jr)\ge2^jr^2/(d+r)$.

{For the sample-based sparsification in Module~\ref{overview:quasi-MHR:2},} divide $[1,B]$ into $J=\lceil(B-1)/(B\eps^c)\rceil$ equal-length bins, each of length at most $d=B\eps^c$.
Within each bin, form a deterministic fine grid with successive ratios at most $1+\eps/16$, including both boundary prices.
{Choose an empirical revenue maximizer on each grid, and denote by $\mathcal C$ the resulting set of at most $J=O(\eps^{-c})$ candidates.}

Conditional on localization, fix a maximizer of $R$ on $[1,B]$ and round it down {within a bin containing it} to a fine-grid price $v$.
This witness does not depend on the reserved samples and satisfies $R(v)\ge(1-\eps/16)R_I^*$.
Apply \eqref{eq:quasi-MHR:local-comparison} only to its bin, with $r=\eps/16$.
Since $d+r\le(B+1)\eps^c$ and $m\ge C\eps^{-(2-c)}\log(1/\delta)$, its empirical maximizer $\widetilde v$ satisfies $R(\widetilde v)\ge R(v)-\eps/16\ge(1-\eps/8)R_I^*$ with probability at least $1-\delta/4$.
Thus $\mathcal C$ contains an approximately optimal price.
The probability guarantee concerns the single fixed witness bin, so no union bound over the bins or their fine-grid prices is needed.

For each $x\in\mathcal C$, a fresh pricing query supplies the reward $xY/B\in[0,1]$ with mean $R(x)/B$.
{For the final comparison, use the bounded-reward Median Elimination procedure from \Cref{thm:quasi-MHR:UB-query}, now applied to $\mathcal C$.} We apply Median Elimination~\cite[Theorem~10]{EMM06} with additive accuracy $\eps/(8B)$ and failure probability $\delta/4$, using $O(J\eps^{-2}\log(1/\delta))=O(\eps^{-(2+c)}\log(1/\delta))$ queries.
Its output has revenue at least $(1-\eps/8)R_I^*-\eps/8\ge(1-\eps/4)R_I^*$.
Returning the corresponding unscaled price and using \Cref{lem:quasi-MHR:localization} gives revenue at least $(1-\eps/4)(1-\eps/16)r^*\ge(1-\eps)r^*$.
Localization, candidate construction, and final selection fail with total probability at most $3\delta/4$.
The localization cost from \Cref{lem:quasi-MHR:localization} is absorbed by the claimed query bound for every $c\in[0,1]$, and the total number of value samples is $m+1=\Theta(\eps^{-(2-c)}\log(1/\delta))$.
\end{proof}

\subsection{Lower Bounds}
\label{subsec:quasi-MHR:LB}

The query-sample tradeoff in \pref{thm:quasi-MHR:UB-tradeoff} cannot be improved.
Fix a constant $c\in[0,1]$.
We construct a base distribution $F^{(0)}$ and $K=\Theta(\eps^{-c})$ hard distributions, where the $i$-th hard distribution $F^{(i)}$ agrees with $F^{(0)}$ outside its own price interval $P^{(i)}$ and raises the revenue on $P^{(i)}$, so that a $(1-\eps)$-optimal price identifies the hard distribution.
As explained in \Cref{subsec:quasi-regular:LB}, a value sample is informative about every hard distribution, whereas a pricing query is informative only about the hard distribution whose interval contains its price.
In our construction, a value sample has KL divergence $O(K\eps^2)$ between $F^{(0)}$ and each $F^{(i)}$, and a pricing query at a price in $P^{(i)}$ has KL divergence $O(\eps^2)$ between $F^{(0)}$ and $F^{(i)}$.
We apply \pref{lem:LB:weighted-information} with equal weights, under which each value sample contributes to all $K$ terms of the weighted sum and each pricing query contributes to only one.
Under the sample budget $\frac{C}{\eps^{2-c}}\log\frac1\delta$ for a small absolute constant $C$, the value samples supply at most half of the KL divergence required by \pref{lem:LB:weighted-information}, so the pricing queries must supply the rest, which requires $\Omega(K\eps^{-2}\log\frac1\delta)=\Omega(\eps^{-(2+c)}\log\frac1\delta)$ queries.
The following theorem formally states this lower bound.

\begin{theorem}[Lower Bound for Quasi-MHR Distributions]
\label{thm:quasi-MHR:LB-tradeoff}
There exists an absolute constant $C>0$ such that the following holds.
Fix $c\in[0,1]$, $0<\delta\le1/4$, and $0<\eps\le1/128$.
If an $(\eps,\delta)$-PAC learning algorithm for quasi-MHR distributions draws at most $\frac{C}{\eps^{2-c}}\log\frac1\delta$ value samples, then there exists a quasi-MHR distribution on which it makes at least $\frac{C}{\eps^{2+c}}\log\frac1\delta$ pricing queries in expectation.
\end{theorem}

To prove \Cref{thm:quasi-MHR:LB-tradeoff}, we first construct the base distribution and the $K$ hard distributions.
Set $K\triangleq\max\{2,\lfloor\eps^{-c}/64\rfloor\}$, $w\triangleq1/(8K)$, and $\rho\triangleq16K\eps$, where $K$ is the number of hard distributions, $w$ is the half-width of their price intervals, and $\rho$ controls the relative change in density.
These choices satisfy $\rho w=2\eps$ and $2^{-7}\eps^{-c}\le K\le2\eps^{-c}$.
They also satisfy $\rho\le\max\{32\eps,\eps^{1-c}/4\}\le1/4$, where the last inequality uses $\eps\le1/128$ and $c\le1$.
For $i\in[K]$, denote by $a_i\triangleq3/2-2iw$, $m_i\triangleq a_i+w$, and $b_i\triangleq a_i+2w$ the endpoints and the midpoint of the interval $P^{(i)}\triangleq(a_i,b_i]$ on which the $i$-th hard distribution $F^{(i)}$ differs from $F^{(0)}$.
The intervals $P^{(i)}$ are pairwise disjoint and lie in $[5/4,3/2]$.

Define the base distribution by
\begin{equation*}
 F^{(0)}(p)\triangleq
 \begin{cases}
  0,&p\le1,\\
  1-1/p,&1<p\le e,\\
  1,&p>e,
 \end{cases}
\end{equation*}
which is the equal-revenue distribution truncated at $e$, with an atom of mass $1/e$ at $e$ and revenue one throughout $[1,e]$.
The $i$-th hard distribution raises the revenue linearly from one at $a_i$ to $1+\rho w/a_i$ at $m_i$ and decreases it linearly back to one at $b_i$.
The corresponding CDF is
\begin{equation*}
 F^{(i)}(p)\triangleq
 \begin{cases}
  1-\rho/a_i-(1-\rho)/p,&a_i<p\le m_i,\\
  1+\rho/a_i-(1+\rho b_i/a_i)/p,&m_i<p\le b_i,\\
  F^{(0)}(p),&p\notin P^{(i)}.
 \end{cases}
\end{equation*}
\begin{figure}[htbp]
\centering
\begin{tikzpicture}[x=2.6cm,y=7cm,font=\small]
 \draw[->] (0,0) -- (3.05,0) node[below]{$p$};
 \draw[->] (0,0) -- (0,0.47) node[left]{$H(p)/p$};
 \node[blue,above] at (2.1,0.36) {$H_0(p)/p$};
 \node[red,left] at (1.18,0.2) {$H_i(p)/p$};
 \draw[gray,densely dotted] (1.2,0) -- (1.2,{ln(1.2)/1.2}) (1.45,0) -- (1.45,{(ln(1.45)-ln(1+0.5*(1.45/1.2-1)))/1.45}) (1.7,0) -- (1.7,{ln(1.7)/1.7});
 \draw[blue,thick] (0,0) -- (1,0);
 \draw[blue,thick,domain=1:2.718,samples=60] plot (\x,{ln(\x)/\x});
 \draw[blue,thick,densely dashed,->] (2.718,{1/2.718}) -- (2.718,0.46);
 \draw[red,thick,domain=1.2:1.45,samples=30] plot (\x,{(ln(\x)-ln(1+0.5*(\x/1.2-1)))/\x});
 \draw[red,thick,domain=1.45:1.7,samples=30] plot (\x,{(ln(\x)-ln(1+0.5*(1.7-\x)/1.2))/\x});
 \draw[<->,>=stealth] (1.45,{(ln(1.45)-ln(1+0.5*(1.45/1.2-1)))/1.45}) -- (1.45,{ln(1.45)/1.45});
 \draw[gray] (1.47,0.222) -- (1.75,0.1) node[right,black]{$\frac{1}{m_i}\log\bigl(1+\frac{\rho w}{a_i}\bigr)$};
 \node[below] at (1,0) {$1$};
 \node[below] at (1.2,0) {$a_i$};
 \node[below] at (1.45,0) {$m_i$};
 \node[below] at (1.7,0) {$b_i$};
 \node[below] at (2.718,0) {$e$};
\end{tikzpicture}
\caption{The ratios $H_0(p)/p$ (blue) and $H_i(p)/p$ (red) for the cumulative hazards of $F^{(0)}$ and $F^{(i)}$, not to scale. On $P^{(i)}=(a_i,b_i]$, $H_i$ lies below $H_0$ by the logarithm of the revenue ratio, which peaks at $\log(1+\rho w/a_i)$ at $m_i$, but $H_i(p)/p$ stays nondecreasing.}
\label{fig:quasi-MHR:LB}
\end{figure}
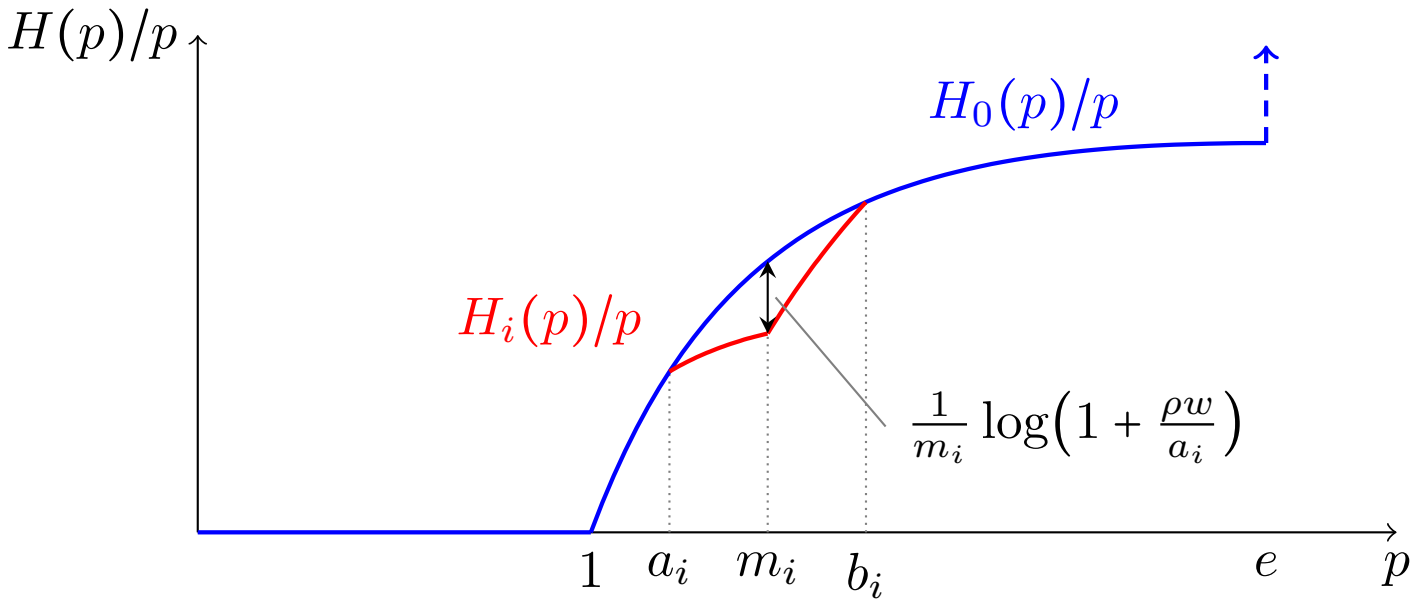

\noindent The branches agree at $a_i,m_i,b_i$, so $F^{(i)}$ agrees with $F^{(0)}$ outside $P^{(i)}$, including the atom at $e$, and assigns the same probability to $P^{(i)}$.

The following lemma verifies that these distributions are quasi-MHR.

\begin{lemma}[Quasi-MHR]
\label{lem:quasi-MHR:quasi-MHR}
The base distribution $F^{(0)}$ and the hard distributions $\{F^{(i)}\}_{i\in[K]}$ are quasi-MHR distributions.
\end{lemma}

\begin{proof}
For $j\in\{0\}\cup[K]$, denote by $Q_j\triangleq1-F^{(j)}$ the sale probability of $F^{(j)}$ and by $H_j\triangleq-\log Q_j$ its cumulative hazard defined in \Cref{def:MHR}.
By \Cref{def:lambda-quasi-regular} with $\regularity=0$, it suffices to show that $H_j(p)/p$ is nondecreasing (see \Cref{fig:quasi-MHR:LB}).
For the base distribution, $H_0(p)/p$ equals $0$ for $p\le1$, $\log p/p$ on $[1,e]$, and $+\infty$ for $p>e$, which is nondecreasing.
Denote by $r_i(p)\triangleq pQ_i(p)$ the revenue under $F^{(i)}$, which equals $1+\rho(p/a_i-1)$ on $(a_i,m_i]$ and $1+\rho(b_i-p)/a_i$ on $(m_i,b_i]$.
Thus $r_i\ge1$, $|r_i'|=\rho/a_i$, and $H_i(p)=\log p-\log r_i(p)$ on $P^{(i)}$.
Differentiating $H_i$ on each half of $P^{(i)}$, we obtain that
\begin{equation*}
 pH_i'(p)-H_i(p)
 =1-\log p+\log r_i(p)-\frac{pr_i'(p)}{r_i(p)}
 \ge1-\log\frac32-\frac65\rho
 >0,
\end{equation*}
where the first inequality uses $1-\log p\ge1-\log\frac32$ since $p\le\frac32$, $\log r_i(p)\ge0$ since $r_i\ge1$, and $pr_i'(p)/r_i(p)\le p\rho/a_i\le\frac65\rho$ since $|r_i'|=\rho/a_i$, $r_i\ge1$, and $p/a_i\le\frac{3/2}{5/4}=\frac65$ since $P^{(i)}\subseteq[5/4,3/2]$, and the last inequality uses $\rho\le\frac14$.
Hence $(H_i(p)/p)'=(pH_i'(p)-H_i(p))/p^2>0$ on each half of $P^{(i)}$, and since $H_i$ is continuous and equals $H_0$ outside $P^{(i)}$, $H_i(p)/p$ is nondecreasing.
\end{proof}

We next show that a $(1-\eps)$-optimal price identifies the changed interval.
Under $F^{(i)}$, revenue outside $P^{(i)}$ equals the base revenue, which is at most one, whereas $r_i(m_i)=1+\rho w/a_i=1+2\eps/a_i\ge1+4\eps/3$.
Since $(1-\eps)(1+4\eps/3)>1$, every $(1-\eps)$-optimal price under $F^{(i)}$ belongs to $P^{(i)}$.

We fix the absolute constant $A\triangleq100$ for the next two lemmas.
The following lemma bounds the KL divergence of a value sample between $F^{(0)}$ and $F^{(i)}$.

\begin{lemma}[KL Divergence of Samples]
\label{lem:quasi-MHR:LB:sample-KL}
For every $i\in[K]$, we have $\DKL(F^{(0)},F^{(i)})\le AK\eps^2$.
\end{lemma}

\begin{proof}
The two distributions have the same atom at $e$, and their densities agree outside $P^{(i)}$.
On the two halves of $P^{(i)}$, the base density is $f_0(p)=1/p^2$, and differentiating the displayed CDF gives $f_i(p)=(1-\rho)/p^2$ and $f_i(p)=(1+\rho b_i/a_i)/p^2$, respectively.
Denote by $z(p)\triangleq f_i(p)/f_0(p)-1$ the relative density change on $P^{(i)}$, so that $|z(p)|\le\rho b_i/a_i\le6\rho/5<1/2$.
Since the two distributions assign the same probability to $P^{(i)}$, $\int_{P^{(i)}}f_0(p)z(p)\,\d p=0$, and the mass of $P^{(i)}$ under $F^{(0)}$ is $1/a_i-1/b_i\le2w$.
As in the proof of \Cref{lem:quasi-regular:LB:sample-KL}, expanding the KL divergence by its definition and substituting $f_i=(1+z)f_0$ on $P^{(i)}$, we obtain that
\begin{align*}
 &\DKL(F^{(0)},F^{(i)})\\
 &=\int_{P^{(i)}}f_0(p)\log\frac{f_0(p)}{f_i(p)}\,\d p
 =\int_{P^{(i)}}f_0(p)\bigl[-\log(1+z(p))+z(p)\bigr]\,\d p\\
 &\le\int_{P^{(i)}}f_0(p)z(p)^2\,\d p
 \le2w\Bigl(\frac65\rho\Bigr)^2
 =\frac{72\cdot32}{25}K\eps^2
 \le AK\eps^2,
\end{align*}
where the second equality uses $\int_{P^{(i)}}f_0(p)z(p)\,\d p=0$, the first inequality uses $-\log(1+z)+z\le z^2$ for $|z|\le\frac12$, the second inequality uses $|z(p)|\le\frac65\rho$ and the mass at most $2w$ of $P^{(i)}$, and the last equality uses $\rho^2w=16K\eps\cdot2\eps=32K\eps^2$.
\end{proof}

The following lemma bounds the KL divergence of a pricing query at a price in $P^{(i)}$.

\begin{lemma}[KL Divergence of Queries]
\label{lem:quasi-MHR:LB:query-KL}
For every $i\in[K]$ and every query price $p\in P^{(i)}$, we have $\DKL(F^{(0)}(p),F^{(i)}(p))\le A\eps^2$.
\end{lemma}

\begin{proof}
For $p\in P^{(i)}$, the sale probabilities are $Q_0(p)=1/p$ and $Q_i(p)=r_i(p)/p$.
Since $1\le r_i(p)\le1+\rho w/a_i$, $\rho w=2\eps$, and $a_i,p\ge5/4$, we have $0\le Q_i(p)-Q_0(p)=(r_i(p)-1)/p\le2\eps$.
Since $p\in P^{(i)}\subseteq[5/4,3/2]$, the base sale probability $Q_0(p)=1/p$ lies in $[2/3,4/5]$.
Hence $Q_i(p)$ lies in $[2/3,4/5+2\eps]\subseteq[2/3,9/10]$, where the inclusion uses $\eps\le1/128$. In addition, we have $Q_i(p)(1-Q_i(p))\ge\min\{\frac23\cdot\frac13,\frac9{10}\cdot\frac1{10}\}=\frac9{100}$.
Expanding the Bernoulli KL divergence by its definition, we obtain that
\begin{align*}
 &\DKL(F^{(0)}(p),F^{(i)}(p))\\
 &=Q_0(p)\log\frac{Q_0(p)}{Q_i(p)}+(1-Q_0(p))\log\frac{1-Q_0(p)}{1-Q_i(p)}\\
 &\le\frac{(Q_i(p)-Q_0(p))^2}{Q_i(p)(1-Q_i(p))}
 \le\frac{(2\eps)^2}{9/100}
 \le A\eps^2,
\end{align*}
where the first inequality is the chi-squared bound from the proof of \Cref{lem:quasi-regular:LB:query-KL}, and the second inequality uses $Q_i(p)-Q_0(p)\le2\eps$ and $Q_i(p)(1-Q_i(p))\ge\frac9{100}$.
\end{proof}

We now prove \Cref{thm:quasi-MHR:LB-tradeoff} by combining the good-price separation with \Cref{lem:quasi-MHR:LB:sample-KL,lem:quasi-MHR:LB:query-KL} through \pref{lem:LB:weighted-information}.

\begin{proof}[Proof of \Cref{thm:quasi-MHR:LB-tradeoff}]
Consider an $(\eps,\delta)$-PAC learning algorithm, with $p^\calA$, $S$, $T$, $T^{(i)}$, $\prob[F^{(j)}]{\cdot}$, and $\expect[F^{(j)}]{\cdot}$ as in the proof of \Cref{thm:quasi-regular:LB-tradeoff}.
As there, the events $\scrE^{(i)}\triangleq\{p^\calA\in P^{(i)}\}$ for $i\in[K]$ are pairwise disjoint, so $\sum\nolimits_{i\in[K]}\prob[F^{(0)}]{\scrE^{(i)}}\le1\le K/2$ since $K\ge2$.
Since every $(1-\eps)$-optimal price under $F^{(i)}$ belongs to $P^{(i)}$, the PAC guarantee gives $\prob[F^{(i)}]{\scrE^{(i)}}\ge1-\delta$.
Applying \pref{lem:LB:weighted-information} with weights $w_i=1$, $d_i=AK\eps^2$ from \Cref{lem:quasi-MHR:LB:sample-KL}, and $b_i=A\eps^2$ from \Cref{lem:quasi-MHR:LB:query-KL}, we obtain
\begin{equation*}
 \sum\nolimits_{i\in[K]}\bigl(AK\eps^2\,\expect[F^{(0)}]{S}+A\eps^2\,\expect[F^{(0)}]{T^{(i)}}\bigr)
 \ge\frac{K}{16}\log\frac1\delta.
\end{equation*}
As in the proof of \Cref{thm:quasi-regular:LB-tradeoff}, $\sum\nolimits_{i\in[K]}T^{(i)}\le T$.
Dividing by $K$ gives
\begin{equation}
\label{eq:quasi-MHR:LB-information}
 AK\eps^2\,\expect[F^{(0)}]{S}+\frac{A\eps^2}{K}\expect[F^{(0)}]{T}
 \ge\frac1{16}\log\frac1\delta.
\end{equation}
Set $C\triangleq2^{-12}/A$.
Since $\expect[F^{(0)}]{S}\le\frac{C}{\eps^{2-c}}\log\frac1\delta$ and $K\le2\eps^{-c}$, the first term on the left is at most $2AC\log\frac1\delta\le\frac1{32}\log\frac1\delta$.
Hence, by $K\ge2^{-7}\eps^{-c}$,
\begin{equation*}
 \expect[F^{(0)}]{T}
 \ge\frac{K}{32A\eps^2}\log\frac1\delta
 \ge\frac{1}{2^{12}A\eps^{2+c}}\log\frac1\delta
 =\frac{C}{\eps^{2+c}}\log\frac1\delta.
\end{equation*}
\end{proof}

Setting $c=0$ and allowing no pricing queries in the proof of \Cref{thm:quasi-MHR:LB-tradeoff} gives the following sample lower bound, which matches the sample upper bound in \Cref{thm:quasi-MHR:UB-sample}.

\begin{theorem}
\label{thm:quasi-MHR:LB-sample}
Fix $0<\delta\le1/4$ and $0<\eps\le1/128$.
For the absolute constant $C$ in \Cref{thm:quasi-MHR:LB-tradeoff}, every $(\eps,\delta)$-PAC learning algorithm for quasi-MHR distributions using only value samples draws at least $\frac{C}{\eps^2}\log\frac1\delta$ samples in expectation on some quasi-MHR distribution.
\end{theorem}

\begin{proof}
For $c=0$, the construction has $K=2$, and $T=0$, so \eqref{eq:quasi-MHR:LB-information} gives $2A\eps^2\,\expect[F^{(0)}]{S}\ge\frac1{16}\log\frac1\delta$, and hence $\expect[F^{(0)}]{S}\ge\frac{1}{32A\eps^2}\log\frac1\delta\ge\frac{C}{\eps^2}\log\frac1\delta$.
\end{proof}

%% file: MHR.tex
\section{MHR Distributions}
\label{sec:MHR}

{In this section, we establish the query and sample upper bounds and the query-sample lower bound for MHR distributions.
We use the same notation and conventions as in \Cref{sec:quasi-regular,sec:quasi-MHR}.}

{\subsection{Overview of the Section}
\label{subsec:MHR:overview}}

\begin{figure}[htbp]
    \centering
    \includegraphics[width=.44\textwidth]{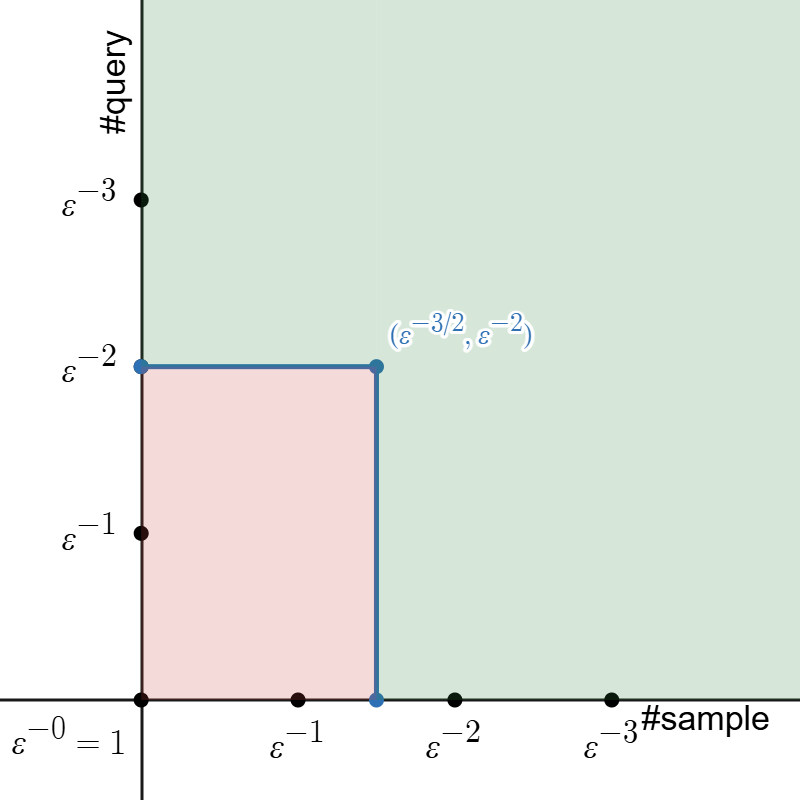}
    \caption{The MHR query-sample tradeoff from \Cref{fig:result-MHR}.
    Both axes are logarithmic; constant factors and $\log(1/\delta)$ are suppressed.
    Green and red denote the learnable and unlearnable regimes.}
    \label{fig:MHR:tradeoff}
\end{figure}

{Our goal is to establish the tight tradeoff shown in \Cref{fig:MHR:tradeoff}, reproduced from \Cref{fig:result-MHR}.
As before, $\samplecount$ and $\querycount$ denote the sample and query budgets, respectively.}

{For the upper bounds, it suffices to establish the two endpoints
\begin{align*}
    (\samplecount,\querycount)
    &=\left(1,\ O\bigl(\eps^{-2}\log(1/\delta)\bigr)\right),\\
    (\samplecount,\querycount)
    &=\left(O\bigl(\eps^{-3/2}\log(1/\delta)\bigr),\ 0\right),
\end{align*}
since an algorithm can simply ignore any additional samples or queries.
We first prove \Cref{thm:MHR:UB-query} in \Cref{subsec:MHR:UB}, giving the single-sample query endpoint.
Next, \Cref{subsec:MHR:UB-sample} presents the sample-only endpoint, \Cref{thm:MHR:UB-sample}; the complete proof appears in the MHR subsection of \Cref{sec:sample-complexity}.

For the lower bound, it suffices to establish the corner
\[
    (\samplecount,\querycount)
    =\left(O\bigl(\eps^{-3/2}\log(1/\delta)\bigr),\
      \Omega\bigl(\eps^{-2}\log(1/\delta)\bigr)\right),
\]
where the constant in the sample budget is sufficiently small and the query lower bound holds in expectation.
We prove this bound in \Cref{subsec:MHR:LB}, as \Cref{thm:MHR:LB-tradeoff}.
It matches the single-sample query upper bound and also gives the matching sample-only lower bound.}

{\subsection{Query Complexity Upper Bound with a Single Sample}
\label{subsec:MHR:UB}}

{We first prove the query upper bound for the MHR family.
Since MHR implies quasi-MHR, we can reuse the localization procedure from \Cref{lem:quasi-MHR:localization}.
Convexity of the cumulative hazard then makes negative log-revenue convex in log-price coordinates.
We construct an unbiased observation of this objective from pricing queries and apply stochastic convex optimization~\cite{CRTA26}.
This avoids the $O(\eps^{-1})$ candidate prices used for quasi-MHR distributions and improves the query bound to $O(\eps^{-2}\log(1/\delta))$.}
\begin{theorem}[Upper Bound for MHR Distributions]
\label{thm:MHR:UB-query}
Fix $0<\eps,\delta\le1/4$.
For the family of MHR distributions, there exists an $(\eps,\delta)$-PAC learning algorithm that first draws one value sample and then makes at most {$O\bigl(\frac{1}{\eps^{2}}\log(\frac{1}{\delta})\bigr)$} pricing queries.
\end{theorem}

{Our algorithm consists of the following two steps.
\begin{enumerate}[label=\textbf{Step \arabic*:},leftmargin=*]
    \item \textbf{Run the quasi-MHR localization procedure.}
    Using the single initial sample and $O(\eps^{-2}\log(1/\delta))$ queries, obtain an interval $I=[a/B,a]$, where $B=3e$, that contains a nearly optimal price and has sale probabilities bounded below by $1/B$.
    \item \textbf{Optimize negative log-revenue.}
    Write prices as $p=ae^z$ and minimize $\phi(z)=H(ae^z)-z$ on $[-\log B,0]$.
    At a selected point $z$, query until the first sale and use a harmonic sum of the failure count to obtain an unbiased observation of $\phi(z)$.
    Run stochastic convex optimization independently $O(\log(1/\delta))$ times and return the price corresponding to the median output.
    Cap the total number of queries in this step at $O(\eps^{-2}\log(1/\delta))$.
\end{enumerate}

The guarantee for Step~1 follows directly from \Cref{lem:quasi-MHR:localization}.
It remains to justify the observations used in Step~2 and bound their total query cost.
We do so in the proof of \Cref{thm:MHR:UB-query} below.}

\begin{proof}[{Proof of \Cref{thm:MHR:UB-query}}]
Recall the sale probability $Q$, cumulative hazard $H$, and optimal revenue $r^*$ used in \pref{sec:quasi-MHR}.
{For Step~1, we apply \Cref{lem:quasi-MHR:localization}, which uses one sample and $O(\eps^{-2}\log(1/\delta))$ queries to return the interval $I=[a/B,a]$ with $B=3e$, satisfying \eqref{eq:quasi-MHR:localization} with probability at least $1-\delta/4$.}
Condition on this successful event and the output $a$ for the optimization analysis.

For $z\in[-\log B,0]$, set $p=ae^z$ and denote by $\phi(z)\triangleq H(p)-z=-\log(\Rprice(p)/a)$ the negative log-revenue objective.
The MHR property makes $H$ convex and nondecreasing, so $H(ae^z)$ and $\phi$ are convex.
The bounds $p/a\in[1/B,1]$ and $Q(p)\in[1/B,1]$ give $0\le\phi(z)\le2\log B$ for every $z\in[-\log B,0]$.
An additive $\eps/8$ approximation to the minimum of $\phi$ therefore gives a multiplicative $e^{-\eps/8}$ approximation to the best revenue on $I$.

To obtain an unbiased oracle at $p=ae^z$, make independent pricing queries until the first sale.
Denote by $N$ the number of failures before that sale, and recall the harmonic numbers $H_n$ defined in the proof of \pref{lem:quasi-regular:preprocess}.
Denote its output by $W(z)\triangleq H_N-z$.
Denote by $q\triangleq Q(p)$ the sale probability; then $\Pr[N\ge j]=(1-q)^j$, and Tonelli's theorem gives
\begin{align*}
 \mathbb E[H_N]=\sum_{j\ge1}\frac{\Pr[N\ge j]}j=\sum_{j\ge1}\frac{(1-q)^j}j
 =-\log q.
\end{align*}
Consequently, $\mathbb E[W(z)]=\phi(z)$.
Thus the oracle is exactly unbiased; taking the logarithm of an empirical sale probability would not have this property.

The noise has a uniform subGaussian bound on the localized interval.
Indeed, $q\ge q_0\triangleq1/B$ and $H_n\le1+\log n$ for $n\ge1$, so $\Pr[H_N>u]\le\exp(-q_0\lfloor e^{u-1}\rfloor)$ for $u\ge1$.
Integrating this tail gives a uniform exponential-square-moment bound for $H_N$, since $B$ is fixed.
The exponential-square-moment characterization of subGaussian variables, followed by centering, gives an absolute constant $\sigma$ for which
\begin{equation}\label{eq:MHR:conditional-subGaussian}
 \mathbb E\left[\exp\{t(W(z)-\phi(z))\}\mid\mathcal F_{s-1}\right]
 \le\exp(\sigma^2t^2/2),
\end{equation}
for every $t\in\mathbb R$.
Here $\mathcal F_{s-1}$ is the $\sigma$-field generated by all feedback before the $s$-th oracle call, and its selected query point $z$ is $\mathcal F_{s-1}$-measurable.
The conditional bound follows because each oracle call uses fresh pricing queries.

We use Algorithm~2 in~\cite[Theorem~4.1]{CRTA26} at failure probability $1/8$: for a convex function $f:[0,1]\to[0,1]$ observed with centered $1$-subGaussian noise, at most $T$ observations give additive error $O(T^{-1/2})$ with probability at least $7/8$.
Its stated noise model is independent of the query points and the past; the same guarantee holds for conditionally centered, conditionally $1$-subGaussian noise.
To check this extension, condition on the history fixing one query batch.
For its $n$ noise variables $\xi_1,\ldots,\xi_n$, successive conditioning gives $\mathbb E[\exp(t\sum_{j=1}^n\xi_j)\mid\text{batch history}]\le\exp(nt^2/2)$.
Consequently, the batch average satisfies $\Pr[|n^{-1}\sum_{j=1}^n\xi_j|>u\mid\text{batch history}]\le2\exp(-nu^2/2)$.
This is the concentration bound used in \cite[Lemma~6.2]{CRTA26}; a conditional union bound over its current grid gives the same event probability.
The interval-shrinkage and budget arguments are deterministic on these events, and their probabilities are combined by conditional expectation.
Thus \eqref{eq:MHR:conditional-subGaussian} supplies every probabilistic hypothesis needed by that algorithm.

Choose an absolute constant $A\ge\max\{2\log B,\sigma\}$ and apply the algorithm to the normalized objective $A^{-1}\phi((x-1)\log B)$ on $[0,1]$, using observations $W((x-1)\log B)/A$.
The bound $0\le\phi\le2\log B$ and \eqref{eq:MHR:conditional-subGaussian} give its bounded-range and noise conditions.
With $T=\lceil C\eps^{-2}\rceil$ for a sufficiently large absolute constant $C$, its output satisfies $\phi(z)\le\min\phi+\eps/8$ with probability at least $7/8$.

Run this fixed-confidence procedure independently an odd number $k=O(\log(1/\delta))$ of times and return the median $\widehat z$ of its outputs.
The sublevel set $\{z:\phi(z)\le\min\phi+\eps/8\}$ is an interval by convexity.
Therefore, the median belongs to this set whenever more than half the runs succeed.
A Chernoff bound makes the probability of failure at most $\delta/4$ for a sufficiently large constant in $k$.
This amplification uses at most $L=kT=O(\eps^{-2}\log(1/\delta))$ oracle observations, without needing to estimate or compare the final objective values.

We now bound the number of pricing queries needed for these $L$ oracle observations.
For the uncapped simulation, the count $N+1$ in each call is conditionally stochastically dominated by a geometric random variable with success probability $q_0$.
Fix $\theta>0$ with $(1-q_0)e^\theta<1$ and denote by $M_0\triangleq q_0e^\theta/(1-(1-q_0)e^\theta)$ the bound on its exponential moment.
The conditional exponential moment of each count is at most $M_0$.
Iterating conditional expectations over at most $L$ calls shows that the total exceeds
\begin{equation*}
 K=\left\lceil\frac{L\log M_0+\log(4/\delta)}\theta\right\rceil
 =O\left(\eps^{-2}\log\frac1\delta\right)
\end{equation*}
with probability at most $\delta/4$.
Impose $K$ as a hard cap: if it is reached before all oracle calls finish, return an arbitrary price.
This cap holds on every run, including unsuccessful localization runs.
On the successful localization event, coupling the capped and uncapped simulations adds at most $\delta/4$ to the failure probability.

On successful localization, successful median amplification, and no truncation, the price $\widehat p=ae^{\widehat z}$ satisfies $\Rprice(\widehat p)\ge e^{-\eps/8}\max_{p\in I}\Rprice(p)\ge(1-\eps/8)(1-\eps/16)r^*\ge(1-\eps)r^*$.
The three failure probabilities sum to at most $3\delta/4$.
Localization and the capped oracle simulation together use $O(\eps^{-2}\log(1/\delta))$ pricing queries, and only the initial value sample is used.
\end{proof}

{\subsection{Sample Complexity Upper Bound}
\label{subsec:MHR:UB-sample}}

{For learning from value samples alone, we use the same empirical revenue maximizer and cutoff $\widehat Q(p)\ge2/(3e)$ as for quasi-MHR distributions.}
The stronger sample bound comes from comparing the empirical errors at competing prices directly.
The quadratic revenue bound of~\cite[Lemma~3.3]{HMR18}, stated in \pref{eq:MHR:sample:quadratic-loss}, implies that a price with relative revenue loss $s$ is within quantile distance $O(\sqrt{s})$ of the monopoly price.
We therefore need a uniform concentration estimate on a small quantile neighborhood.

\begin{restate}[\Cref{thm:MHR:UB-sample}]
Fix $0<\eps,\delta\le1/4$.
For the family of MHR distributions, there exists an $(\eps, \delta)$-PAC learning algorithm that draws at most {$O\bigl(\frac{1}{\eps^{3/2}}\log(\frac{1}{\delta})\bigr)$} many samples and makes no pricing queries.
\end{restate}

We reuse the nested-interval estimate in \pref{eq:quasi-MHR:nested-concentration}, proved in the query-sample tradeoff analysis.
The resulting local bound, \pref{eq:MHR:sample:local-concentration}, controls differences of empirical relative errors within quantile distance $d$ using $O((d/t^2+1/t)\log(1/\delta))$ samples for accuracy $t$.
For a revenue-loss range $[s,2s)$, we take $d=O(\sqrt{s})$ and $t=\Theta(s)$, giving a failure probability bounded by $C\exp(-cns^{3/2})$ for universal constants $C,c>0$ and sample size $n$.
Summing over the ranges with $s=2^j\eps$ gives the stated bound without an extra $\log(1/\eps)$ factor.
We prove the quadratic-loss comparison, the local concentration estimate, and the complete revenue guarantee in the MHR proof in \pref{sec:sample-complexity}.

\subsection{Lower Bounds}
\label{subsec:MHR:LB}

To show that the query complexity in \pref{thm:MHR:UB-query} is optimal, we construct two MHR distributions with disjoint sets of $(1-\eps)$-optimal prices.
Since MHR distributions are defined by convexity of the cumulative hazard, we perturb the cumulative hazard of a base distribution near its optimal price by a bump of height $\Theta(\eps)$ and width $\Theta(\sqrt\eps)$ that keeps it convex.
The width is the same as in the MHR sample lower bound of~\cite[Theorem~4.12]{HMR18}, which perturbs the density of the uniform distribution on $[1,2]$.
The density then changes by a factor $1\pm O(\sqrt\eps)$ on an interval of mass $O(\sqrt\eps)$, so a value sample has KL divergence $O(\eps^{3/2})$, whereas the sale probability at each price changes by $O(\eps)$, so a pricing query has KL divergence $O(\eps^2)$.
By \pref{lem:LB:weighted-information} with a single alternative distribution, a PAC algorithm must collect KL divergence at least $\frac1{16}\log\frac1\delta$ from its value samples and pricing queries.
Under the sample budget, the value samples supply at most half of it, so the pricing queries must supply the rest, which requires $\Omega(\eps^{-2}\log\frac1\delta)$ queries.
The following theorem formally states this lower bound.

\begin{theorem}[Lower Bound for MHR Distributions]
\label{thm:MHR:LB-tradeoff}
There exists an absolute constant $C>0$ such that the following holds.
Fix $0<\delta\le1/4$ and $0<\eps\le1/6400$.
If an $(\eps,\delta)$-PAC learning algorithm for MHR distributions draws at most $\frac{C}{\eps^{3/2}}\log\frac1\delta$ value samples, then there exists an MHR distribution on which it makes at least $\frac{C}{\eps^2}\log\frac1\delta$ pricing queries in expectation.
\end{theorem}

To prove \Cref{thm:MHR:LB-tradeoff}, we first construct the two distributions through their cumulative hazards.
Set $d\triangleq10\sqrt\eps$, so that $d\le1/8$.
Denote by $\psi$ the continuously differentiable function
\begin{equation*}
 \psi(t)\triangleq
 \begin{cases}
  0,&t\notin[-2,4],\\
  -(t+2)^2/4,&-2\le t\le0,\\
  t^2/2-t-1,&0\le t\le2,\\
  -(4-t)^2/4,&2\le t\le4,
 \end{cases}
\end{equation*}
which satisfies $-3/2\le\psi\le0$, $|\psi'|\le1$, and $\psi''\ge-1/2$ on each piece.
On $p\ge0$, define $F^{(j)}(p)\triangleq1-e^{-H_j(p)}$ for $j\in\{0,1\}$, where $H_0(p)\triangleq p^2/2$ and $H_1(p)\triangleq p^2/2+d^2\psi((p-1)/d)$.
By \Cref{def:MHR}, $H_j$ is the cumulative hazard of $F^{(j)}$, and $Q_j\triangleq e^{-H_j}$ is its sale probability.
The two distributions differ only on $J\triangleq[1-2d,1+4d]\subseteq[3/4,3/2]$.
The following lemma verifies that they are MHR.

\begin{lemma}[MHR]
\label{lem:MHR:LB:MHR}
The distributions $F^{(0)}$ and $F^{(1)}$ are MHR distributions.
\end{lemma}

\begin{proof}
By \Cref{def:MHR}, it suffices that the cumulative hazards $H_0$ and $H_1$ are convex.
The base hazard $H_0(p)=p^2/2$ is convex.
Differentiating $H_1$ gives the continuous derivative $H_1'(p)=p+d\psi'((p-1)/d)$ and $H_1''(p)=1+\psi''((p-1)/d)\ge\frac12$ on each piece of $\psi$, so $H_1$ is convex.
\end{proof}

We next show that the two sets of $(1-\eps)$-optimal prices are disjoint.
For $p>0$, denote by $\phi_j(p)\triangleq H_j(p)-\log p$ the negative logarithm of the revenue $pQ_j(p)$, and by $r_j^*$ the optimal revenue under $F^{(j)}$.
A price is $(1-\eps)$-optimal under $F^{(j)}$ exactly when $\phi_j(p)-\inf_{x>0}\phi_j(x)\le\log(1/(1-\eps))$.
Since $\phi_0'(1)=0$ and $\phi_0''(p)=1+p^{-2}\ge1$, every $(1-\eps)$-optimal price $p$ under $F^{(0)}$ satisfies $(p-1)^2/2\le\phi_0(p)-\phi_0(1)\le\log(1/(1-\eps))\le2\eps$, and hence $|p-1|\le2\sqrt\eps=d/5$.
Set $p_b\triangleq1+d/5$ and $p_a\triangleq1+d/3$.
Since $\psi'(1/5)=-4/5$, we have $\phi_1'(p_b)=p_b-p_b^{-1}-4d/5\le2d/5-4d/5<0$, so the convex function $\phi_1$ is decreasing on $(0,p_b]$, and every $(1-\eps)$-optimal price $p$ under $F^{(0)}$ satisfies $\phi_1(p)\ge\phi_1(p_b)$.
Since $\phi_0'(1+s)\le2s$ for $s\ge0$, we have $\phi_0(p_a)-\phi_0(p_b)\le\int_{d/5}^{d/3}2s\,\d s=16d^2/225$.
Together with $\psi(1/5)-\psi(1/3)=22/225$, the definition of $H_1$ gives
\begin{equation*}
 \phi_1(p_b)-\phi_1(p_a)
 =\phi_0(p_b)-\phi_0(p_a)+d^2\bigl(\psi(1/5)-\psi(1/3)\bigr)
 \ge d^2\left(\frac{22}{225}-\frac{16}{225}\right)
 =\frac83\eps
 >\log\frac1{1-\eps}.
\end{equation*}
Consequently, every $(1-\eps)$-optimal price $p$ under $F^{(0)}$ has revenue $e^{-\phi_1(p)}\le e^{-\phi_1(p_b)}<(1-\eps)e^{-\phi_1(p_a)}\le(1-\eps)r_1^*$ under $F^{(1)}$, and the two sets are disjoint.

We fix the absolute constant $A\triangleq15\cdot10^4$ for the next two lemmas.
The following lemma bounds the KL divergence of a value sample between $F^{(0)}$ and $F^{(1)}$.

\begin{lemma}[KL Divergence of Samples]
\label{lem:MHR:LB:sample-KL}
We have $\DKL(F^{(0)},F^{(1)})\le A\eps^{3/2}$.
\end{lemma}

\begin{proof}
The densities $f_j=H_j'e^{-H_j}$ of $F^{(j)}$ agree outside $J$, and on $J$ we have $f_1(p)/f_0(p)=(1+d\psi'((p-1)/d)/p)e^{-d^2\psi((p-1)/d)}$.
Denote by $z(p)\triangleq f_1(p)/f_0(p)-1$ the relative density change on $J$.
Since $|\psi'|\le1$, $-3/2\le\psi\le0$, and $p\ge3/4$ on $J$, the ratio $f_1(p)/f_0(p)$ lies in $[1-4d/3,(1+4d/3)e^{3d^2/2}]$, so $|z(p)|\le(1+4d/3)e^{3d^2/2}-1\le2d\le\frac14$, where the last two inequalities use $d\le1/8$.
Since $f_0(p)=pe^{-p^2/2}\le1$, the mass of $J=[1-2d,1+4d]$ under $F^{(0)}$ is at most $6d$, and since the two distributions agree at both endpoints of $J$, $\int_Jf_0(p)z(p)\,\d p=0$.
As in the proof of \Cref{lem:quasi-regular:LB:sample-KL}, expanding the KL divergence by its definition and substituting $f_1=(1+z)f_0$ on $J$, we obtain that
\begin{align*}
 \DKL(F^{(0)},F^{(1)})
 &=\int_Jf_0(p)\log\frac{f_0(p)}{f_1(p)}\,\d p
 =\int_Jf_0(p)\bigl[-\log(1+z(p))+z(p)\bigr]\,\d p\\
 &\le\int_Jf_0(p)z(p)^2\,\d p
 \le6d(2d)^2
 =24\cdot10^3\eps^{3/2}
 \le A\eps^{3/2},
\end{align*}
where the second equality uses $\int_Jf_0(p)z(p)\,\d p=0$, the first inequality uses $-\log(1+z)+z\le z^2$ for $|z|\le\frac12$, the second inequality uses $|z(p)|\le2d$ and the mass at most $6d$ of $J$, and the last equality uses $d=10\sqrt\eps$.
\end{proof}

The following lemma bounds the KL divergence of a pricing query at a price in $J$.

\begin{lemma}[KL Divergence of Queries]
\label{lem:MHR:LB:query-KL}
For every query price $p\in J$, we have $\DKL(F^{(0)}(p),F^{(1)}(p))\le A\eps^2$.
\end{lemma}

\begin{proof}
Since $Q_0$ and $Q_1$ are decreasing and agree at both endpoints of $J\subseteq[3/4,3/2]$, both $Q_0(p)$ and $Q_1(p)$ lie in $[Q_0(3/2),Q_0(3/4)]=[e^{-9/8},e^{-9/32}]\subseteq[1/4,4/5]$.
Since $x(1-x)$ is concave, its minimum over $[1/4,4/5]$ is attained at an endpoint, so $Q_1(p)(1-Q_1(p))\ge\min\{\frac14\cdot\frac34,\frac45\cdot\frac15\}=\frac4{25}$.
Since $|e^{-x}-e^{-y}|\le|x-y|$ for $x,y\ge0$, we also have $|Q_1(p)-Q_0(p)|\le|H_1(p)-H_0(p)|=d^2|\psi((p-1)/d)|\le\frac32d^2$.
Expanding the Bernoulli KL divergence by its definition, we obtain that
\begin{align*}
 \DKL(F^{(0)}(p),F^{(1)}(p))
 &=Q_0(p)\log\frac{Q_0(p)}{Q_1(p)}+(1-Q_0(p))\log\frac{1-Q_0(p)}{1-Q_1(p)}\\
 &\le\frac{(Q_0(p)-Q_1(p))^2}{Q_1(p)(1-Q_1(p))}
 \le\frac{(3d^2/2)^2}{4/25}
 \le15d^4
 =A\eps^2,
\end{align*}
where the first inequality is the chi-squared bound from the proof of \Cref{lem:quasi-regular:LB:query-KL}, the second inequality uses the two bounds above, and the last equality uses $d=10\sqrt\eps$.
\end{proof}

We now prove \Cref{thm:MHR:LB-tradeoff} by combining the disjointness of the good-price sets with \Cref{lem:MHR:LB:sample-KL,lem:MHR:LB:query-KL} through \pref{lem:LB:weighted-information}.

\begin{proof}[Proof of \Cref{thm:MHR:LB-tradeoff}]
Consider an $(\eps,\delta)$-PAC learning algorithm, with $p^\calA$, $S$, $T$, $\prob[F^{(j)}]{\cdot}$, and $\expect[F^{(j)}]{\cdot}$ as in the proof of \Cref{thm:quasi-regular:LB-tradeoff}.
Denote by $\scrE^{(1)}$ the event that $p^\calA$ is $(1-\eps)$-optimal under $F^{(1)}$, and by $T^{(1)}$ the number of pricing queries with prices in $J$.
As in the proof of \Cref{thm:regular:LB-tradeoff}, the PAC guarantee gives $\prob[F^{(1)}]{\scrE^{(1)}}\ge1-\delta$, and since the two sets of $(1-\eps)$-optimal prices are disjoint, it also gives $\prob[F^{(0)}]{\scrE^{(1)}}\le\delta\le1/2$.
Applying \pref{lem:LB:weighted-information} with $I=\{1\}$, the price interval $P^{(1)}=J$, the weight $w_1=1$, and the bounds $d_1=A\eps^{3/2}$ and $b_1=A\eps^2$ from \Cref{lem:MHR:LB:sample-KL,lem:MHR:LB:query-KL}, and using $T^{(1)}\le T$, we obtain
\begin{equation*}
 A\eps^{3/2}\expect[F^{(0)}]{S}+A\eps^2\expect[F^{(0)}]{T}
 \ge\frac1{16}\log\frac1\delta.
\end{equation*}
Set $C\triangleq1/(32A)$.
Since $\expect[F^{(0)}]{S}\le\frac{C}{\eps^{3/2}}\log\frac1\delta$, the first term on the left is at most $AC\log\frac1\delta=\frac1{32}\log\frac1\delta$, and hence $\expect[F^{(0)}]{T}\ge\frac{1}{32A\eps^2}\log\frac1\delta=\frac{C}{\eps^2}\log\frac1\delta$.
\end{proof}

%% file: Paper/apx-DKW.tex
\begin{theorem}[A Multiplicative-Accuracy DKW Inequality]
\label{thm:multiplicative-DKW}
We consider independent uniform samples $U_1,\ldots,U_n\sim\Unif(0,1)$,
with empirical distribution function
$\empCDF(x)
\eqdef\frac1n\sum\nolimits_{i\in[n]}\indicator[U_i\le x]$ for all $x\in[0,1]$.
For every quantile threshold $q\in(0,1)$, relative-error tolerance
$\eps'\in(0,1)$, and sample size $n\ge1$, we have
\begin{align*}
    \prob{\sup\nolimits_{x\in[q,1]}
        \left|\frac{\empCDF(x)}{x}-1\right|>\eps'}
    \le
    2\exp\!\left(
        -\frac{nq(\eps')^2}{2(1+\eps'/3)}
    \right)~.
\end{align*}
Consequently, for every failure probability $\delta\in(0,1)$,
a sample size satisfying
$n\ge
\frac{2(1+\eps'/3)}{q(\eps')^2}
\log\frac2\delta$ ensures that
\begin{align*}
    \prob{
        (1-\eps')x\le\empCDF(x)\le(1+\eps')x
        \text{ for every }x\in[q,1]
    }
    \ge1-\delta~.
\end{align*}
In particular, the sufficient sample size is $O\!\left(q^{-1}(\eps')^{-2}\log(1/\delta)\right)$
for $0<\delta\le1/2$.
\end{theorem}

\begin{corollary}
\label{cor:multiplicative-DKW-sale-probability}
Fix any arbitrary value distribution $F$ on $[0,\infty)$ with its
sale-probability function
$\Qsale(p)\eqdef1-F(p)=\prob[v\sim F]{v\ge p}$.
Let $\empQsale(p)\eqdef\frac1n\sum\nolimits_{i\in[n]}\indicator[v_i\ge p]$ be the empirical sale-probability function constructed from $n$ independent samples
$v_1,\ldots,v_n\sim F$.
For every quantile threshold $q\in(0,1)$, relative-error tolerance
$\eps'\in(0,1)$, and failure probability $\delta\in(0,1)$,
a sample size satisfying
\begin{align*}
    n
    \ge
    \frac{2(1+\eps'/3)}{q(\eps')^2}
    \log\frac2\delta~
\end{align*}
ensures that, with probability at least $1-\delta$, the following bounds
hold simultaneously for all prices $p\ge0$:
\begin{alignat*}{2}
    (1-\eps')\Qsale(p)
    &\le\empQsale(p)\le(1+\eps')\Qsale(p)
    \quad
    &&\text{whenever }\Qsale(p)\ge q~,\\
    \empQsale(p)
    &\le(1+\eps')q
    \quad
    &&\text{whenever }\Qsale(p)<q~.
\end{alignat*}
\end{corollary}
\begin{proof}[Proof of \Cref{cor:multiplicative-DKW-sale-probability}]
We use the generalized inverse
$\salequantile(u)\eqdef\sup\{p\ge0:\Qsale(p)\ge u\}$
to represent the value samples as $v_i=\salequantile(U_i)$, where the
auxiliary samples $U_1,\ldots,U_n$ are independent and uniform on $(0,1)$.
The left-continuity of the sale-probability function $\Qsale$ ensures that
$\indicator[v_i\ge p]=\indicator[U_i\le\Qsale(p)]$.
Thus, the empirical CDF $\empCDF$ of the auxiliary uniform samples
satisfies $\empQsale(p)=\empCDF(\Qsale(p))$ simultaneously for all prices
$p\ge0$, including when the value distribution has atoms.

We apply \Cref{thm:multiplicative-DKW} with the stated parameters and
sample size to obtain, with probability at least $1-\delta$,
\begin{align*}
    (1-\eps')x
    \le
    \empCDF(x)
    \le
    (1+\eps')x
    \quad\text{for every }x\in[q,1]~.
\end{align*}
On this event, we obtain the first conclusion by substituting
$x=\Qsale(p)$ whenever the sale probability satisfies $\Qsale(p)\ge q$.
For every remaining price, we use monotonicity of the empirical CDF
to obtain
\begin{align*}
    \empQsale(p)
    =
    \empCDF(\Qsale(p))
    \le
    \empCDF(q)
    \le
    (1+\eps')q~.
\end{align*}
Both conclusions therefore hold simultaneously on the same event. This completes the proof.
\end{proof}

We establish the \Cref{thm:multiplicative-DKW} using the following intermediate results.
For each threshold $t\in(0,1]$, we define the relative empirical process
$\relprocess(t)\eqdef\empCDF(t)/t$ and the backward filtration
\begin{align*}
    \calF_t
    \eqdef
    \sigma\!\left(\empCDF(u):u\in[t,1]\right)~.
\end{align*}
The information represented by this filtration increases as the threshold
decreases: we have $\calF_t\subseteq\calF_s$ whenever $0<s<t\le1$.

\begin{lemma}[Reverse-Martingale Property]
\label{lem:DKW-martingale}
The relative empirical process $\relprocess$ is a martingale when the
threshold decreases.
Specifically, for every pair of thresholds $0<s<t\le1$, we have
\begin{align*}
    \expect{\relprocess(s)\mid\calF_t}
    =\relprocess(t)
    \quad\text{almost surely}~.
\end{align*}
\end{lemma}

\begin{proof}[Proof of \Cref{lem:DKW-martingale}]
We consider fixed thresholds $0<s<t\le1$.
The information in $\calF_t$ determines the sample count
$K\eqdef n\empCDF(t)$ in $[0,t]$, together with the unordered sample
locations in $(t,1]$.

We first condition further on which observations lie in $[0,t]$ and
on the values of all remaining observations.
Conditional on this information, the $K$ observations in $[0,t]$ are
independent and uniform on $[0,t]$.
Their count in $[0,s]$ therefore has the binomial distribution
$\Bin(K,s/t)$.
Since this conditional law depends on the revealed information only
through the count $K$, which is already measurable with respect to
$\calF_t$, the same conditional law holds given $\calF_t$.
We obtain
\begin{align*}
    \expect{\relprocess(s)\mid\calF_t}
    =
    \frac1{ns}\,
    \expect{n\empCDF(s)\mid\calF_t}
    =
    \frac1{ns}\,n\empCDF(t)\frac{s}{t}
    =\relprocess(t)~.
\end{align*}
The bound $0\le\relprocess(t)\le1/t$ ensures integrability at every
positive threshold.
This completes the proof.
\end{proof}

\begin{proposition}[{Doob's Maximal Inequality \cite{D19}}]
\label{prop:DKW-Doob}
For every integrable discrete-time submartingale
$\expprocess_0,\ldots,\expprocess_m$ and positive threshold $a$, we have
$\prob{\max\nolimits_{0\le j\le m}\expprocess_j\ge a}
\le
\frac{\expect{\max\{\expprocess_m,0\}}}{a}$.
In particular, the numerator is $\expect{\expprocess_m}$ when the
submartingale is nonnegative.
\end{proposition}
\wtdelete{\begin{proof}[Proof of \Cref{prop:DKW-Doob}]
We define the first-crossing event
\begin{align*}
    \firstcrossing_j
    \eqdef
    \{\expprocess_j\ge a\}
    \cap\bigcap\nolimits_{i\in[0:j-1]}\{\expprocess_i<a\}~,
    \quad j=0,\ldots,m~,
\end{align*}
where the intersection is unrestricted when $j=0$.
These events are disjoint, and each event $\firstcrossing_j$ is measurable
with respect to the information available at time $j$.
We use the submartingale property to obtain
\begin{align*}
    a \prob{\max\nolimits_{0\le j\le m}\expprocess_j\ge a}
    &=
    a\sum\nolimits_{j\in[0:m]}\prob{\firstcrossing_j}
    \le
    \sum\nolimits_{j\in[0:m]}
    \expect{\expprocess_j\indicator[\firstcrossing_j]}\\
    &\le
    \sum\nolimits_{j\in[0:m]}
    \expect{\expprocess_m\indicator[\firstcrossing_j]}
    =
    \expect{\expprocess_m
        \indicator\!\left[
            \bigcup\nolimits_{j\in[0:m]}\firstcrossing_j
        \right]}
    \le
    \expect{\max\{\expprocess_m,0\}}~.
\end{align*}
This completes the proof.
\end{proof}}

\begin{lemma}[Bernstein MGF Bound]
\label{lem:DKW-mgf}
For every quantile threshold $q\in(0,1)$ and exponential parameter
$\lambda\in\R$ satisfying $|\lambda|<3nq$, we have
\begin{align*}
    \expect{e^{\lambda(\relprocess(q)-1)}}
    \le
    \exp\!\left(
        \frac{\lambda^2}
        {2nq\left(1-\frac{|\lambda|}{3nq}\right)}
    \right)~.
\end{align*}
\end{lemma}

\wtdelete{\begin{proof}[Proof of \Cref{lem:DKW-mgf}]
We define the independent Bernoulli variables
$\bernindicator_i\eqdef\indicator[U_i\le q]$
and the centered variables
$\centeredindicator_i\eqdef(\bernindicator_i-q)/q$,
for $i=1,\ldots,n$.
These variables satisfy
\begin{align*}
    \relprocess(q)-1
    =\frac1n\sum\nolimits_{i\in[n]}\centeredindicator_i~,
    \quad
    \expect{\centeredindicator_i}=0~,
    \quad
    \expect{\centeredindicator_i^2}
    =\frac{1-q}{q}\le\frac1q~,
    \quad
    |\centeredindicator_i|\le\frac1q~.
\end{align*}
We therefore obtain the following moment bound for every integer
$k\ge2$:
\begin{align*}
    \bigl|\expect{\centeredindicator_i^k}\bigr|
    \le
    \expect{|\centeredindicator_i|^k}
    \le
    q^{-(k-2)}\expect{\centeredindicator_i^2}
    \le
    q^{-(k-1)}~.
\end{align*}
We expand the exponential in a Taylor series and use the factorial
bound $k!\ge2\cdot3^{k-2}$ for every $k\ge2$, to obtain
\begin{align*}
    \expect{e^{(\lambda/n)\centeredindicator_i}}
    =
    1+\sum\nolimits_{k=2}^{\infty}
    \frac{(\lambda/n)^k\expect{\centeredindicator_i^k}}{k!}
    &\le
    1+\sum\nolimits_{k=2}^{\infty}
    \frac{(|\lambda|/n)^k}{k!\,q^{k-1}}\\
    &\le
    1+\frac{\lambda^2}{2n^2q}
    \sum\nolimits_{j=0}^{\infty}
    \left(\frac{|\lambda|}{3nq}\right)^j\\
    &=
    1+\frac{\lambda^2}
    {2n^2q\left(1-\frac{|\lambda|}{3nq}\right)}
    \le
    \exp\!\left(
        \frac{\lambda^2}
        {2n^2q\left(1-\frac{|\lambda|}{3nq}\right)}
    \right)~.
\end{align*}
The boundedness of $\centeredindicator_i$ justifies exchanging expectation
and the Taylor series.
We use the moment bound in the first inequality, the factorial bound in
the second, and the elementary inequality $1+u\le e^u$ in the last.
We finally use independence to conclude that
\begin{align*}
    \expect{e^{\lambda(\relprocess(q)-1)}}
    =
    \prod\nolimits_{i\in[n]}
    \expect{e^{(\lambda/n)\centeredindicator_i}}
    \le
    \exp\!\left(
        \frac{\lambda^2}
        {2nq\left(1-\frac{|\lambda|}{3nq}\right)}
    \right)~.
\end{align*}
This completes the proof.
\end{proof}}

\begin{proof}[Proof of \Cref{thm:multiplicative-DKW}]
We fix the quantile threshold $q\in(0,1)$ and relative-error tolerance
$\eps'\in(0,1)$.
For an exponential parameter $\lambda\in\R$, we define the nonnegative
process
$\expprocess_\lambda(t)\eqdef
\exp(\lambda(\relprocess(t)-1))$,
for $t\in[q,1]$.
We apply conditional Jensen's inequality and
\Cref{lem:DKW-martingale} to obtain, for every pair of thresholds
$q\le s<t\le1$,
\begin{align*}
    \expect{\expprocess_\lambda(s)\mid\calF_t}
    =
    \expect{
        e^{\lambda(\relprocess(s)-1)}
        \mid\calF_t
    }
    \ge
    \exp\!\left(
        \lambda\bigl(
            \expect{\relprocess(s)\mid\calF_t}-1
        \bigr)
    \right)
    =
    \expprocess_\lambda(t)~.
\end{align*}
Thus, the process $\expprocess_\lambda$ is a nonnegative submartingale
when the threshold decreases.
The bound $0\le\relprocess(t)\le1/q$ on $[q,1]$ ensures that this process
is integrable.

We first control the maximum over finite sets of thresholds.
For every integer $m\ge0$, we define the finite grid
\begin{align*}
    \calD_m
    \eqdef
    \left\{
        q+(1-q)\frac{j}{2^m}:j=0,\ldots,2^m
    \right\}~.
\end{align*}
We order the grid points from $1$ down to $q$, so the terminal threshold
is always $q$.
For every positive exponential parameter $\lambda$,
\Cref{prop:DKW-Doob} gives
\begin{align*}
    \prob{
        \max\nolimits_{t\in\calD_m}(\relprocess(t)-1)>\eps'
    }
    &\le
    e^{-\lambda\eps'}
    \expect{e^{\lambda(\relprocess(q)-1)}}~,\\
    \prob{
        \max\nolimits_{t\in\calD_m}(1-\relprocess(t))>\eps'
    }
    &\le
    e^{-\lambda\eps'}
    \expect{e^{-\lambda(\relprocess(q)-1)}}~.
\end{align*}

We next extend these bounds to the entire interval.
The empirical CDF $\empCDF$, and hence the relative empirical process
$\relprocess$, is right-continuous.
The grids are nested, their union is dense in $[q,1]$, and both endpoints
belong to every grid.
We consequently have
\begin{align*}
    \sup\nolimits_{t\in[q,1]}(\relprocess(t)-1)
    &=
    \sup\nolimits_{m\ge0}
    \max\nolimits_{t\in\calD_m}(\relprocess(t)-1)~,\\
    \sup\nolimits_{t\in[q,1]}(1-\relprocess(t))
    &=
    \sup\nolimits_{m\ge0}
    \max\nolimits_{t\in\calD_m}(1-\relprocess(t))~.
\end{align*}
We can therefore pass to the limit through the increasing exceedance
events.
For either choice of sign and every positive exponential parameter
$\lambda<3nq$, \Cref{lem:DKW-mgf} yields
\begin{align*}
    \prob{
        \sup\nolimits_{t\in[q,1]}
        \bigl(\pm(\relprocess(t)-1)\bigr)>\eps'
    }
    \le
    e^{-\lambda\eps'}
    \expect{e^{\pm\lambda(\relprocess(q)-1)}}
    \le
    \exp\!\left(
        -\lambda\eps'
        +\frac{\lambda^2}
        {2nq\left(1-\frac{\lambda}{3nq}\right)}
    \right)~.
\end{align*}
Here, the two choices of sign give the upper- and lower-deviation bounds
separately.

We choose the exponential parameter
$\lambda\eqdef nq\eps'/(1+\eps'/3)$.
This choice satisfies $\lambda<3nq$, and direct substitution gives
\begin{align*}
    -\lambda\eps'
    +\frac{\lambda^2}
    {2nq\left(1-\frac{\lambda}{3nq}\right)}
    =
    -\frac{nq(\eps')^2}{2(1+\eps'/3)}~.
\end{align*}
We combine the upper and lower deviations to obtain
\begin{align*}
    \prob{
        \sup\nolimits_{t\in[q,1]}
        \left|\frac{\empCDF(t)}{t}-1\right|>\eps'
    }
    \le
    2\exp\!\left(
        -\frac{nq(\eps')^2}{2(1+\eps'/3)}
    \right)~.
\end{align*}
The stated lower bound on the sample size makes this failure probability
at most $\delta$.
This completes the proof.
\end{proof}

%% file: Paper/apx-sample-UB.tex
\subsection{Quasi-Regular Distributions}

\begin{theorem}[Upper Bound for Quasi-Regular Distributions]
\label{thm:quasi-regular:UB-sample}
For the family of quasi-regular distributions, there exists an $(\eps, \delta)$-PAC learning algorithm that draws at most $O\bigl(\frac{1}{\eps^{3}}\log(\frac{1}{\delta})\bigr)$ many samples.
\end{theorem}

\begin{proof}[Proof of \Cref{thm:quasi-regular:UB-sample}]
Given a quasi-regular value distribution $F$, define its sale-probability
function $\Qsale(p) \eqdef 1-F(p)$, revenue-price curve
$\Rprice(p) \eqdef p\Qsale(p)$, and optimal revenue
$r^{*} \eqdef \sup\nolimits_{p\ge0}\Rprice(p)$.

We set the accuracy parameter $\eta\eqdef\eps/8$ and choose the sample size
$n
\eqdef
\left\lceil
    \frac{2(1+\eta/3)}{\eta^3}\log\frac2\delta
\right\rceil
=
O\!\left(\eps^{-3}\log\frac2\delta\right)$
We draw independent value samples $v_1,\ldots,v_n\sim F$ and construct
the empirical sale-probability function
$\empQsale(p)\eqdef\frac1n\sum\nolimits_{i\in[n]}\indicator[v_i\ge p]$
and empirical revenue-price curve
$\empRprice(p)\eqdef p\empQsale(p)$.
We apply \Cref{cor:multiplicative-DKW-sale-probability} with the quantile
threshold $q=\eta$ and relative-error tolerance $\eps'=\eta$.
With probability at least $1-\delta$, we obtain the following bounds
simultaneously for all prices $p\ge0$:
\begin{equation}
\begin{alignedat}{2}
    (1-\eta)\Qsale(p)
    &\le\empQsale(p)\le(1+\eta)\Qsale(p)
    \quad
    &&\text{whenever }\Qsale(p)\ge\eta~,\\
    \empQsale(p)
    &\le(1+\eta)\eta
    \quad
    &&\text{whenever }\Qsale(p)<\eta~.
    \label{eq:quasi-regular:sample:concentration}
\end{alignedat}
\end{equation}
    

We next show that excluding prices with small sale probabilities preserves
nearly optimal revenue.
Specifically, for every quantile threshold $a\in(0,1)$, we have
\begin{align}
    \sup\nolimits_{p:\Qsale(p)\ge a}\Rprice(p)
    \ge (1-a)r^{*}~. 
    \label{eq:quasi-regular:sample:quantile-cutoff}
\end{align}
Any price $p$ with $\Qsale(p)\ge a$ already satisfies the restriction.
For a positive-revenue price $p$ with $\Qsale(p)<a$, we define the lower price
$\lowerprice \eqdef \frac{1-a}{a}\Rprice(p)<p$.
We show that this lower price has sale probability at least $a$, while
retaining at least a $(1-a)$-fraction of the original revenue.
When the sale probability $\Qsale(\lowerprice)$ is less than one, we obtain
\begin{align*}
    \frac{\lowerprice\Qsale(\lowerprice)}
         {1-\Qsale(\lowerprice)}
    \ge \frac{p\Qsale(p)}{1-\Qsale(p)}
    \ge \Rprice(p)~,
    \quad
    \Qsale(\lowerprice)
    \ge \frac{\Rprice(p)}{\lowerprice+\Rprice(p)}
    =a~.
\end{align*}
Using the quasi-regularity condition that the ratio
$p\Qsale(p)/(1-\Qsale(p))$ is nonincreasing and rearranging,
we therefore have
\begin{align*}
    \Rprice(\lowerprice)
    =\lowerprice\Qsale(\lowerprice)
    \ge a\lowerprice
    =(1-a)\Rprice(p)~.
\end{align*}
We conclude Eqn.~\eqref{eq:quasi-regular:sample:quantile-cutoff} by taking the supremum over the original prices.

Our algorithm returns a price $\empprice$ maximizing empirical revenue
among prices satisfying the empirical cutoff $\empQsale(p)\ge2\eta$.
The zero price is feasible, and we can choose a maximizer from the finite
set $\{0,v_1,\ldots,v_n\}$: every feasible price can be increased to the
smallest sample at or above it without changing its empirical sale
probability.

We now condition on the simultaneous bounds in
Eqn.~\eqref{eq:quasi-regular:sample:concentration}.
We first observe that every price $p$ with sale probability
$\Qsale(p)\ge4\eta$ is feasible for the empirical optimization, since
\begin{align*}
    \empQsale(p)
    \ge (1-\eta)\Qsale(p)
    \ge 4\eta(1-\eta)
    \ge 2\eta~.
\end{align*}
We also observe that every empirically feasible price has true sale
probability at least $\eta$.
Otherwise, the second bound in
Eqn.~\eqref{eq:quasi-regular:sample:concentration} would give
$\empQsale(p)\le(1+\eta)\eta<2\eta$, contradicting feasibility.
Thus, we can apply the relative-error bounds both to the returned price
and to every price with sale probability at least $4\eta$.
We obtain
\begin{align*}
    \Rprice(\empprice)
    \ge \frac{\empRprice(\empprice)}{1+\eta}
    \ge \frac{1}{1+\eta}
        \sup\nolimits_{p:\Qsale(p)\ge4\eta}\empRprice(p)
    &\ge \frac{1-\eta}{1+\eta}
        \sup\nolimits_{p:\Qsale(p)\ge4\eta}\Rprice(p)
    \ge \frac{(1-\eta)(1-4\eta)}{1+\eta}r^{*}\\
    &\ge (1-6\eta)r^{*}
    \ge (1-\eps)r^{*}~.
\end{align*}
We use Eqn.~\eqref{eq:quasi-regular:sample:concentration} in the first and third
inequalities, empirical optimality in the second, and
Eqn.~\eqref{eq:quasi-regular:sample:quantile-cutoff} with the quantile threshold
$a=4\eta$ in the fourth.
We obtain the last two inequalities from the elementary bound
$(1-\eta)/(1+\eta)\ge1-2\eta$ and our choice of the accuracy parameter
$\eta=\eps/8$.

We therefore obtain an $(\eps,\delta)$-PAC learning algorithm for quasi-regular distributions using $O(\eps^{-3}\log(2/\delta))$ samples.
This completes the proof.
\end{proof}

\subsection{Regular Distributions}

\begin{theorem}[Upper Bound for Regular Distributions]
\label{thm:regular:UB-sample}
For the family of regular distributions, there exists an $(\eps, \delta)$-PAC learning algorithm that draws at most $O\bigl(\frac{1}{\eps^{3}}\log(\frac{1}{\delta})\bigr)$ many samples.
\end{theorem}

\begin{proof}[Proof of \Cref{thm:regular:UB-sample}]
Given a regular value distribution $F$, define its revenue-price curve
$\Rprice(p) \eqdef p(1-F(p))$ and optimal revenue
$r^{*} \eqdef \sup_{p\ge0}\Rprice(p)$.
We observe that the distribution $F$ is also quasi-regular, so we apply the sample-based algorithm established in \Cref{thm:quasi-regular:UB-sample} without modification.

Specifically, we use the accuracy parameter $\eta \eqdef \eps/8$ and a
sufficiently large sample size $n=O(\eps^{-3}\log(2/\delta))$.
We construct the empirical sale-probability function $\empQsale$ and
empirical revenue-price curve $\empRprice$ as in the proof of \Cref{thm:quasi-regular:UB-sample},
and return a price $\empprice$ maximizing empirical revenue among prices
satisfying the empirical cutoff $\empQsale(p)\ge2\eta$.
\Cref{thm:quasi-regular:UB-sample} guarantees that this algorithm uses no pricing queries and
that the returned price satisfies
\begin{align*}
    \Pr\bigl[\Rprice(\empprice)\ge(1-\eps)r^{*}\bigr]
    \ge 1-\delta~.
\end{align*}
This guarantee applies to every regular distribution because it holds
for the entire quasi-regular family.

We therefore obtain an $(\eps,\delta)$-PAC learning algorithm for regular
distributions using $O(\eps^{-3}\log(2/\delta))$ samples.
This completes the proof.
\end{proof}

\subsection{Quasi-MHR Distributions}

\begin{theorem}[Upper Bound for Quasi-MHR Distributions]
\label{thm:quasi-MHR:UB-sample}
For the family of quasi-MHR distributions, there exists an $(\eps, \delta)$-PAC learning algorithm that draws at most $O\bigl(\frac{1}{\eps^{2}}\log(\frac{1}{\delta})\bigr)$ many samples.
\end{theorem}

\begin{proof}[Proof of \Cref{thm:quasi-MHR:UB-sample}]
Given a quasi-MHR value distribution $F$, define its sale-probability
function $\Qsale(p) \eqdef 1-F(p)$, revenue-price curve
$\Rprice(p) \eqdef p\Qsale(p)$, and optimal revenue
$r^{*} \eqdef \sup\nolimits_{p\ge0}\Rprice(p)$.

We recall from Step~1 of the proof of \Cref{thm:quasi-MHR:UB-query} that restricting attention to prices with sale probability at least $1/e$ does not decrease the optimal revenue.
In particular, we have
\begin{align}
    \sup\nolimits_{p:\Qsale(p)\ge1/e}\Rprice(p)
    =r^{*}~.
    \label{eq:quasi-MHR:sample:quantile-cutoff}
\end{align}

We set the accuracy parameter $\eta \eqdef \eps/8$ and choose the sample
size $n \eqdef \left\lceil
\frac{9e^{2}}{2\eta^{2}}\log\frac{2}{\delta}
\right\rceil$.
We draw independent value samples $v_1,\ldots,v_n\sim F$ and construct
the empirical sale-probability function
$\empQsale(p) \eqdef \frac1n\sum\nolimits_{i\in[n]}\indicator[v_i\ge p]$
and empirical revenue-price curve
$\empRprice(p) \eqdef p\empQsale(p)$.
We apply the standard additive-accuracy DKW inequality \cite{M90},
as discussed in \Cref{subsec:overview}, to obtain the following bound with
probability at least $1-\delta$:
\begin{align}
    \sup\nolimits_{p\ge0}
    \bigl|\empQsale(p)-\Qsale(p)\bigr|
    \le\frac{\eta}{3e}~.
    \label{eq:quasi-MHR:sample:concentration}
\end{align}

Our algorithm returns a price $\empprice$ maximizing empirical revenue
among prices satisfying the empirical cutoff $\empQsale(p)\ge2/(3e)$,
with ties broken by choosing the smallest price.
The zero price is feasible, and the argument in the proof of \Cref{thm:quasi-regular:UB-sample} shows that we can choose a maximizer from the finite set $\{0,v_1,\ldots,v_n\}$.

We now condition on the simultaneous bound in
Eqn.~\eqref{eq:quasi-MHR:sample:concentration}.
We first observe that every price $p$ with sale probability
$\Qsale(p)\ge1/e$ is feasible for the empirical optimization, since
$\empQsale(p)\ge1/e-\eta/(3e)\ge2/(3e)$.
We also observe that every empirically feasible price has true sale
probability at least $1/(3e)$, since
$\Qsale(p)\ge\empQsale(p)-\eta/(3e)\ge(2-\eta)/(3e)\ge1/(3e)$.
We can therefore convert the additive-error bound into a relative-error
bound throughout the relevant range:
\begin{align}
    \frac{\bigl|\empQsale(p)-\Qsale(p)\bigr|}{\Qsale(p)}
    \le\frac{\eta}{3e\Qsale(p)}
    \le\eta
    \quad\text{whenever }\Qsale(p)\ge\frac1{3e}~.
    \label{eq:quasi-MHR:sample:relative-error}
\end{align}
This bound applies both to the returned price and to every price with
sale probability at least $1/e$.

We combine empirical optimality with
Eqs.~\eqref{eq:quasi-MHR:sample:quantile-cutoff}
and \eqref{eq:quasi-MHR:sample:relative-error} to obtain
\begin{align*}
    \Rprice(\empprice)
    \ge\frac{\empRprice(\empprice)}{1+\eta}
    \ge\frac1{1+\eta}
        \sup\nolimits_{p:\Qsale(p)\ge1/e}\empRprice(p)
    & \ge\frac{1-\eta}{1+\eta}
        \sup\nolimits_{p:\Qsale(p)\ge1/e}\Rprice(p)\\
    &=\frac{1-\eta}{1+\eta}r^{*}
    \ge(1-2\eta)r^{*}
    \ge(1-\eps)r^{*}~.
\end{align*}
We use Eqn.~\eqref{eq:quasi-MHR:sample:relative-error} in the first and
third inequalities, empirical optimality in the second, and
Eqn.~\eqref{eq:quasi-MHR:sample:quantile-cutoff} in the equality.
We obtain the last two inequalities from the elementary bound
$(1-\eta)/(1+\eta)\ge1-2\eta$ and our choice of the accuracy parameter
$\eta=\eps/8$.

We therefore obtain an $(\eps,\delta)$-PAC learning algorithm for quasi-MHR distributions using $O(\eps^{-2}\log(2/\delta))$ samples. 
This completes the proof.
\end{proof}

\subsection{MHR Distributions}
The proof for MHR distributions is a bit different from previous three distribution families, where we do not apply the multiplicative DKW Inequality.
\begin{theorem}[Upper Bound for MHR Distributions]
\label{thm:MHR:UB-sample}
For the family of MHR distributions, there exists an $(\eps, \delta)$-PAC learning algorithm that draws at most $O\bigl(\frac{1}{\eps^{3/2}}\log(\frac{1}{\delta})\bigr)$ many samples.
\end{theorem}

\begin{proof}[Proof of \Cref{thm:MHR:UB-sample}]
Given an MHR value distribution $F$, define its sale-probability function
$\Qsale(p) \eqdef 1-F(p)$, revenue-price curve
$\Rprice(p) \eqdef p\Qsale(p)$, and optimal revenue
$r^{*} \eqdef \sup\nolimits_{p\ge0}\Rprice(p)$.
The MHR condition gives either bounded support or an exponentially decreasing upper tail, so the optimal revenue is finite and achieved at a monopoly price $\monopolyprice$.
We denote the corresponding monopoly quantile by
$\monopolyquantile \eqdef \Qsale(\monopolyprice)$.

We recall that the monopoly quantile satisfies $\monopolyquantile\ge1/e$.
We also use the quadratic revenue bound \cite[Lemma~3.3]{HMR18}, which is stated in \Cref{subsec:overview}.
Specifically, we define the relative revenue loss of a price $p$ by
$\revenueloss(p) \eqdef 1-\Rprice(p)/r^{*}$ and obtain
\begin{align}
    \revenueloss(p)
    \ge\frac14\bigl(\Qsale(p)-\monopolyquantile\bigr)^2~,
    \quad
    \bigl|\Qsale(p)-\monopolyquantile\bigr|
    \le2\sqrt{\revenueloss(p)}
    \quad\text{whenever }\Qsale(p)>0~.
    \label{eq:MHR:sample:quadratic-loss}
\end{align}
The revenue-quantile curve $\Rquantile$ satisfies
$\Rprice(p)\le\Rquantile(\Qsale(p))$, so the same bound covers prices
outside the support as well.

We choose the sample size
$n \eqdef \lceil C\eps^{-3/2}\log(2/\delta)\rceil$,
where the universal constant $C>0$ will be sufficiently large.
From independent value samples $v_1,\ldots,v_n\sim F$, we construct the
empirical sale-probability function
$\empQsale(p) \eqdef \frac1n\sum\nolimits_{i\in[n]}\indicator[v_i\ge p]$
and empirical revenue-price curve
$\empRprice(p) \eqdef p\empQsale(p)$.
Our algorithm is the same as in the proof of
\Cref{thm:quasi-MHR:UB-sample}: we return a price $\empprice$ maximizing
empirical revenue among prices satisfying the empirical cutoff
$\empQsale(p)\ge2/(3e)$, with ties broken by choosing the smallest price.
We can choose this maximizer from the finite set $\{0,v_1,\ldots,v_n\}$,
as established in the proof of \Cref{thm:quasi-regular:UB-sample}.

We first consider the event
$\calE \eqdef
\{\sup\nolimits_{p\ge0}|\empQsale(p)-\Qsale(p)|\le1/(3e)\}$.
We apply the standard additive-accuracy DKW inequality \cite{M90},
as in the proof of \Cref{thm:quasi-MHR:UB-sample}, to obtain
\begin{align*}
    \Pr[\calE^{c}]
    \le2\exp\!\left(-\frac{2n}{9e^2}\right)
    \le\frac{\delta}{2}~,
\end{align*}
where the last inequality holds for a sufficiently large choice of the
constant $C$.
On the event $\calE$, the monopoly price $\monopolyprice$ is feasible
because
$\empQsale(\monopolyprice)
\ge\monopolyquantile-1/(3e)\ge2/(3e)$.
We also observe that every empirically feasible price has true sale
probability at least $1/(3e)$, and that the empirical relative sale
probability at the monopoly price satisfies
$\empQsale(\monopolyprice)/\monopolyquantile\ge2/3$.

We next establish a local concentration bound for the difference between
empirical relative errors.
This is the additional estimate needed for the sharper sample bound
outlined in \Cref{subsec:overview}.
Specifically, for every quantile radius $d\in(0,1]$ and accuracy level
$t\in(0,1)$, we show that there are universal constants $C_0,c_0>0$ such
that
\begin{align}
    \Pr\!\left[
        \sup\nolimits_{p\ge0:\ \Qsale(p)\ge1/(3e), 
            |\Qsale(p)-\monopolyquantile|\le d}
        \left|
            \frac{\empQsale(p)}{\Qsale(p)}
            -\frac{\empQsale(\monopolyprice)}{\monopolyquantile}
        \right|>t
    \right]
    \le C_0\exp\!\left(-\frac{c_0nt^2}{d+t}\right)~.
    \label{eq:MHR:sample:local-concentration}
\end{align}
We prove this bound without conditioning on the event $\calE$.
We begin with the identity
\begin{align*}
    \frac{\empQsale(p)}{\Qsale(p)}
    -\frac{\empQsale(\monopolyprice)}{\monopolyquantile}
    =
    \frac{
        [\empQsale(p)-\empQsale(\monopolyprice)]
        -[\Qsale(p)-\monopolyquantile]
    }{\Qsale(p)}
    +
    \frac{\monopolyquantile-\Qsale(p)}
         {\Qsale(p)\monopolyquantile}
    \bigl(\empQsale(\monopolyprice)-\monopolyquantile\bigr)~.
\end{align*}
Throughout the range in
Eqn.~\eqref{eq:MHR:sample:local-concentration}, the absolute values of
the two coefficients on the right are bounded by $3e$ and $3e^2d$,
respectively.
We observe that the first numerator is, up to sign, the error in the
empirical probability of the interval $[p,\monopolyprice)$ or
$[\monopolyprice,p)$.
These intervals have probability at most $d$, and they form a nested
family on each side of the monopoly price.

We therefore apply the uniform interval bound in
Eqn.~\eqref{eq:quasi-MHR:nested-concentration}, established in the proof
of \Cref{thm:quasi-MHR:UB-tradeoff}, separately to these two families,
with probability bound $V=d$ and error tolerance $t/(6e)$.
We also apply Hoeffding's inequality to the error at the monopoly price,
with tolerance $t/(6e^2d)$.
Unless one of these deviations occurs, the two terms in the identity
have absolute values at most $t/2$ each.
The resulting probability bounds have exponents proportional to
$nt^2/(d+t)$ and $nt^2/d^2$, respectively.
Since the radius satisfies $d\le1$, we have $d^2\le d+t$, and a union
bound gives Eqn.~\eqref{eq:MHR:sample:local-concentration}.

We now compare the returned price with the monopoly price.
On the event $\calE$, empirical optimality gives
$\empRprice(\empprice)\ge\empRprice(\monopolyprice)>0$.
Together with the lower bound $\Qsale(\empprice)\ge1/(3e)$, this implies
that the returned price has relative revenue loss
$\revenueloss(\empprice)<1$.
We consequently obtain
\begin{align}
    \frac{\empQsale(\empprice)}{\Qsale(\empprice)}
    -\frac{\empQsale(\monopolyprice)}{\monopolyquantile}
    =
    \frac{\empRprice(\empprice)}{\Rprice(\empprice)}
    -\frac{\empRprice(\monopolyprice)}{r^{*}}
    \ge
    \frac{\revenueloss(\empprice)}
         {1-\revenueloss(\empprice)}
    \frac{\empQsale(\monopolyprice)}{\monopolyquantile}
    \ge\frac{\revenueloss(\empprice)}{2}~.
    \label{eq:MHR:sample:empirical-comparison}
\end{align}
We use empirical optimality and the identity
$\Rprice(\empprice)=(1-\revenueloss(\empprice))r^{*}$
in the first inequality.
We use the bound
$\empQsale(\monopolyprice)/\monopolyquantile\ge2/3$
in the second.
Thus, a substantial revenue loss requires a substantial difference
between the empirical relative errors.

We group the possible revenue losses using the thresholds
$\lossthreshold_j \eqdef 2^j\eps$, for integers $j\ge0$.
We consider an index $j$ with $\lossthreshold_j<1$.
On the event $\calE$, if the returned price satisfies
$\lossthreshold_j\le\revenueloss(\empprice)<2\lossthreshold_j$,
then Eqs.~\eqref{eq:MHR:sample:quadratic-loss}
and \eqref{eq:MHR:sample:empirical-comparison} imply
\begin{align*}
    \bigl|\Qsale(\empprice)-\monopolyquantile\bigr|
    \le2\sqrt{2\lossthreshold_j}~,
    \quad
    \frac{\empQsale(\empprice)}{\Qsale(\empprice)}
    -\frac{\empQsale(\monopolyprice)}{\monopolyquantile}
    \ge\frac{\lossthreshold_j}{2}~.
\end{align*}
We apply Eqn.~\eqref{eq:MHR:sample:local-concentration} with the radius
$d=\min\{1,2\sqrt{2\lossthreshold_j}\}$ and accuracy level
$t=\lossthreshold_j/4$.
Our parameter choices satisfy $d+t\le4\sqrt{\lossthreshold_j}$,
so we obtain
\begin{align}
    \prob{\calE\cap
        \{\lossthreshold_j\le\revenueloss(\empprice)
        <2\lossthreshold_j\}}
    \le
    C_0\exp\!\left(
        -\frac{c_0n(\lossthreshold_j/4)^2}
              {d+\lossthreshold_j/4}
    \right)
    \le
    C_0\exp\!\left(
        -\frac{c_0}{64}n\lossthreshold_j^{3/2}
    \right)~.
    \label{eq:MHR:sample:loss-range}
\end{align}

We finally sum these bounds over the possible loss ranges.
Since every loss in $[\eps,1)$ belongs to one of these ranges,
we have
\begin{align*}
    \Pr\!\left[
        \Rprice(\empprice)<(1-\eps)r^{*}
    \right]
    &\le
    \Pr[\calE^{c}]
    +\sum\nolimits_{\substack{j\ge0\\\lossthreshold_j<1}}
    \Pr\!\left[
        \calE\cap
        \{\lossthreshold_j\le\revenueloss(\empprice)
        <2\lossthreshold_j\}
    \right]\\
    &\le
    \frac{\delta}{2}
    +C_0\sum\nolimits_{j=0}^{\infty}
    \exp\!\left(
        -\frac{c_0}{64}n\eps^{3/2}2^{3j/2}
    \right)
    \le\delta~.
\end{align*}
We justify the last inequality using the bound $2^{3j/2}\ge j+1$.
The infinite sum is therefore at most
$\frac{
    \exp\!\left(-c_0n\eps^{3/2}/64\right)
}{
    1-\exp\!\left(-c_0n\eps^{3/2}/64\right)
}$.
Our choice of the sample size
$n=\lceil C\eps^{-3/2}\log(2/\delta)\rceil$,
with a sufficiently large universal constant $C$, makes this expression
multiplied by $C_0$ at most $\delta/2$.
In particular, the summation introduces no additional
$\log(1/\eps)$ factor.

We therefore obtain an $(\eps,\delta)$-PAC learning algorithm for MHR
distributions using $O(\eps^{-3/2}\log(2/\delta))$ samples.
This completes the proof.
\end{proof}

%% file: impossibility.tex
\newcommand{\querybudget}{T}
\newcommand{\probparam}{\eta}
\newcommand{\outputprice}{\price_{\learnalg}}
\newcommand{\simprices}{\calP_{0}}
\newcommand{\pricethreshold}{\tau}
\newcommand{\valuescale}{a}

\section{The Impossibility of Learning without Sample}
\label{sec:impossibility}

We show that pricing-query access alone cannot provide a nontrivial multiplicative revenue guarantee with any query budget (no access to a single sample). 
This impossibility result already holds for uniform distributions on arbitrarily small intervals, which satisfy the MHR condition. 
In particular, the result continues to hold when the learner knows that the value distribution is supported on $[0, 1]$.


\begin{theorem}
\label{thm:impossibility}
For every finite query budget $\querybudget\geq0$, every possibly randomized learning algorithm $\learnalg$ that uses no value samples and at most $\querybudget$ adaptive pricing queries, and every  $\probparam\in\left(0,1\right)$, there exists an absolutely continuous MHR distribution $\prior$, supported on $\left[0,1\right]$ and with optimal revenue $\optrevenue>0$, such that
\begin{align*}
    \prob[\prior,\learnalg]{\Rprice\left(\outputprice\right)=0}
    &>1-\probparam~.
\end{align*}
The same distribution satisfies
$\expect[\prior,\learnalg]{\Rprice\left(\outputprice\right)}
<\probparam\optrevenue$.
Consequently, for every $\eps,\failureprob\in\left(0,1\right)$, no algorithm using zero value samples and a finite query budget is $\left(\eps,\failureprob\right)$-PAC for the MHR, regular, quasi-MHR, or quasi-regular family.
\end{theorem}

\begin{proof}
We fix an algorithm $\learnalg$, a finite query budget $\querybudget$, and $\probparam\in\left(0,1\right)$. We first simulate the algorithm against the feedback rule $\queryfeedback^{\roundidx}=\indicator\left[\price^{\roundidx}=0\right]$, which corresponds to a buyer whose value is always zero. Every zero-price query results in a sale, whereas every strictly positive price results in no sale. The simulated execution therefore depends only on the algorithm's internal randomness.

We denote the set of all query prices and the final output price in the simulated execution by $\simprices$. This set contains at most $\querybudget+1$ elements. We define its smallest strictly positive price, capped at one, by
\begin{align*}
    \pricethreshold
    &\triangleq\min\nolimits\left(
        \left\{1\right\}\cup
        \left(\simprices\cap\left(0,\infty\right)\right)
    \right)~.
\end{align*}
The inclusion of $1$ ensures that the minimum is well-defined, and the finiteness of the set $\simprices$ implies that the threshold $\pricethreshold$ is strictly positive almost surely. The probability $\prob[\learnalg]{\pricethreshold>\valuescale}$ therefore converges to one as the scale $\valuescale$ decreases to zero. We can consequently choose a deterministic scale $\valuescale\in\left(0,1\right)$, independently of the realized internal randomness, such that
$\prob[\learnalg]{\pricethreshold>\valuescale}
>1-\probparam$.

We take $\prior$ to be the uniform distribution on $\left[0,\valuescale\right]$, which is an MHR distribution. 
We couple the actual execution under the distribution $\prior$ with the simulated execution through the same internal randomness. On the event $\left\{\pricethreshold>\valuescale\right\}$, every simulated query price is either zero or strictly greater than $\valuescale$. Each such query produces exactly the simulated feedback: a zero price always results in a sale, whereas a price greater than $\valuescale$ never does. Identical preceding histories and internal randomness therefore generate the same next query and feedback. Induction over the query rounds shows that the two executions, including their stopping decisions and final outputs, coincide on this event.

The common output price on this event is also either zero or strictly greater than $\valuescale$, and hence yields zero revenue. We therefore obtain
\begin{align*}
    \prob[\prior,\learnalg]{\Rprice\left(\outputprice\right)=0}
    &\geq\prob[\learnalg]{\pricethreshold>\valuescale}>1-\probparam~.
\end{align*}
Every output price earns revenue between zero and the optimal revenue $\optrevenue$. The expected revenue consequently satisfies
\begin{align*}
    \expect[\prior,\learnalg]{\Rprice\left(\outputprice\right)}
    &\leq\optrevenue\,\prob[\prior,\learnalg]{\Rprice\left(\outputprice\right)>0}
    <\probparam\optrevenue~.
\end{align*}

For the PAC impossibility, we fix accuracy and failure parameters $\eps,\failureprob\in\left(0,1\right)$ and choose the probability parameter $\probparam<1-\failureprob$. The approximation event requires strictly positive revenue because the optimal revenue $\optrevenue$ is positive and the accuracy parameter satisfies $\eps<1$. We thus have
\begin{align*}
    \prob[\prior,\learnalg]{\Rprice\left(\outputprice\right)\geq\left(1-\eps\right)\optrevenue}
    &\leq\prob[\prior,\learnalg]{\Rprice\left(\outputprice\right)>0}
    <\probparam<1-\failureprob~.
\end{align*}
This violates the $\left(\eps,\failureprob\right)$-PAC requirement for the MHR family. The impossibility extends to the regular, quasi-MHR, and quasi-regular families because each contains the MHR family.
\end{proof}